\documentclass[preprint2]{aastex}
\usepackage{lscape}
\usepackage{epsfig}
\usepackage{natbib}
\shorttitle{VSOP AGN Survey Results V.}
\shortauthors{Dodson et al.}

\newcommand{\spfig}[1]  {
\includegraphics{#1}
\vspace{45mm}
}

\newcommand{\etal}{et~al.}
\newcommand{\uv}{({\em u,v})}

\begin{document}
\title{The VSOP 5~GHz Active Galactic Nucleus Survey: V. \\ Imaging
Results for the Remaining 140 sources}

\author{R.~Dodson\altaffilmark{1,2},                
        E.~B.~Fomalont\altaffilmark{3},             
        K.~Wiik\altaffilmark{4,1},                  
        S.~Horiuchi\altaffilmark{5,6,7,8},            
        H.~Hirabayashi\altaffilmark{1,9},           
        P.~G.~Edwards\altaffilmark{10,1},            
        Y.~Murata\altaffilmark{1,9},                
        Y.~Asaki\altaffilmark{1,9},                 
        G.~A.~Moellenbrock\altaffilmark{1,11},      
        W.~K.~Scott\altaffilmark{12},               
        A.~R.~Taylor\altaffilmark{12}               
        L.~I.~Gurvits\altaffilmark{13},             
        Z.~Paragi\altaffilmark{13,14},              
        S.~Frey\altaffilmark{15,14},                
        Z.-Q.~Shen\altaffilmark{16,1},              
        J.~E.~J.~Lovell\altaffilmark{10,1,17},          
        S.~J.~Tingay\altaffilmark{6,7,18},               
        M.~J.~Rioja\altaffilmark{2},                
        S.~Fodor\altaffilmark{6,19},                
        M.~L.~Lister\altaffilmark{20},              
        L.~Mosoni\altaffilmark{21,15},              
        G.~Coldwell\altaffilmark{22},               
        B.~G.~Piner\altaffilmark{19},               
        J.~Yang\altaffilmark{13,16}                 
}

\altaffiltext{1}{Institute of Space and Astronautical Science, 
                 Japan Aerospace Exploration Agency,
                 3-1-1 Yoshinodai,
                 Sagamihara, Kanagawa 229-8510, Japan}

\altaffiltext{2}{Observatorio Astron\'omico Nacional, 
	          Apartado 112, E-28803, 
		  Alcala de Henares, Espa\~na}

\altaffiltext{3}{National Radio Astronomy Observatory, 520 Edgemont Road,
                 Charlottesville, VA 22903, USA}

\altaffiltext{4}{Tuorla Observatory,  University of Turku,
		 V\"ais\"al\"antie 20, FIN-21500 Piikki\"o, Finland}

\altaffiltext{5}{National Astronomical Observatory, 2-21-1 Osawa, 
                 Mitaka, Tokyo 181-8588, Japan}

\altaffiltext{6}{Jet Propulsion Laboratory, 4800 Oak Grove Drive,
                 Pasadena, CA 91109, USA}

\altaffiltext{7}{Centre for Astrophysics and Supercomputing,
                  Swinburne University of Technology,  P.O. Box 218, Hawthorn,
                  Vic. 3122, Australia}

\altaffiltext{8}{Canberra Deep Space Communication Complex,
                  PO Box 1035, Tuggeranong, ACT 2901, Australia}

\altaffiltext{9}{Department of Space and Astronautical Science,
                 School of Physical Sciences,
		 The Graduate University for Advanced Studies,
		 3-1-1 Yoshinodai, Sagamihara, Kanagawa 229-8510, Japan}

\altaffiltext{10}{Australia Telescope National Facility,
                 Commonwealth Scientific and Industrial Research Organization,
                 P. O. Box 76, Epping NSW 2122, Australia}

\altaffiltext{11}{National Radio Astronomy Observatory, 
                 P.O. Box 0, Socorro, NM 87801, USA}

\altaffiltext{12}{Physics and Astronomy Department, University of Calgary,
                 2500 University Dr. NW,
                 Calgary, Alberta, Canada, T2N 1N4}

\altaffiltext{13}{Joint Institute for VLBI in Europe, P.O. Box 2,
                 7990 AA, Dwingeloo, Netherlands} 

\altaffiltext{14}{MTA Research Group for Physical Geodesy and
                  Geodynamics, P.O. Box 91,
                  H-1521 Budapest, Hungary}

\altaffiltext{15}{F\"{O}MI Satellite Geodetic Observatory, P.O. Box 585,
                  H-1592, Budapest, Hungary}

\altaffiltext{16}{Shanghai Astronomical Observatory, 
                  Chinese Academy of Sciences, 80 Nandan Lu, 
                  Shanghai 200030, China} 

\altaffiltext{17}{School of Mathematics and Physics, 
                  University of Tasmania, Private Bag 21, 
                  Hobart, Australia}

\altaffiltext{18}{Department of Imaging and Applied Physics,
                  Curtin University of Technology, Perth, Australia.} 

\altaffiltext{19}{Physics Department, Whittier College,
                  13406 East Philadelphia, P.O. Box 634, Whittier,
                  CA 90608-4413, USA} 

\altaffiltext{20}{Department of Physics,
                  Purdue University,
                  West Lafayette, IN 47907, USA}

\altaffiltext{21}{MTA Konkoly Observatory,
                 P.O. Box 67, H-1525, Budapest, Hungary}

\altaffiltext{22}{Observatorio Astron\'omico de la 
                  Universidad Nacional de C\'{o}rdoba, 
                  Argentina }


\begin{abstract}
In February 1997, the Japanese radio astronomy satellite HALCA was
launched to provide the space-bourne element for the VLBI Space
Observatory Programme (VSOP) mission. Approximately twenty-five
percent of the mission time was dedicated to the VSOP Survey of bright
compact Active Galactic Nuclei (AGN) at 5~GHz.
This paper, the fifth in the series, presents images and models for
the remaining 140 sources not included in Paper III, which contained 102
sources.  For most sources, the plots of the \uv\ coverage, the
visibility amplitude versus \uv\ distance, and the high resolution
image are presented.  Model fit parameters to the major radio
components are determined, and the brightness temperature of the core
component for each source is calculated.  The brightness temperature
distributions for all of the sources in the VSOP AGN survey are
discussed.

\end{abstract}
\keywords{galaxies: active ---
          radio continuum: galaxies --- 
          surveys}
\section{Introduction}

The radio astronomy satellite HALCA (Highly Advanced Laboratory for
Communications and Astronomy) was launched by the Institute of Space
and Astronautical Science in February 1997 to participate in Very Long
Baseline Interferometry (VLBI) observations with arrays of ground
radio telescopes.  HALCA provides the longest baselines of the VSOP,
an international endeavor that has involved over 28 ground radio
telescopes, five tracking stations and three correlators
\citep{hir98,hir00a}.  HALCA was placed in an orbit with an apogee
height above the Earth's surface of 21,400\,km, a perigee height of
560\,km, and an orbital period of 6.3~hours.

During the seven years of HALCA's mission lifetime, about 75\% of
observing time was used for projects selected by international
peer-review from open proposals submitted by the astronomical
community in response to Announcements of Opportunity.  This part of
the mission's scientific programme constituted the General Observing
Time (GOT). The remaining observing time was devoted to a mission-led
survey of active galactic nuclei at 5\,GHz: the VSOP Survey Program.
The major goal of the Survey was to determine the statistical
properties of the sub-milliarcsecond structure of a complete sample of
AGNs.  \citep{hir00b, fom00a}. 
Following the end of the formal international mission period in
February 2002, the Japanese-dominated effort continued survey
observations until October 2003, when HALCA lost its attitude control
capability. This occurred well after the end of the original planned
mission lifetime.

This paper is the fifth in the series of VSOP Survey related papers.
\citet{sco04} (henceforth P-III) contains the results for 102 sources
which were observed and reduced before 2001 October. \citet{hor04}
(henceforth P-IV) analyzed the cumulative visibilities of those
sources to obtain the `typical source structure'. This paper contains
the additional 140 survey sources which were successfully observed by
VSOP and
completes the survey programme observing results.  The brightness
temperature properties of the entire sample of sources are discussed.

\section{The Observations}


The VSOP mission and the 5 GHz AGN Survey are
fully discussed in \citet{hir98,fom00b,hir00a,hir00b}.
Briefly, in order to be included in the VSOP Survey, a
source was required to have:
\newline $\bullet$ a total flux density at 5 GHz, $S_5\geq
5.0$~Jy\newline \centerline{or\phantom{aaaaaaaaaaaaaaaaaaaaaaa}}
$\bullet$ a total flux density at 5 GHz, $S_5\geq 0.95$~Jy and\newline
$\bullet$ a spectral index $\alpha \geq -0.45$ ($S \propto
\nu^\alpha$) and\newline $\bullet$ a galactic latitude $|b|\geq
10^\circ$.

The finding surveys from which sources were selected were primarily
the Green Bank GB6 Catalog for the northern sky \citep{gre96}, and the
Parkes-MIT-NRAO (PMN) Survey \citep{law86,gri93} for the southern sky.
The 402 sources satisfying these criteria comprise the VSOP source
list \citep{hir00b}.  

As this source list was compiled from single dish catalogues, some
of the selected sources would not be detectable by HALCA due to
insufficient correlated flux density on baselines longer than about 1000
km.  Therefore, sources with declination $>-44^\circ$ were observed in
a VLBA pre-launch survey (VLBApls, 
Fomalont et al. 2000b) and a cutoff criterion, a minimum flux density
of $0.32$\,Jy at 140~M$\lambda$, was established for inclusion of a
source in the VSOP Survey \citep{fom00a}. For sources south of
$-44^\circ$ this cutoff could not be determined, so all sources were
included for HALCA observations.  We find that 14 of these 24 southern
sources observed have no detectable flux density on baselines to
HALCA. Of the 402 sources in the complete sample, 294 were selected
for VSOP observations, and this sample is designated as the VSOP
Source Sample (VSS) \citep{hir00b,edw02}.



Observations of the VSS began in August 1997, with the final
observations being made in October 2003 
{when a second of the four HALCA momentum wheels became
non-functional. Despite heroic attempts to heat up and free this
reaction wheel through out 2004 no further observations were
possible}.
Of the VSS sample of 294 sources all but 29 were observed. Details of
the final status, the latest values of total density flux at 5~GHz,
the redshift, relevant references, best fit (or lower limit) observer
frame brightness temperatures of the core, detected area, and flux
density on the longest baselines, can be found in
Table~1. 

A typical VSOP Survey observation used $\sim$3 ground telescopes and
HALCA, co-observing for up to $\sim$6\,hours.  Ground radio telescopes
that made the largest contributions to the Survey Program observations
include the VLBA (USA), Mopra (Australia), Hartebeesthoek (South
Africa), Sheshan (China), Hobart (Australia), Kashima (Japan), Usuda
(Japan), Ceduna (Australia), Kalyazin (Russia), Noto (Italy).  Other
participants were the Green Bank 43\,m (USA), ATCA (Australia),
Effelsberg (Germany), Arecibo (Puerto Rico), Torun (Poland), Onsala
(Sweden), VLA (USA), Jodrel Bank Mk2 (UK) and Medicina (Italy). Further
details are available in \citet{evn_survey}.

The VSOP survey observations were made at 5 GHz, with two
left-circularly polarized 16\,MHz IF bandwidths, sampled with two bits
\citep{hir00a}. 
GOT observations of survey sources which were made with a similar
configuration, were also included (see P-III for discussion of this).
Data were usually correlated at either the DRAO Penticton
correlator \citep{car99} or the NAOJ Mitaka correlator \citep{shi98}, with one
non-GOT experiment processed at the Socorro correlator \citep{nap94} 
along with two dozen GOT extractions 
\citep[see][for details]{hir00b,sco04}
and a test experiment
in the data presented here. After correlation, the data were sent to
ISAS for distribution to the Survey Reduction Team members. The subset
of those members who contributed the reductions that appear in this
paper are represented in the author list. 


\section{Data Reduction}

Analysis of the data has been described elsewhere \citep{lov04}
and hence will only be briefly outlined here.  The data were imported
into AIPS, amplitude calibrated (with the measured or expected system
temperature and, if needed, the autocorrelation normalised) then
fringe fitted.  A check of the amplitude calibration of the ground
telescopes was made by observing a nearby calibrator source for about
five minutes during the experiment.  Because HALCA could not slew
quickly between sources, its amplitude calibration could not be checked
using a bright astronomical source. 
It was found, however, to be quite stable. 
%
Nevertheless, the amplitude calibration was entirely derived from the
measured or expected gains for both HALCA and a number of ground
antennae, which can be very uncertain.
In P-III, great effort was spent in measuring an absolute correction
for the flux scale of the survey, in comparison with that of the
VLBApls. In the dataset presented here, we have fewer GOTs and a
smaller overlap with the VLBApls, so such an approach was not
possible. Therefore we have assumed that the same correction (that the
survey experiments underestimate the flux densities by a factor of
$0.83\pm0.05$, compared to the VLBApls) can be made for these
experiments. The amplitude scale errors were estimated as 20\%, via
the same comparison.
After satisfactory delay and rate calibration, the data for all
spectral channels were summed to a single channel per 16~MHz sub-band
(i.e. two) and exported to DIFMAP \citep{she97} for self calibration
and model fitting. Scripts were used as much as possible to ensure
that the methods were standardized.

The results of, and supporting documentation for, the data reduction
can be found on the project web site
(http://www.vsop.isas.jaxa.jp/survey). The raw and calibrated data are
available from ISAS on request.

Most VSS sources have been imaged with previous ground VLBI
observations (including the VLBApls, which was specifically designed
to add the lower resolution data to the VSOP Survey), and consistency
of the VSOP image with these other images was used to constrain the
cleaning and modeling.  Where no supporting information was available,
the models are relatively conservative.

For the entire VSOP survey programme, 265 of the 294 sources were
observed. The observations that are reported in this paper are listed
in Table~2, which includes source names, experiment code, Ground Radio
Telescopes, Tracking stations and Correlator used, time over which
fringes were detected and the optical ID and redshift.
For fifty of the observed sources, fringes to the spacecraft were not
detected.  Many of the sources were significantly resolved on
shorter ground-only baselines, so that the lack of space fringes (RMS
detection is typically 0.1 Jy) is consistent with the resolving
structure seen on shorter baselines.  However, for twenty three of the
observed sources where space fringes were not detected, ground
observations suggested that the space baselines (typically greater
than 150 M$\lambda$) should have been detected.  These observations
are not included in the table of results, since the ground-only data
provides no additional information about the source structure than is
published elsewhere.

\section{The Results}
\subsection{The \uv\ Coverage, Visibility Amplitudes and Images}

The graphical results for most of the 140 additional survey sources
are given in Fig.~1, 
which shows the \uv\ coverage, the visibility amplitude versus
projected \uv\ distance, and the image displayed in contour form, with
one row per source. The \uv\ distance is plotted in M$\lambda$, the
flux density in janskys, and the image in milliarcseconds (mas).  The
J2000 name of the source is listed at the top left of each row,
followed by the experiment name and the observation date. The peak
brightness (P) in janskys per beam appears above the image plot,
followed by the noise level ($\sigma$), and the beam (B) major and
minor axes and the major axis position angle.
The image $\sigma$ is estimated from a Gaussian fit to the pixel flux
density distribution, but in a few cases had to be manually increased due to
high sidelobes in the image. The lowest contours are at $-$3 (dashed line) and
3 (solid line) times $\sigma$,
doubling thereafter.
The images presented are all made with uniform weighting, which
highlights the highest resolution structure. The weighting scheme for
each source was selected to give the clearest image. Further details are to
be found on the VSOP Survey web site. 
The \uv\ coverage among the sources varies considerably, and this had
significant impact on the quality, resolution and dynamic range of the
images.  For sources which were so heavily resolved that no space
fringes are detected and with limited ground baseline coverage, no
graphical results are shown, although the indication of overall
angular size is still presented in Table~3. Sources for which no
conclusions could be drawn (J1218$-$4600, J1424$-$4913 and J2358$-$1020)
are not included.


During deconvolution we cross-checked our images with any other
ground-based images of the source available, usually the VLBA
Pre-Launch Survey \citep{fom00b}.  Although most AGNs are variable
with time and frequency, these other images provided reasonable
constraints to the VSOP source images, for example, the source extent
and component locations.  The space baselines have considerably higher
noise compared to those on the ground, because of the higher system
temperature and smaller size of the HALCA antenna. Therefore the space
data were up-weighted to approximately the same significance as the
ground data, in order to emphasize the highest resolution structure in
the source. This was typically a factor of ten to twenty.

The VSOP data indicated the strength and approximate angular size of a
core component, even if, in some cases, most of the most extended
emission, shown with lower-resolution images, was resolved out in the
VSOP data.  In general, the Survey datasets have a typical image
fidelity of 20:1, and features less than 5\% of the peak brightness
should be treated with caution.


\subsection{Model-Fitting and Brightness Temperature Determination}

Once we obtained the best image for a source, consistent with the
quality and quantity of the \uv\ coverage, we estimated the parameters
for the major components of the source structure by Gaussian fitting:
the integrated flux density, the centroid position and the major, minor axis
and orientation of the major axis.  These parameters are listed in
Table~3.
In all cases, the radio core could be identified (it
was usually the most compact component and often at one edge of the
radio emission).  Occasionally, high resolution ground images at 15
and 23 GHz were also used to identify the location of the core.  The
radio core component is listed as the first component for sources
which contain several components.

The brightness temperature of a component in the observer's frame is
given by: 
\begin{equation} T_b=\frac{S\lambda^2}{2 k_b \Omega}
\label{eqtb} 
\end{equation} 
where $S$ is the component flux density
at wavelength $\lambda$, $k_b$ is Boltzmann's constant, and
$\Omega\approx 1.13 \cdot\rm{(\theta_{maj})(\theta_{min})}$ is the (Gaussian)
weighted solid angle
subtended by the component (which we have expressed in terms of the
full width at half maximum of the component major and minor axes).  To
convert to brightness temperature in the source frame, equation
(\ref{eqtb}) is multiplied by $(1+z)$, where $z$ is the source redshift.

For the radio core only, we carefully determined the best-fit angular
size and its allowable range. Although several quasi-analytical
methods are available (eg. Difmap: \citet{she97}, DIFWRAP:
\citet{lov00}), we relied on the ad-hoc method of varying critical
parameters and estimating the range of angular size and flux density
for each core component that is consistent with the data and its
scatter (cf \citet{lov00}).  
Two brightness temperatures are given for the radio cores in
Table~3.
First is the measured brightness temperature in the observer's frame,
assuming that the core is a Gaussian-shaped component of the specified
parameters.  Sources for which only lower limits could be derived
(i.e. unresolved sources) are marked with a $>$. The second brightness
temperature is the lower limit of the brightness temperature (using
the maximum angular size) converted to the {\it source frame}.  If no
redshift is available, zero redshift is assumed.  These values are the
lowest possible temperature compatible with the data. We list these as
we feel that these are a more useful value than upper limits or best fits which
depend critically on the interpretation of the very highest resolution
data, which have the highest noise and the most sparse coverage.

\subsection {Discussion of the Brightness Temperature Distributions}

Histograms depicting the brightness temperature distribution in both
the source and observer's frame for the sources presented here are
shown in Fig.~2. 
Here we have used the values for the best fits to the models
(following the style of P-III), rather than using the lower limits,
in order to combine both datasets. (Sources with no measured redshift
are not included in the plot of source-frame brightness temperatures.)

Most cores have $T_b>10^{11}$~K, with approximately 56\% of the
sources having a measured brightness temperature in excess of
$10^{12}$~K in the source frame, and approximately 30\% of the sources
having a measured brightness temperature greater than $10^{12}$~K in
the observer's frame.

We find that overall, the mean brightness temperatures in our data are
slightly lower than
those in P-III. This could be expected, given that our data include
many more of the weaker sources (the median total flux density of sources in
this paper is less than half of those in P-III),
and a significant number with no space fringes, none of which were
included in P-III. 


Comparing this result with other similar datasets \citep{tin01,kov05}
we note that, as discussed in P-III, the distribution in \citet{tin01}
matches the one presented here once corrected by the factor
$(1+z)^{1/2}/0.56$. This factor comes from the corrections for
changing their results from an optically thick core model to a
Gaussian, and from co-moving frame to the source frame. The
distribution presented in \citet{kov05} is from VLBA observations at
15~GHz.
They find also a median value of $10^{12}$K, but the distribution
towards $10^{13}$ and beyond is largely made up of lower limits,
rather than actual measurements as we have here. We have compared the
$T_b$ for the source in common with \citet{kov05}
by selecting the data with the closest observation dates. 
The $T_b$ in the VSS tend to be higher, 
as expected since the majority of the
brightness temperatures in \citet{kov05} are lower limits,
with a median ratio of 2.4. 
Detailed
comparison of individual sources, in particular those with very
different $T_b$, will be presented in a future paper \citep{VLBA2cm2}. 


When looking at individual sources, we believe that the most useful
number that can be provided is the lower limit to the brightness
temperatures, which is that which {\em must} be generated by any
proposed model or theory, rather than the brightest possible which
{\em could} be required by any proposed model. The former will not
produce the highest measured temperatures, nor the most extended
distribution. However, it will provide limits and distributions which
must be achieved or exceeded. To probe the distribution of brightness
temperatures, we prefer the approach taken in P-IV, where a cumulative
visibility distribution was produced from all sources and a
measurement of typical core sizes was fitted to these data. By this approach
the very high errors at the extremes have reduced contributions.

\subsection{Comments on Individual Sources}

Survey sources that were not successfully scheduled are included in Table~1
with references (where they exist) to other VLBI
observations marked with a $\dagger$ on the experiment name.

The complete list of VSOP Survey observations is presented in Table~2. 
In addition, short notes on the sources are given.  A general
comparison is made with VLBI images from other observations, primarily
the VLBApls \citep{fom00b}, U. S. Naval Observatory Database (USNO)
\citep{fey96, fey97, fey00}, a space VLBI Survey of Pearson-Readhead
sources (VSOPPR) \citep{lis01}, the VLBA 2~cm Survey (VLBA2cm1)
\citep{kel98, zen02, kel04, kov05}, MOJAVE \citep{lis05} and { VLBA2cm2
\citep{VLBA2cm2}}, results from VSOP observations of southern sources
(VSOPsth) \citep{tin02}, and the VLBA Calibrator Survey (VCS)
\citep{vcs}. Any significant differences are noted. The largest
linear extent, or upper limit, is given.



\noindent
  J0013$+$4051 --- The core is $<0.2$\,mas in size, with extended emission
to the north-west as observed in other VLBA images. \\
%
  J0105$+$4819 --- The core is 0.2\,mas in size.
The more extended emission in other VLBA images that
surrounds the core is undetected with the VSOP data. \\
%
  J0108$+$0135 --- The core is $0.2$\,mas in size and the location of the
extended component is in agreement with other VLBA images. \\  
%
  J0116$-$1136 --- Not detected on space baselines.  The VSOP
ground-only image and other VLBA images give a core size of about 0.5
mas in size. \\
%
  J0121$+$0422 --- The core is 0.4\,mas in size.  There is faint radio
emission to the east, also seen with 15 GHz VLBA observations
\citep{kel04} \\   
%
  J0125$-$0005 --- The core is 0.3\,mas in size and there is extended emission
to the west. \\  
%
  J0141$-$0928 ---  Redshifts of both 0.733 and 0.44 have been
reported \citep{sto97}.  The core is 0.4\,mas in size and nearly
circular \citep{vcs} \\ 
%
  J0149$+$0555 --- The core is 0.4\,mas in size and the extended emission is similar to
that seen with other VLBA images \citep{vcs}\\  
%
  J0152$+$2207 --- The core is 0.4\,mas in size and extended to the north.  A
faint extended component to the north, seen with other VLBA images,
is present. \citep{fom00b, kel04} \\  
%
  J0204$+$1514 --- The core is $<0.2$\,mas in size.  The faint emission to the
south-east is in the opposite direction of most of the jet emission,
but the source structure evolution is complicated.\\
%
  J0204$-$1701 --- The core is 0.6\,mas in size and there is an extended
component to the south.\\   
%
  J0217$+$0144 --- The core is unresolved, $<0.1$\,mas in size.  The faint
extended structure from lower resolution VLBA images is not detected. \\  
%
  J0224$+$0659 --- The core is 0.3\,mas in size.  Emission to the
west is also detected. \\  
%
  J0231$+$1322 --- The core is about 0.4\,mas in size.  Faint emission to the
north-east is also detected. \\  
%
  J0237$+$2848 --- The core is $<0.2$\,mas in size.  Very extended
  emission to the north is detected. \\  
%
  J0239$+$0416 --- The core is $<0.2$\,mas in size.  Extended emission to the
north-west is detected.\\  
%
  J0242$+$1101 --- The core is $<0.2$\,mas in size.  The emission to
  the south-east is clearly extended.\\
%
  J0253$-$5441 --- The core is 0.3\,mas in size, in agreement with
  \cite{ojh05}.  The fainter component to the west is too weak to be
  detected by VSOP.\\   
%
  J0259$+$0747 --- The core is 0.8\,mas in size.  The faint emission 3\,mas to
  the south, seen in VLBA images, is not detected. \\  
%
  J0303$-$6211 --- The core is 0.6\,mas in size, extended east-west.
There is a faint component 1.5\,mas to the east. \\
%
%
  J0309$-$6058 --- Two small-diameter components, separated by 0.5\,mas,
 are detected.  The component to the north-east has the higher brightness
temperature and is assumed to be the core. \\  
%
  J0312$+$0133 --- The core is $<0.2$\,mas in size.  Faint emission to
  the east, seen with the VLBA, is just detected with VSOP. \\  
%
  J0321$+$1221 --- The core is $<0.15$\,mas in size.  The extended
component seen by VSOP is the inner part of a 20\,mas jet seen by the VLBApls.\\   
%
  J0336$+$3218 --- The core is 0.4\,mas in size.  Because of the lack
  of short spacings, none of the extended structure is detected by VSOP. \\  
%
  J0339$-$0146 --- The core is 0.2\,mas in size.  The extended emission
  to the north-east is seen with other VLBA images. \\  
%
  J0359$+$5057 --- The core is about 1.0\,mas in size.  The extended
  emission to the north-east, which extends 20\,mas from the core in
  the VLBA images, is just detected by VSOP.\\   
%
  J0402$-$3147 --- The core is 0.3\,mas in size.  The faint emission to
  the west is seen by the VLBApls. \\
%
  J0403$+$2600 --- The core is 0.35\,mas in size. The extended emission
  to the west seen by the VLBApls is detected with VSOP.  The model does
  not include the emission 15\,mas to the north. \\   
%
  J0406$-$3826 ---  The core is $<0.25$\,mas in size.  The faint
  component to the west is found in the VLBApls, although the VCS image at
8 GHz has the fainter component to the east.  This source is one of the
most extreme of the Intra Day Variable sources \citep{ked_97}. \\
%
  J0414$+$0534 --- A gravitationally lensed object with two major
  emission centers separated by 400\,mas \citep{fom00b}.  No image is shown, but
  the model fit of the Shanghai to Kashima baseline (20 M$\lambda$)
  suggests a size of 1.6\,mas for the more compact (probably southern)
  component. \\ 
%
  J0416$-$1851 --- The core is 0.5\,mas in size.  However, the flux
  density is about a factor 5 to 10 less than seen by other VLBI
  observations.\\   
%
  J0424$+$0036 --- The core is 0.2\,mas in size.  The faint emission to
  the north is associated with more extensive emission seen in other
  VLBA images. \\   
%
  J0424$-$3756 --- The core is 0.5\,mas in size.  The faint, extended
  component to the east is observed in other VLBA images. \\  
%
  J0428$-$3756 --- The core is $<0.5$\,mas in size.  The faint,
  extended component to the east is observed in the VLBApls. \\  
%
  J0433$+$0521 --- 
The VSOP observations suggest that the core is $<0.3$
mas in size. Only some of the extended structure to the west is imaged
with VSOP.\\
%
  J0437$-$1844 --- The core is $0.7$\,mas in size and the extended
emission to the north-west is observed with other VLBA images.  The
space baselines are short, so the angular resolution is relatively low.\\  
%
  J0442$-$0017 --- Not detected on space baselines and no image is
shown.  The approximate size is 1.1\,mas.  \\
%
  J0449$+$1121 --- The VSOP data are consistent with a core $<0.2$\,mas in size and extended
emission to the east.\\  
%
  J0450$-$8101 --- The core is $<0.2$\,mas in size.  There is extended emission
on both sides of the core. \\
%
  J0508$+$8432 --- A redshift of 0.112 was initially reported for
this object, but subsequently an absorption system at 1.34 places
a lower limit on the redshift \citep{sto97}. The core is $0.35$\,mas in size.\\
%
  J0509$+$0541 --- The core is $<0.4$\,mas in size.  The extended
  emission to the south and east is seen in the VLBApls. \\  
%
  J0513$-$2159 --- The core is $1.1$\,mas in size with no detection on space
  baselines.  No image is given. \\  
%
  J0522$-$3627 --- The brightest component is identified as the core
and is 0.4\,mas in size.  Most of the extended emission is to the
north-west, with fainter emission to the south-east which is identified as the
core in \citet{tin02b}.  This component is $<0.2$\,mas in size. \\
%
  J0530$+$1331 --- The space baselines suggest that the core
  component lies at the extreme south-west of the emission, and contains
  about 0.22\,Jy within a size $<0.3$\,mas.  Only a small part of the
  extended emission is contained in the other two model components. \\  
%
  J0541$-$0541 --- The core is $1.2$\,mas in size, with no detection on space
  baselines.  No image is given. \\   
%
  J0607$+$6720 --- The emission is about 0.6\,mas in size.  It is best
  fit by two small components each with a size $<0.2$\,mas. \\  
%
  J0614$+$6046 --- The core is 0.25\,mas in size.\\  
%
  J0626$+$8202 --- The core is 0.3\,mas in size.  The extended emission
  is complex and approximated by the additional two components. This
  source shows very interesting sub-mas structure.\\  
%
  J0646$+$4451 --- The core is identified with the fainter component
  to the west, based on the extended structure from VLBA images.  The
  core is $<0.2$\,mas in size.\\   
%
  J0648$-$3044 --- The core is $0.4$\,mas in size.  There is extended
  emission to the east, in agreement with other VLBA images. \\  
%
  J0713$+$4349 --- The space baselines suggest a core component of
  0.20\,Jy with a size $<0.2$\,mas.  No image is given since there is
  little ground baseline data. \\   
%
  J0739$+$0137 --- The core is 0.2\,mas in size.  There is extended emission to
  the north-west. \\   
%
  J0743$-$6726 --- The core is 0.7\,mas in size.  There is a slightly
  extended component 29\,mas to the east.  This structure agrees with
  \citet{ojh05}. \\   
%
  J0745$+$1011 --- The core is 0.4\,mas in size.  There is emission
  both north and south of the core.  The VLBA images show that the
  source structure is variable and changes with frequency. \\  
%
  J0750$+$1231 --- The core is 0.6\,mas in size.  There is extended emission to the
east. \\  
%
  J0808$+$4950 --- The core is $0.15$\,mas in size.  Most of the
  extended emission is to the south-east, as seen in other VLBA images. \\  
%
  J0820$-$1258 --- The core is $<0.3$\,mas in size.  There is extended emission to the east. \\  
%
  J0825$+$0309 --- The core is $<0.1$\,mas in size.  There is limited
  data so the faint extended emission to the north, seen in other VLBA
  images, is not present. \\   
%
  J0831$+$0429 --- The core is 0.5\,mas in size.  There is extended emission to
the east.\\   
%
  J0842$+$1835 --- The southern, fainter component is the core with an
angular size $0.3$\,mas.  The northern component is only the inner part
of the jet emission which extends over 10\,mas from the core.\\  
%
  J0854$+$5757 --- The core is $0.3$\,mas in size.  Most of the
extended emission is south of the core, with a hint of emission to the
north.\\   
%
  J0909$+$4253 --- The core is $<0.15$\,mas in size.  There is extended emission
to the south. \\  
%
  J0921$-$2618 --- Not detected on space baselines and no image is
given.  The approximate angular size of the emission is 0.9\,mas in
position angle $-17^\circ$, which agrees with the VLBApls results. \\  
%
  J0948$+$4039 --- The core is $0.15$\,mas in size.  There is extended emission
  to the south-east. \\   
%
  J0956$+$2515 --- The core is $<0.2$\,mas in size.  The extended emission,
  some of which overlaps the core, lies to the west. \\  
  J0958$+$4725 --- The core is 0.3\,mas in size.  There is no
  significant extended emission.\\   
%
  J0958$+$6533 --- The core is $<0.2$\,mas in size.  There is extended
  emission to the north-west which agrees with \citet{vcs}. \\
%
  J1014$+$2301 --- No detection on space baselines.  No image is
  shown and the size of the emission of 0.8\,mas.\\  
%
  J1035$-$2011 --- The core is $>0.15$\,mas in size.  The faint
  extended emission is not detected in these observations.  \\  
%
  J1041$+$0610 --- The core is 0.5\,mas in size.  There is extended
  emission to the south-east. \\  
%
  J1048$-$1909 --- The core is $<0.1$\,mas in size.  There is extended
  emission to the south.\\  
%
  J1051$+$2119 --- The core is $0.2$\,mas in size.  There is extended emission
  to the east. \\   
%
  J1051$-$3138 --- The core is $<0.2$\,mas in size.  There is extended
  emission near and south-west of the core. \\  
%
  J1058$+$1951 --- The source contains 0.07\,Jy at 180 M$\lambda$
  north-south, which is consistent with the VLBApls, with 0.4\,Jy in a
  0.9\,mas north-south component.\\
%
  J1058$-$8003 --- The core is 0.4\,mas in size.  There is no
  indication of extended emission. \\
%
  J1118$-$4634 --- The core of 0.27\,Jy is 0.7\,mas in size.  The total
  flux density is about 1.0\,Jy \citep {tin03}, hence there may be
  additional extended emission.\\
%
  J1125$+$2610 --- The core is 0.8\,mas in size, elongated in the
direction of the more extended emission seen by the VLBApls.\\
%
  J1127$-$1857 --- The core is $<0.2$\,mas in size.  There is extended emission
  to the south.\\
%
  J1130$-$1449 --- The core is $<0.2$\,mas in size.  There is extended
emission to the east. See \citet{tin02}.\\
%
  J1150$-$0023 --- No detection on space baselines and no image is given.
The core region contains about 0.13\,Jy in a component $<0.6$\,mas.  See
VLBApls for a model of the extended emission.\\
%
  J1153$+$4931 --- No detection on space baselines and no image is given.
The core emission is about 2.5\,mas in the north-south direction. This
does not agree with the VLBApls data. \\
%
  J1153$+$8058 --- The core is $0.3$\,mas in size.  There is faint
emission south of the core. \\
%
  J1159$+$2914 --- The core is 0.3\,mas in size.  There is extended
emission to the north-east. \\
%
  J1218$-$4600 --- No detectable flux density on ground baselines greater
than $\sim 100$\,M$\lambda$.  Hence, there is no image and model. \\
%
  J1224$+$0330 --- The core is $<0.25$\,mas in size.  There is extended
emission to the west. \\
%
  J1224$+$2122 --- The core is $<0.4$\,mas in size.  There is extended
emission to the north.\\
%
  J1257$-$3155 --- The core is $<0.3$\,mas in size in the position
angle of the extended emission seen by the VLBApls.  There is a possible
amplitude scaling error, so the core flux density may be a factor of
two higher. \\
%
  J1305$-$1033 --- The core is $<0.2$\,mas in size.  No additional
structure is seen with VSOP because of lack of short spacings. \\
%
  J1316$-$3338 --- The core is $0.3$\,mas in size.  There is extended
structure to the south-west and west.\\
%
  J1351$-$1449 --- Not detected on space baselines. No image is
given.  The core size is 1.0\,mas. \\
%
  J1357$+$1919 --- We believe that the northern most component is the
core with a size of 0.7\,mas.  The component 2\,mas south-east has a
higher brightness temperature, but is probably a bright spot of the
jet.  More extended emission occurs further to the south-east. \\
%
  J1405$+$0415 --- The structure seen in the image with the full VSOP
GOT 5-GHz data \citep{yan06} is consistent with our result: the jet is
extended $\sim$15\,mas to the West. \\
%
  J1415$+$1320 --- The core is $0.4$\,mas in size.  There is extended
emission at the same position as the core, and also to the
north-west. \\
%
  J1424$-$4913 --- No detectable flux density on global baselines, greater
than $\sim 140$\,M$\lambda$. \\
%
  J1427$-$4206 --- The core is 0.25\,mas in size.  There is extended
emission to the north and east \citep{tin02}. \\
%
  J1436$+$6336 --- The core is $<0.25$\,mas in size.  There is a
stronger component north of the core, and weaker emission to the south
(VLBApls).\\
%
  J1504$+$1029 --- The core is $<0.2$\,mas in size.  There is extended
emission to the east.\\
%
  J1516$+$0015 --- The core is $<0.2$\,mas in size.  Only space
baselines are available, hence none of the extended structure (to the
north-west) is seen. \\
%
  J1522$-$2730 --- The core is $0.25$\,mas in size.  There is extended
structure to the west. \\
%
  J1550$+$0527 --- The core is $<0.1$\,mas in size.  There is extended
structure to the north.  \\
%
  J1557$-$0001 --- The core is $0.20$\,mas in size.  No other
significant extended structure is seen with VSOP. \\
%
  J1602$+$3326 --- The core is $<0.2$\,mas in size.  There is extended
emission somewhat south-east of the core, but it is poorly defined
because of the lack of short spacings.\\
%
  J1608$+$1029 --- The core is 0.5\,mas in size.  There is extended
emission to the north-west. \\
%
  J1625$-$2527 --- The core is 0.5\,mas in size.  There is no
significant extended structure. \\
%
  J1647$-$6437 --- The core is 0.9\,mas in size.  There is little \uv
coverage, but no significant extended structure is seen. \\
%
  J1658$+$4737 --- Only space baselines.  The core is 0.3\,mas in size.
The large scale emission to the north is not seen. \\
%
  J1726$-$6427 --- There are no detections on space baselines, and no image is
given.  The Hartebeesthoek--Hobart baseline (145 M$\lambda$) detects only 0.04
Jy. \\
%
  J1743$-$0350 --- The core is $<0.15$\,mas in size and contains about 20\% of
the emission.  The extended emission is somewhat south and east of the
core. \\
%
  J1744$-$5144 --- There are no detections on space baselines,  and no image is given.
The Hartebeesthoek--Hobart baseline (145 M$\lambda$) detects only 0.07\,Jy. \\  
%
  J1753$+$4409 --- There are only space baselines and no image is given.
The core is $<0.15$\,mas in size.\\
%
  J1809$-$4552 --- The core is 0.5\,mas in size and extended in PA
  $-64^\circ$. \\
%
  J1824$+$1044 --- The core is 0.15\,mas in size and there is some extended
structure to the north. \\
%
  J1837$-$7108 --- The core is 0.2\,mas in size and there is extended
structure. We model this as to the south, however there is some
ambiguity as to whether it is actually to the south or the north.  \\
%
  J1842$+$7946 --- There are no detections on space baselines.  The
image and model show a core $0.3$\,mas in size and extended emission to
the north-west up to 5\,mas away. \\
%
  J1911$-$2006 --- The core is $<0.2$\,mas in size.  The flux density in the
extended emission is higher and closer to the core than that in the
VLBApls. \\
%
  J1912$-$8010 --- There are no detections on space baselines and no
image is given.  The core is 0.6\,mas in size, in general agreement
with \citet{ojh05}. \\
%
  J1925$+$2106 --- The core is 0.5\,mas in size.  There is extended
emission to the west.  \\
%
  J1927$+$7358 --- \citet{lis01} show the complete VSOP image.  With
more limited data, we assume that the core is the component at the
northern edge of the source (component D in Lister \etal ). Its flux density
is 0.4\,Jy with an angular size $<0.4$\,mas. Only the two other
components just south of the core are listed in our table. The extended
emission to the south is not included.  \\
%
  J1932$-$4546 --- There are no detections on space baselines and no
image is given.  Hartebeesthoek--Hobart at 140 M$\lambda$ has 0.1\,Jy
correlated flux density.  If one assumes a total flux density of
0.74\,Jy \citep{tin03} in a circular Gaussian, the core size is 1.1
mas.  \\
%
  J1937$-$3958 --- The core is $0.3$\,mas in size.  There are no
baselines less than 120 M$\lambda$ so the extended structure seen with
the VLBApls is not present. \\
%
  J1939$-$6342 --- \citet{ojh04} shows two components, separated by
40\~mas.  The source is not detected on space baselines and no image
is given.  Hartebeesthoek--Hobart at 140 M$\lambda$ has 0.1\,Jy
correlated flux density.  If one assumes a total flux density of 1.0\,Jy 
in the core component (the western of the 40-mas double), the core
angular size is $1.1$\,mas.  \\
%
  J1940$-$6907 --- There are no detections on space baselines.  One
component fit to the sparse ground data suggest a core size of $0.9$
mas. \\
%
  J1955$+$5131 --- The core is 0.3\,mas in size.  There is extended
emission toward the north-west.\\
%
  J2009$-$4849 --- There are no detections on space baselines.  The
core is 1.2\,mas in size.  The faint component to the west seen by
\citet{ojh05} is below the VSOP detection level. \\
%
  J2011$-$1546 --- The core is $<0.3$\,mas in size.  There is a
component just north of the core and a more extended component 2\,mas
north of the core. \\
%
  J2123$+$0535 --- The core is $0.3$\,mas in size.  There is extended
emission to the north-east. \\
%
  J2139$+$1423 --- The core is $<0.3$\,mas in size.  There is extended
emission slightly to the east and to the south.\\
%
  J2148$+$0657 --- The core is 0.9\,mas in size with no indication of
more compact emission.  There is faint extended emission to the
south-east. \\
%
  J2151$+$0552 --- The core is $<0.3$\,mas in size.  There is extended
emission over the core and to the west. \\
%
  J2152$-$7807 --- There are no detections on space baselines.  The
  sparse data suggest a component of 0.9\,mas in size, elongated in
  position angle $33^\circ$. \\
%
  J2207$-$5346 --- The core component is $0.15$\,mas in size.  There is
extended emission towards the east. \\  
%
  J2218$-$0335 --- The core component is 1.3\,mas in size.  Any compact
core in this source is less 0.1\,Jy. \\
%
  J2232$+$1143 --- The core component is 0.4\,mas in size.  The
complicated emission extends well to the south.  \\
%
  J2236$+$2828 --- The core component is 0.5\,mas in size.  Any further
compact component is less than 0.3\,Jy. \\
%
  J2239$-$5701 --- There are no detections on space baselines and no
image is given.  Hartebeesthoek--Hobart at 140 M$\lambda$ has 0.4\,Jy
correlated flux density.  If one assumes a total flux density of 0.7\,Jy \citep{tin03}
in a circular Gaussian, the core size is 0.6\,mas. \\
%
  J2246$-$1206 --- The core component is $<0.2$~mas in size.  There is extended
emission to the north.  \\
%
  J2258$-$2758 --- The core component is $0.2$~mas in size.  There is extended
emission to the south and east.  \\
%
  J2336$-$5236 --- There are no detections on space baselines and no
image is given.  Hartebeesthoek--Hobart at 140 M$\lambda$ has 0.15\,Jy
correlated flux density.  If one assumes a total flux density of 1.63\,Jy \citep{tin03} in
a circular Gaussian, the core size is 1.2\,mas. \\
%
  J2357$-$5311 --- The core component is 0.3\,mas in size.  There is
extended emission to the south-west, in agreement with the lower
resolution image of \citet{she_98}. \\
%
  J2358$-$1020 --- There are space fringes implying a core component
  of 0.2\,Jy within 0.2\,mas.  The ground data has amplitude scaling
  errors, so there is no image nor model. \\
%
\section{Summary and Discussion}

We have presented images, models and comments of the 140 sources which
were observed as part of the VSOP Survey project that were not covered
in P-III. We have combined the brightness temperature measurements and
limits found for the entire sample to produce the T$_b$ distribution
for the VSS.

We find that about half of the AGN sample of sources reported upon in
this paper have significant radio emission in the core component, with
$T_b \ge 10^{12}$~K in the source frame.  Since the maximum
brightness temperature one is able to determine using only
ground-based arrays is of the order of $10^{12}$~K, our results
confirm the necessity of using space VLBI to explore the extremely
high brightness temperature regime.  In addition, our Survey results
clearly show that by using space VLBI with higher sensitivity, and
somewhat higher resolution, the radio cores of many AGN can be
successfully imaged.

Because of the variability of many of the sources in the Survey
sample, detailed spectral indices of the core components are difficult
to determine. However, many of the sources were observed with the VLBA
at 15 GHz as part of the VLBA2cm2 survey, and the spectral properties
of the cores will be reported elsewhere \citep{VLBA2cm2}.

It was not possible to slew the HALCA satellite during the observing
runs, therefore HALCA was not able to participate in scans of fringe
finders, or flux calibrators. It is the absence of these which forces
us to label a number of experiments with no space fringes as failures,
when it could be the effects of the source structure.  The design of
VSOP-2 will allow fringe checks, and also phase referencing
experiments, to be performed \citep{hir00c}.

{
The completion of this survey has been a Quixotic 
endeavor, and possibly: 
``vino a dar en el m\'as estra\~no pensamiento que jam\'as dio loco en el
mundo; y fue que le pareci\'o convenible y necesario, as\'i para el
aumento de su honra como para el servicio de su rep\'ublica, hacerse
caballero andante'' \citep{quixote}.
}

\acknowledgements

We gratefully acknowledge the VSOP Project, which is led by the
Institute of Space and Astronautical Science of the Japan Aerospace
Exploration Agency, in cooperation with many organizations and radio
telescopes around the world.  RD, KJW, GAM and JEJL acknowledge
support from the Japan Society for the Promotion of Science. RD
acknowledges support as a Marie-Curie fellow via EU FP6 grant
MIF1-CT-2005-021873. ZS acknowledges support from NNSFC (10573029 and
10625314). WKS wishes to acknowledge support from the Canadian Space
Agency. SH acknowledges support through an NRC/NASA-JPL Research
Associateship. SF, ZP and LM acknowledge the OTKA T046097 grant
received from the Hungarian Scientific Research Fund. SF and GC
acknowledge the ASTRON and JIVE summer studentship programme. JY
acknowledges the CAS-UNAW cooperation programme in Radio Astronomy.
This research has made use of data from the University of Michigan
Radio Astronomy Observatory (UMRAO) which is supported by funds from
the University of Michigan and the NSF, the United States Naval
Observatory (USNO) Radio Reference Frame Image Database (RRFID), and
the NASA/IPAC Extragalactic Database (NED) which is operated by the
Jet Propulsion Laboratory, California Institute of Technology, under
contract with the National Aeronautics and Space Administration.  The
NRAO is a facility of the National Science Foundation, operated under
cooperative agreement by Associated Universities, Inc. The Australia
Telescope Compact Array is part of the Australia Telescope which is
funded by the Commonwealth of Australia for operation as a National
Facility managed by CSIRO. Meetings of the VSOP Survey Working Group
were supported by EC FP6 Integrated Infrastructure Initiative,
RadioNet (contract RII3-CT-2003-505818), JIVE and the OAN.


\clearpage
\begin{deluxetable}{llrclccrccc}
\label{tab:vss}
\tablecolumns{9}
\tablecaption{VSOP AGN Survey Source List}
\tabletypesize{\footnotesize}
\tablehead {

\multicolumn{1}{c} {(1)} &
\multicolumn{1}{c} {(2)} &
\multicolumn{1}{c} {(3)} &
\multicolumn{1}{c} {(4)} &
\multicolumn{1}{c} {(5)} &
\multicolumn{1}{c} {(6)} &
\multicolumn{1}{c} {(7)} &
\multicolumn{1}{c} {(8)} &
\multicolumn{1}{c} {(9)} \\
\multicolumn{1}{c} {IAU Name} &
\multicolumn{1}{c} {S$_5$} &
\multicolumn{1}{c} {ID} &
\multicolumn{1}{c} {z} &
\multicolumn{1}{c} {Exp} &
\multicolumn{1}{c} {Paper} &
\multicolumn{1}{c} {T$_b$} &
\multicolumn{1}{c} {area} &
\multicolumn{1}{c} {Jy -- M$\lambda$} \\
\multicolumn{1}{c} {} &
\multicolumn{1}{c} {[Jy]} &
\multicolumn{1}{c} {} &
\multicolumn{1}{c} {} &
\multicolumn{1}{c} {} &
\multicolumn{1}{c} {} &
\multicolumn{1}{c} {[10$^{12}$ K]} &
\multicolumn{1}{c} {[mas$^2$]} &
\multicolumn{1}{c} {} \\

}
\startdata
 J0006$-$0623 & $2.69^{\dagger}$  &      G &0.347$^{*}$& vs03t & III & $>$0.5  & 0 &0.52 -- 380\\
 J0013$+$4051 & 1.0 &      G &0.256& vs13a &   V & 1.3     & 0.01 &0.23 -- 460\\
 J0019$+$7327 & $0.78^{*}$  &    QSO &1.781& vs07a & III & $>$0.2  & 0 &0.12 -- 420\\
 J0042$+$2320 & 1.6 &   VisS &   & vs07u & III & 0.3     & 0.04 &0.19 -- 460\\
 J0050$-$0929 & 1.9 &    QSO &   & vs06i$^\dagger$ & 1,2  & &  &--\\
 J0105$+$4819 & 1.1 & RadioS &   & vs08q &   V & 0.3     & 0.04 &0.13 -- 540\\
 J0106$-$4034 & $3.40^{\dagger}$  &    QSO &0.584& vs03s & III & $>$3.8  & 0 &1.36 -- 500\\
 J0108$+$0135 & $3.60^{*}$  &    QSO &2.107& vs02t &   V & 1.7     & 0.04 &0.54 -- 500\\
\enddata \footnotesize \tablecomments {
Column (2): Total flux density, from \citet{tin03} where marked with a
$\dagger$, where marked with a * from UMRAO, otherwise as given by NED. 
Column (3): ID class listed in NED,
where QSO -- quasar, G -- galaxy, RadioS -- radio source, and VisS --
visual source. 
Column (4): Redshift measurement from \cite{ver06}, except where
marked with a *, in which case it is from NED.
Column (5): VSOP experiment code (if not observed marked with $\dagger$),
Column (6): Paper in which the results are to be found (III or V), 
 or, if not observed, other images to be found in:
 1=\citet{fom00b}, 
 2=\citet{kel98},
 3=\citet{fey00}, 
 4=\citet{zen02}, 
 5=\citet{ojh04}, 
 6=\citet{lis05}
Column (7): Observer frame brightness temperature, or lower limit,
Column (8): Detected core area in mas$^2$ (point sources being 0).
Column (9): Flux density (in Jy) remaining at the longest baselines, and
the length of that baseline (in M$\lambda$). 
}
\end{deluxetable}

\clearpage

\topmargin 4cm

\begin{landscape}
\begin{deluxetable}{cllcllccclll}
\label{tab:obs}
\tabletypesize{\scriptsize}
\tablecaption{Survey Experiment Details\label{tab1}}
\tablewidth{0pt}
\tablehead{
\multicolumn{1}{c} {(1)} &
\multicolumn{1}{c} {(2)} &
\multicolumn{1}{c} {(3)} &
\multicolumn{1}{c} {(4)} &
\multicolumn{1}{c} {(5)} &
\multicolumn{1}{c} {(6)} &
\multicolumn{1}{c} {(7)} &
\multicolumn{1}{c} {(8)} &
\multicolumn{1}{c} {(9)} &
\multicolumn{1}{c} {(10)} &
\multicolumn{1}{c} {(11)} \\
\multicolumn{2}{c}{Source Names} &
\colhead{Obs.} &
\colhead{Obs.\ Date} &
\colhead{GRTs} &
\colhead{TSs} &
\colhead{Time On} &
\colhead{HALCA Time} &
\colhead{Corr.} &
\colhead{ID} &
\colhead{$z$} \\
\colhead{J2000} &
\colhead{B1950} &
\colhead{Code} &
\colhead{} &
\colhead{} &
\colhead{} &
\colhead{Src.} &
\colhead{On Src.} &
\colhead{} &
\colhead{} &
\colhead{} &
\colhead{} \\
\colhead{} &
\colhead{} &
\colhead{} &
\colhead{(yyyy-mmm-dd)} &
\colhead{} &
\colhead{} &
\colhead{(hh:mm)} &
\colhead{(hh:mm)} &
\colhead{} &
\colhead{} &
\colhead{} 
}
\startdata
 J0013$+$4051 & 0010$+$405 & vs13a &2002 Jul 23& KRN & U &04:40 &04:00 & P & g &0.256\\ 
 J0105$+$4819 & 0102$+$480 & vs08q &2002 Jan 17& URP & T & 03:10 & 03:10 & S & e &        \\ 
 J0108$+$0135 & 0106$+$013 & vs02t &2002 Jul 13& NHR & U &03:00 &02:30 & M & q &2.107\\ 
 J0116$-$1136 & 0113$-$118 & vs06h &2001 Jul 15&SHC&U&03:50 &00:00 & P& q &0.672\\ 
 J0121$+$0422 & 0119$+$041 & vs05y &1998 Aug 01&GH&R&03:15 &02:00 &P& q &0.637\\ 
 J0125$-$0005 & 0122$-$003 & vs05x &2002 Jul 16& NR & U &02:20 &02:20 & M & q &1.070\\ 
 J0141$-$0928 & 0138$-$097 & vs10i &2002 Jan 11&CKMS& T&05:45 &04:20 & P& b &0.733$^{*}$\\ 
 J0149$+$0555 & 0146$+$056 & vs06z &2001 Jul 27& CRMS & UR &09:00 &05:00 & P & q &2.345\\
\enddata
\tablecomments{
Col. $(1)$: J2000 name,\newline
Col. $(2)$: Other names,\newline
Col. $(3)$: Experiment code, if appended by an asterisk (*) also a GOT
experiment.\newline
Col. $(4)$: Starting date of the observation. If experiments from
different epochs were
merged both dates are provided.\newline
Col. $(5)$: Ground radio telescopes
to which fringes were successfully obtained. Lower case letters indicate
VLBA GRTs: b$=$Brewster, f$=$Fort Davis, h$=$Hancock, k$=$Kitt Peak, l$=$Los Alamos, m$=$Mauna Kea,
n$=$North Liberty, o$=$Owens Valley, p$=$Pie Town,
s$=$St. Croix. Upper Case letters indicate other GRTs: X$=$Arecibo,
A$=$ATCA, C$=$Ceduna, E$=$Effelsburg, I=Medicina,
G$=$Green Bank $140^\prime$, H$=$Hartebeesthoek, T$=$Hobart, J$=$Jodrell Bank MkII, R$=$Kalyazin, K$=$Kashima, N$=$Noto,
M$=$Mopra, O$=$Onsala, S$=$Sheshan, P$=$Torun, U$=$Usuda, Y$=$VLA\newline
Col. $(6)$: Tracking Stations to which fringes were successfully obtained:
G$=$Goldstone, N$=$Green Bank, R$=$Robledo,
T$=$Tidbinbilla, U$=$Usuda\newline
Col. $(7)$: Approximate time on source over which fringes were found\newline
Col. $(8)$: Approximate time on source over which fringes were found on space baselines\newline
Col. $(9)$: Correlator: M$=$Mitaka, P$=$Penticton, S$=$Socorro\newline
Col. $(10)$:Optical classification:
q~--~quasar, b~--~ BL Lac object,
g~--~AGN other 
than b and q (e.g. Seyfert galaxy),
e~--~empty field or unidentified optical counterpart;
from \cite{ver06} unless appended by an asterisk (*) in which case it
is from NED\newline
Col. $(11)$: Redshift from \cite{ver06} unless appended by an
asterisk (*) in which case it is from NED.\newline
}
\end{deluxetable}
\end{landscape}

\begin{landscape}
\begin {deluxetable}{lrrrrrrcl|lrrrrrrcl}
\label{tab:parms}
\tablecaption{Source Component Parameters}
\tabletypesize{\scriptsize}
\tablewidth{0cm}
\tablehead{

\colhead{Source}& 
\colhead{S\phm{*}}& 
\colhead{Rad}& 
\colhead{$\phi$}& 
\colhead{$\theta_{{\rm maj}}$}& 
\colhead{$\theta_{{\rm min}}$}& 
\colhead{P.A.}& 
\colhead{$T_b / 10^{12}$}&
\colhead{$T_b / 10^{12}$}&

\colhead{Source}& 
\colhead{S\phm{*}}& 
\colhead{Rad}& 
\colhead{$\phi$}& 
\colhead{$\theta_{\rm maj}$}& 
\colhead{$\theta_{\rm min}$}& 
\colhead{P.A.}& 
\colhead{$T_b / 10^{12}$}&
\colhead{$T_b / 10^{12}$}\\

\colhead{}&
\colhead{(Jy)\phm{*}}&
\colhead{(mas)}&
\colhead{($^\circ$)}&
\colhead{(mas)}&
\colhead{(mas)}&
\colhead{($^\circ$)}&
\colhead{(K)}&
\colhead{(K)}&

\colhead{}&
\colhead{(Jy)\phm{*}}&
\colhead{(mas)}&
\colhead{($^\circ$)}&
\colhead{(mas)}&
\colhead{(mas)}&
\colhead{($^\circ$)}&
\colhead{(K)}&
\colhead{(K)} 
}
\startdata 
 J0013$+$4051   & 0.19 & 0.1 & 143 & 0.1 & 0.1 & ... & 1.29 & 0.32  
 &  J0259$+$0747   & 0.44 & 0.1 & 170 & 0.8 & 0.2 & -28 & 0.14 & 0.18 \\ 
                & 0.11 & 0.4 & -34 & 0.2 & 0.2 & ... &&
 &  J0303$-$6211   & 1.68 & 0.0 & ... & 0.6 & 0.4 & 76 & 0.38 & 0.32* \\ 
 J0105$+$4819   & 0.29 & 0.0 & ... & 0.2 & 0.2 & ... & 0.32 & 0.18*  
 &                 & 0.15 & 1.1 & 82 & 0.0 & 0.0 & ... &&\\
 J0108$+$0135   & 1.35 & 0.1 & -5 & 0.2 & 0.2 & ... & 1.71 & 4.03  
 &  J0309$-$6058   & 0.33 & 0.3 & 35 & 0.1 & 0.1 & ... & 0.92 & 0.25* \\ 
                & 0.73 & 0.6 & -108 & 0.4 & 0.4 & ... &&
 &                 & 0.63 & 0.2 & -143 & 0.2 & 0.2 & ... &&\\
 J0116$-$1136   & 0.65 & 0.0 & ... & 0.8 & 0.2 & -35 & 0.21 & 0.25  
 &  J0312$+$0133   & 0.23 & 0.0 & ... & 0.1 & 0.1 & ... & 1.14 & 0.50 \\ 
                & 0.05 & 1.0 & -41 & 0.6 & 0.6 & ... &&
 &                 & 0.05 & 1.1 & 91 & 1.1 & 0.3 & 90 &&\\
 J0121$+$0422   & 0.76 & 0.0 & ... & 0.4 & 0.1 & -55 & 0.79 & 0.86  
 &  J0321$+$1221   & 0.20 & 0.9 & -167 & 0.0 & 0.0 & ... & $>$ & 1.68 \\
\enddata
\tablecomments{Col. $(2)$: The integrated flux density of the component,
\newline
Col. $(3)$: The radial distance of the component center from the
center of the map\newline
Col. $(4)$: The position angle of the center of the component,
measured counterclockwise from an imaginary vertical line from the map center
towards the north\newline
Col. $(5)$: The FWHM of the major axis of the component\newline
Col. $(6)$: The FWHM of the minor axis of the component\newline
Col. $(7)$: The position angle of the major axis of the component,
measured from the north towards the east\newline
Col. $(8)$: The measured observer frame brightness temperature of the
component, where resolved, in units of $10^{12}$K. This can be
converted to source frame temperature by multiplying by 1+z, where z
is to be found in Table 2.\newline 
Col. $(9)$: The lower limit to the brightness temperature of the
component, in units of $10^{12}$K, in the source frame, unless
appended by an asterisk (*), in which case it is in the observer
frame.\newline
}
\end{deluxetable}
\end{landscape}


\topmargin 2cm

\clearpage
\onecolumn

\begin{figure}
\spfig{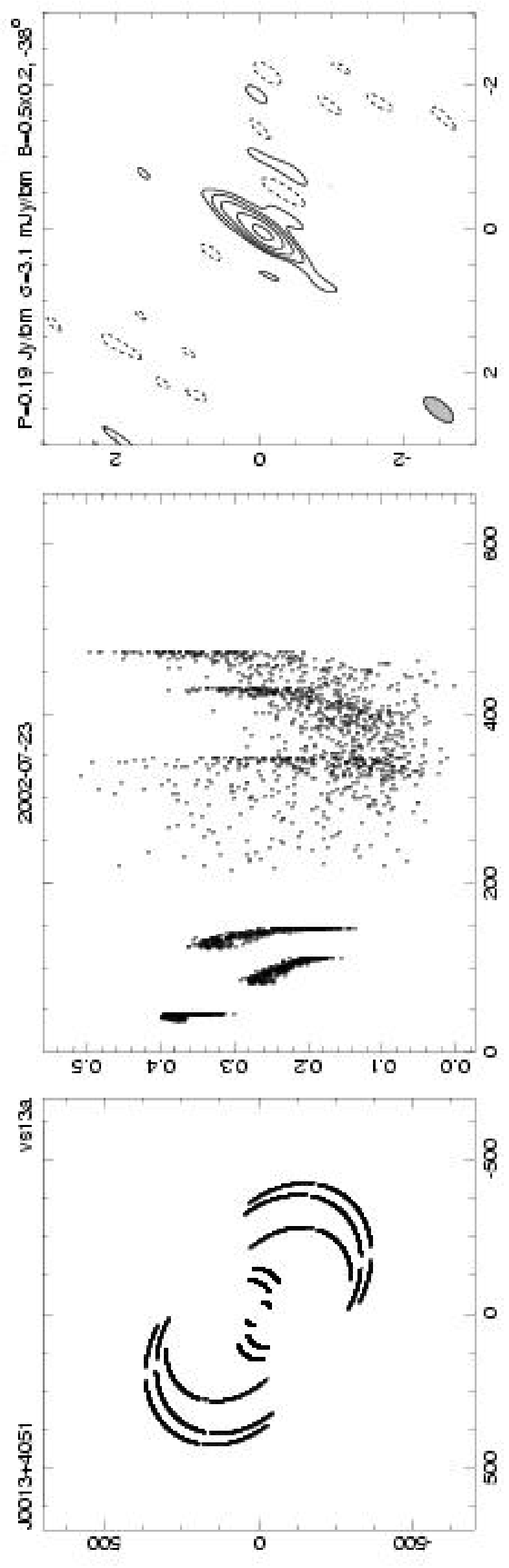}
\spfig{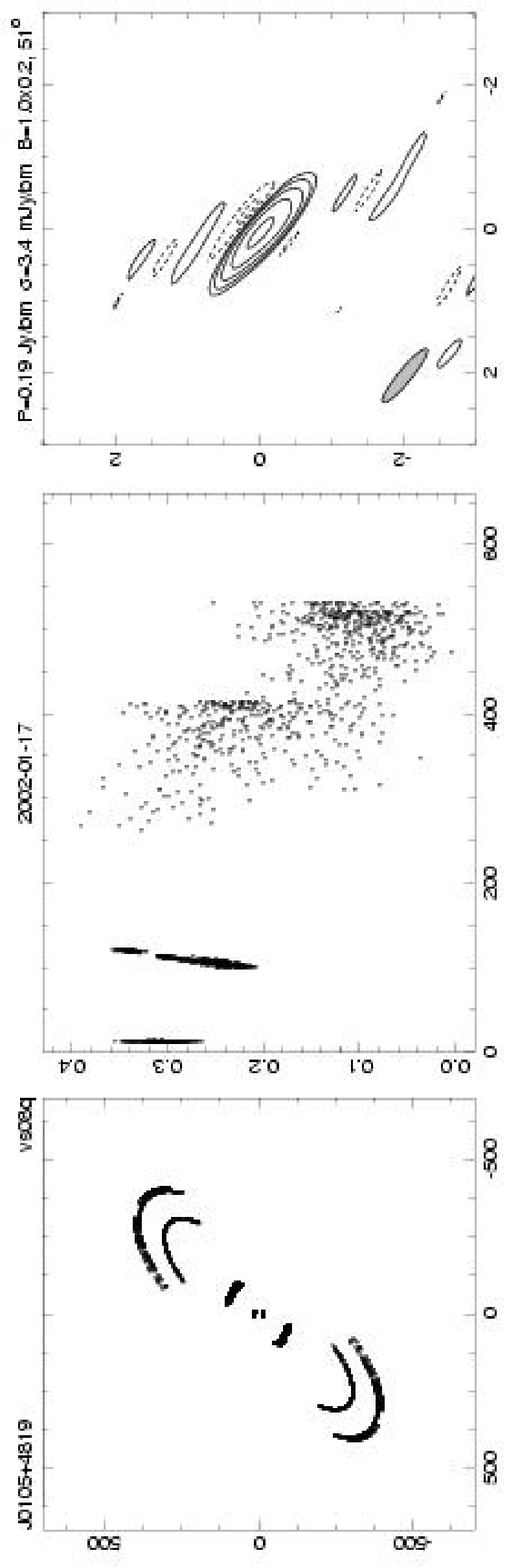}

{Fig. 1. -- Images of the Survey sources: For each source three
  separate panels are presented horizontally across the page. The
  first panel shows a plot of the \uv\ coverage, with {\em u} on the
  horizontal axis and {\em v} on the vertical axis. Both axes are
  measured in units of M$\lambda$. The second panel shows a plot of
  the amplitude of the visibilities (in Janskys) vs. \uv\ radius, with
  the latter again measured in M$\lambda$ (time-averaged to 150 seconds). For
  both of these plots only data that were actually used to make the
  final image are shown. Finally, a contour plot of the cleaned image
  is shown on the right. The contour levels double with each level and
  start at three times the image RMS, listed above the image, along with an
  additional negative contour, equal in magnitude to the minimum
  positive contour level. The peak flux, minimum contour level, and
  synthesized HPBW in milliarcsecond are shown on the top border.}

\end{figure}
\clearpage
\begin{figure}
\spfig{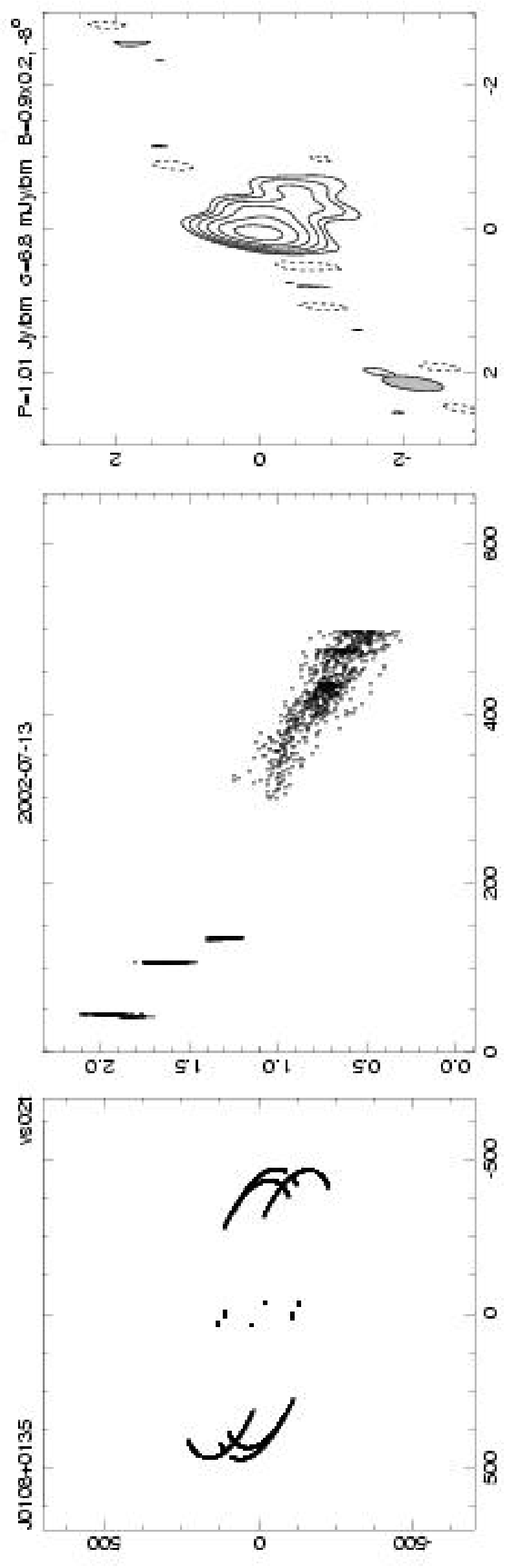}
\spfig{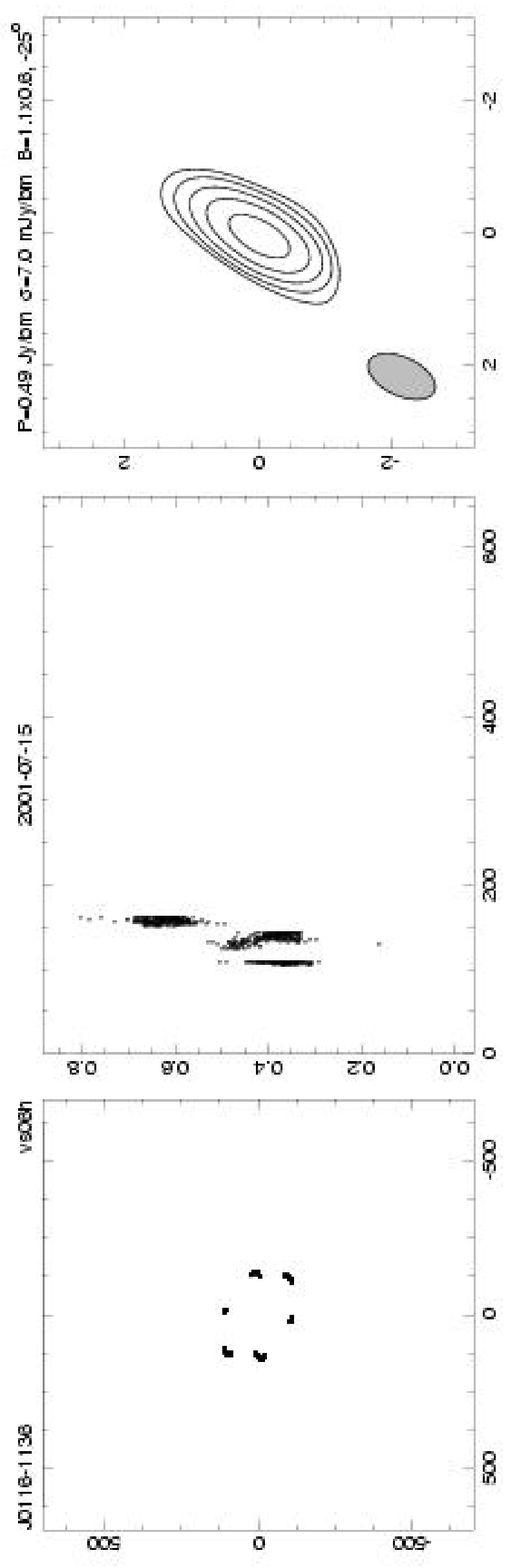} \typeout{no image?}
\spfig{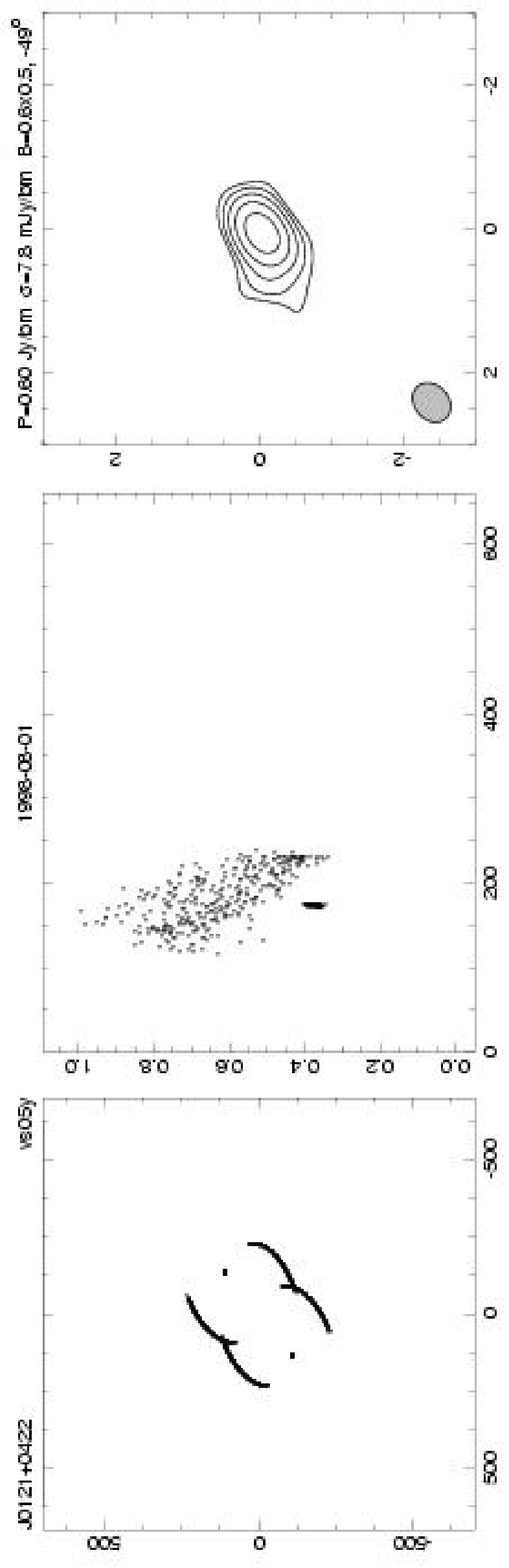}
\spfig{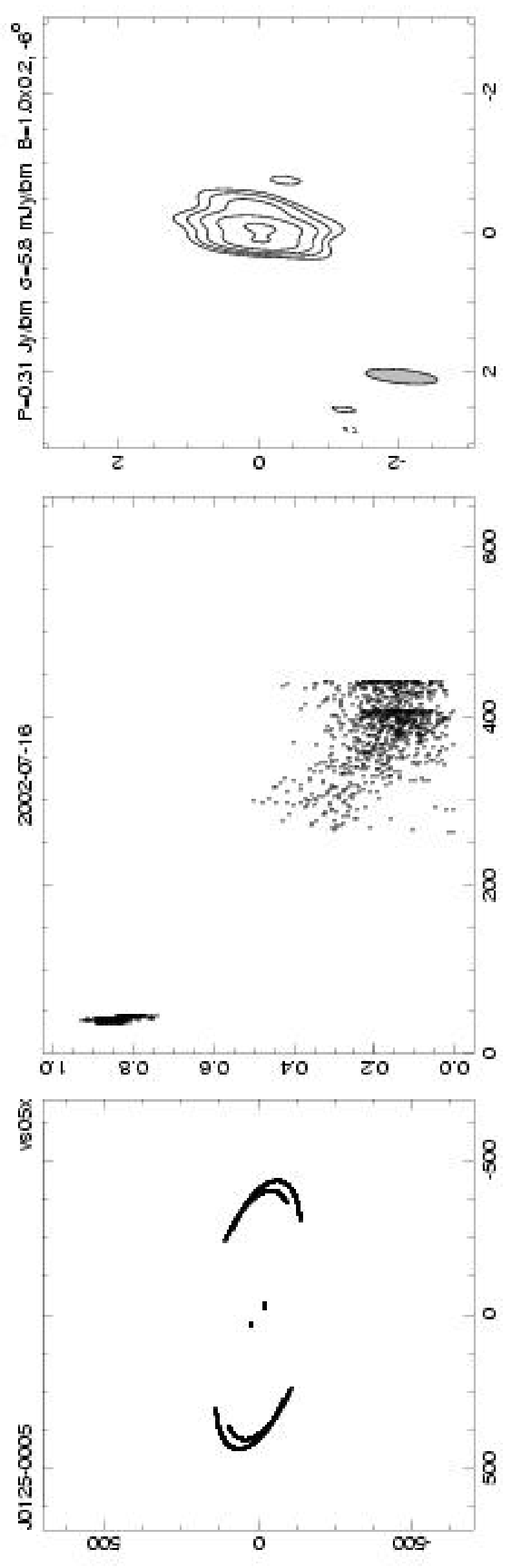}
\spfig{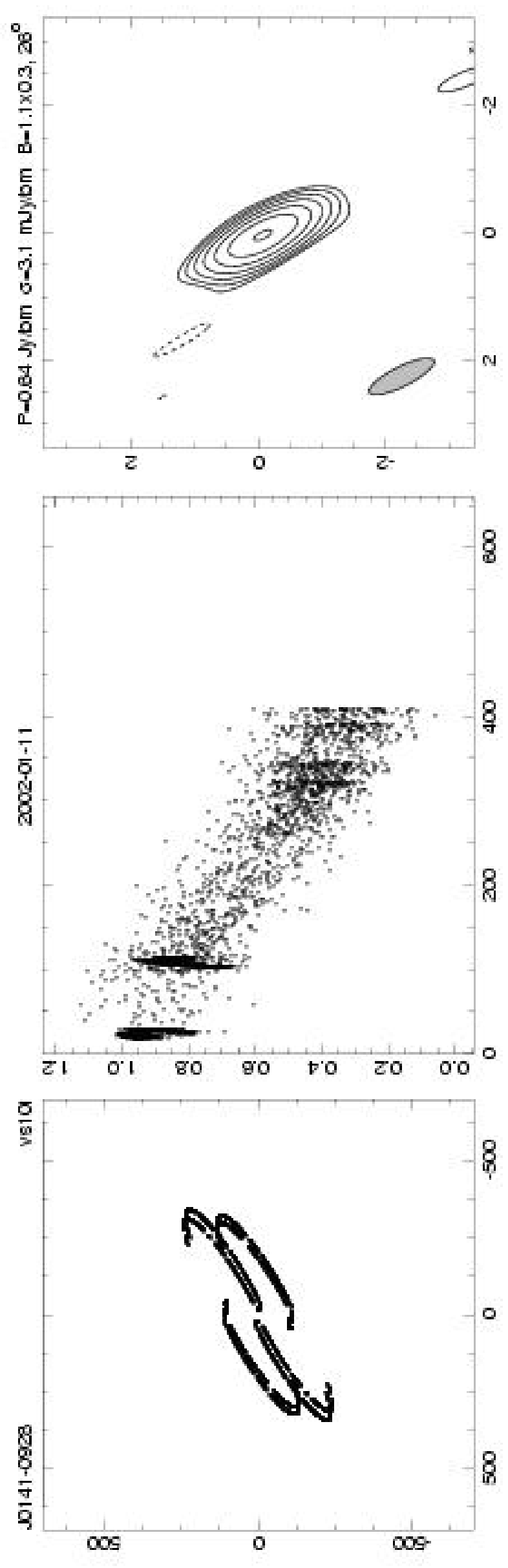}

{Fig. 1. -- {\em continued}}
\end{figure}
\clearpage
\begin{figure}
\spfig{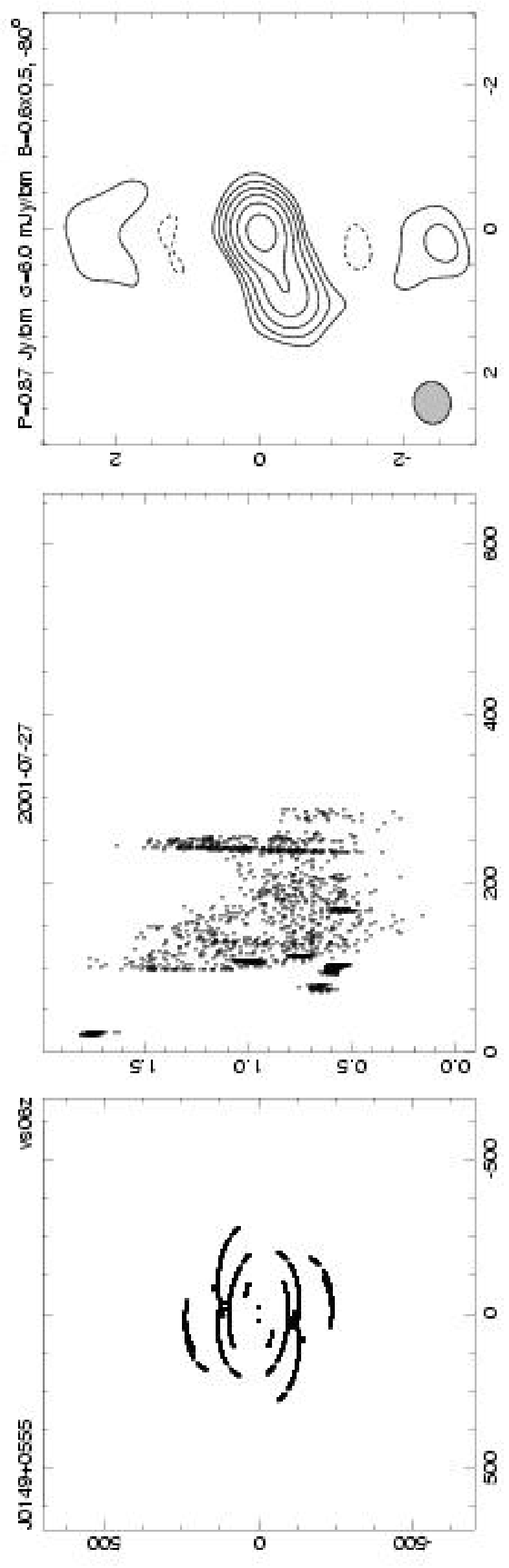}
\spfig{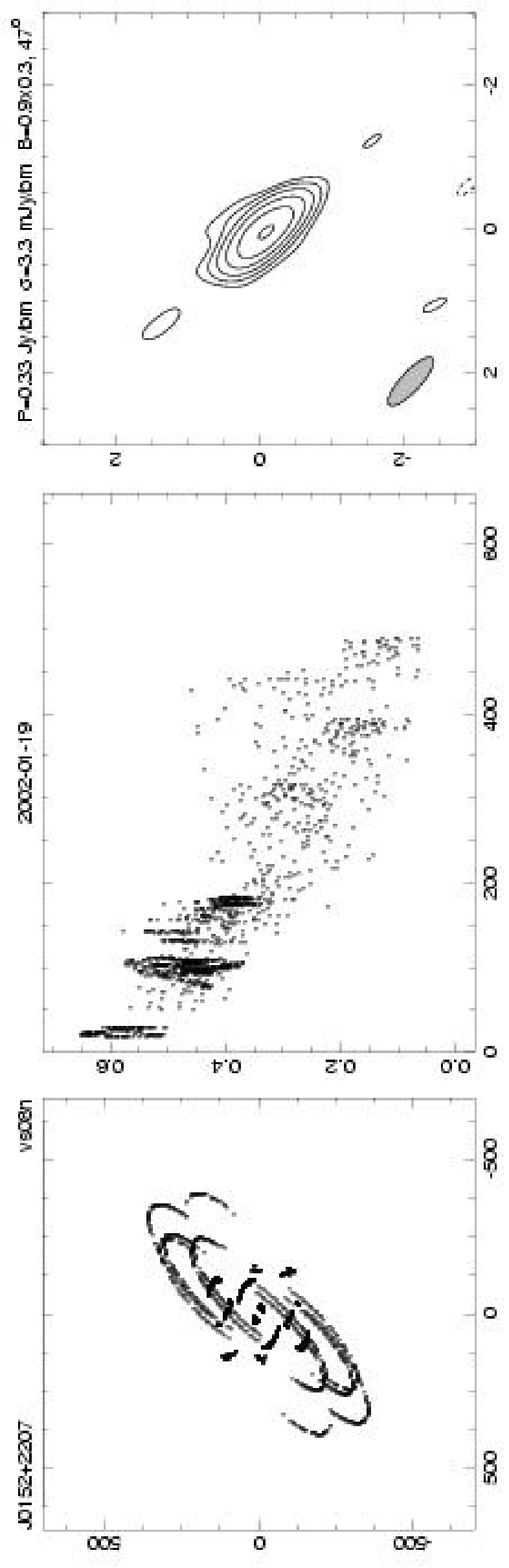}
\spfig{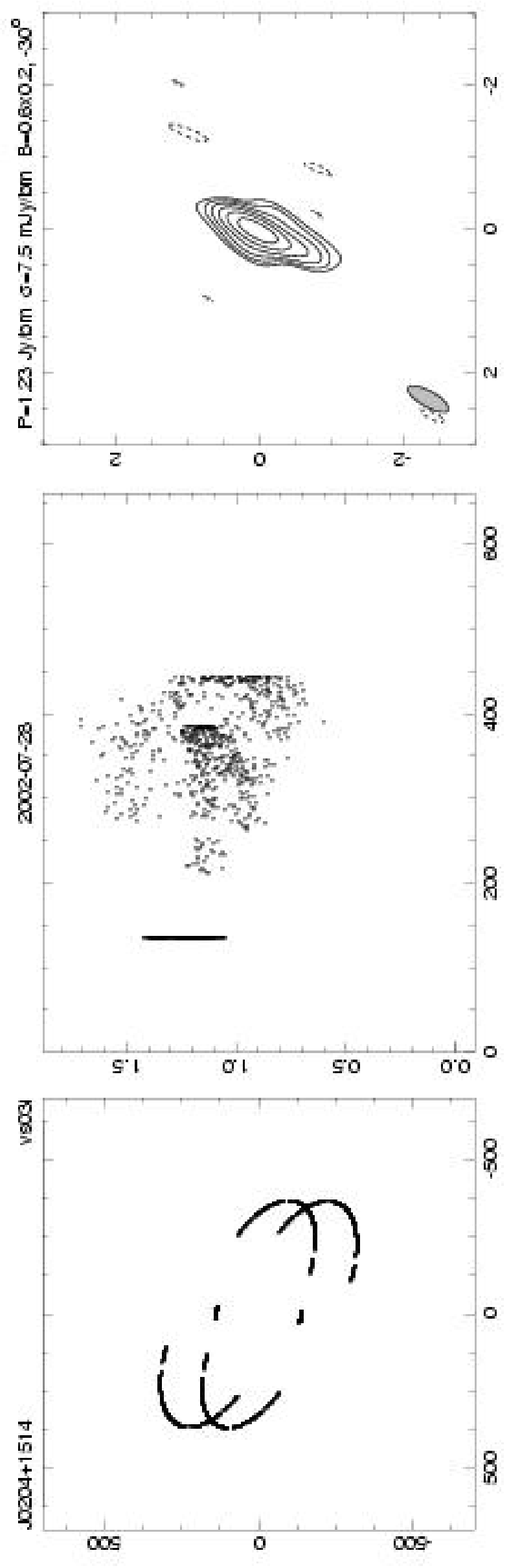}
\spfig{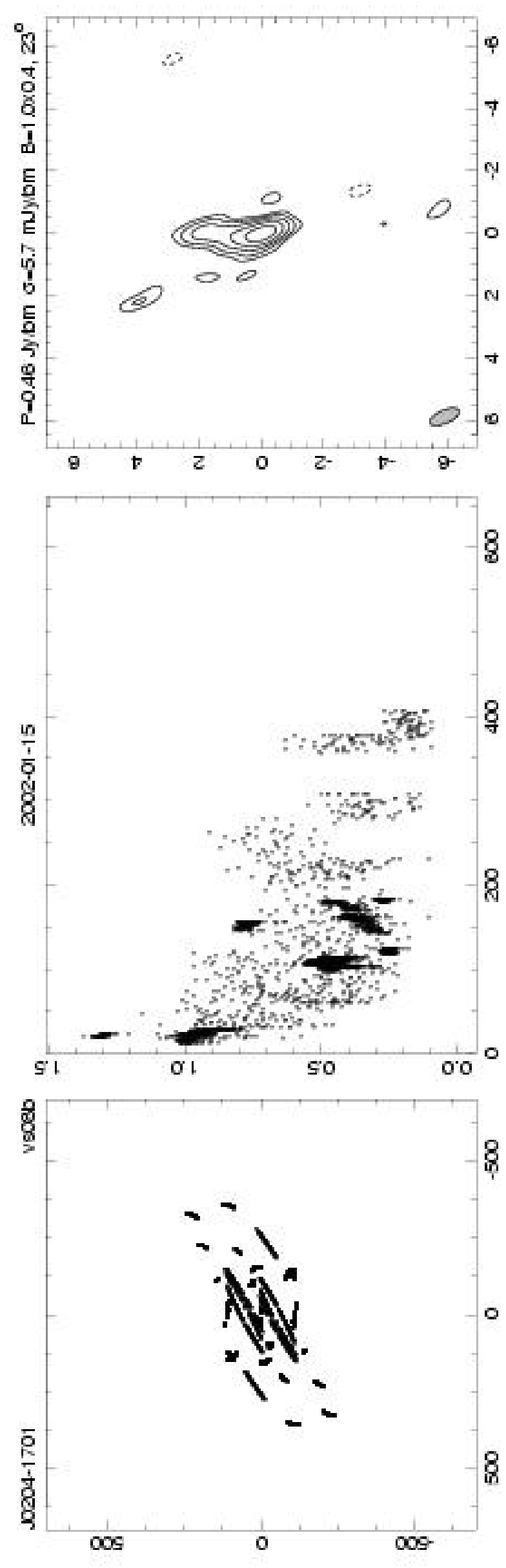}
\spfig{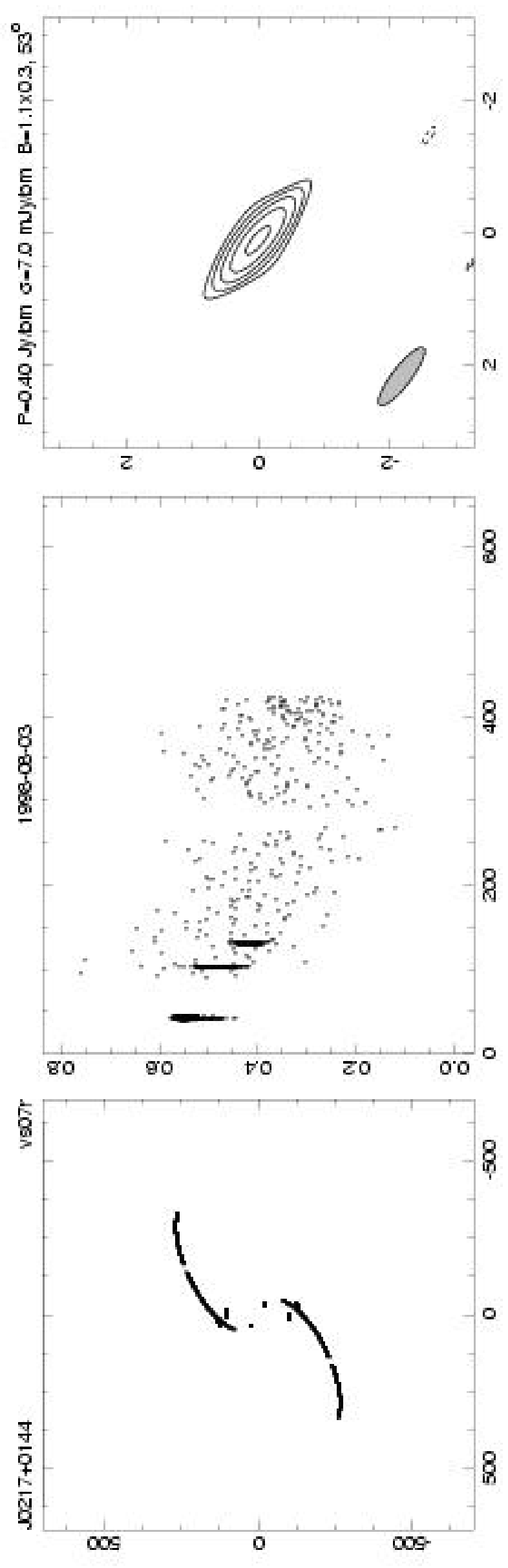}

{Fig. 1. -- {\em continued}}
\end{figure}
\clearpage
\begin{figure}
\spfig{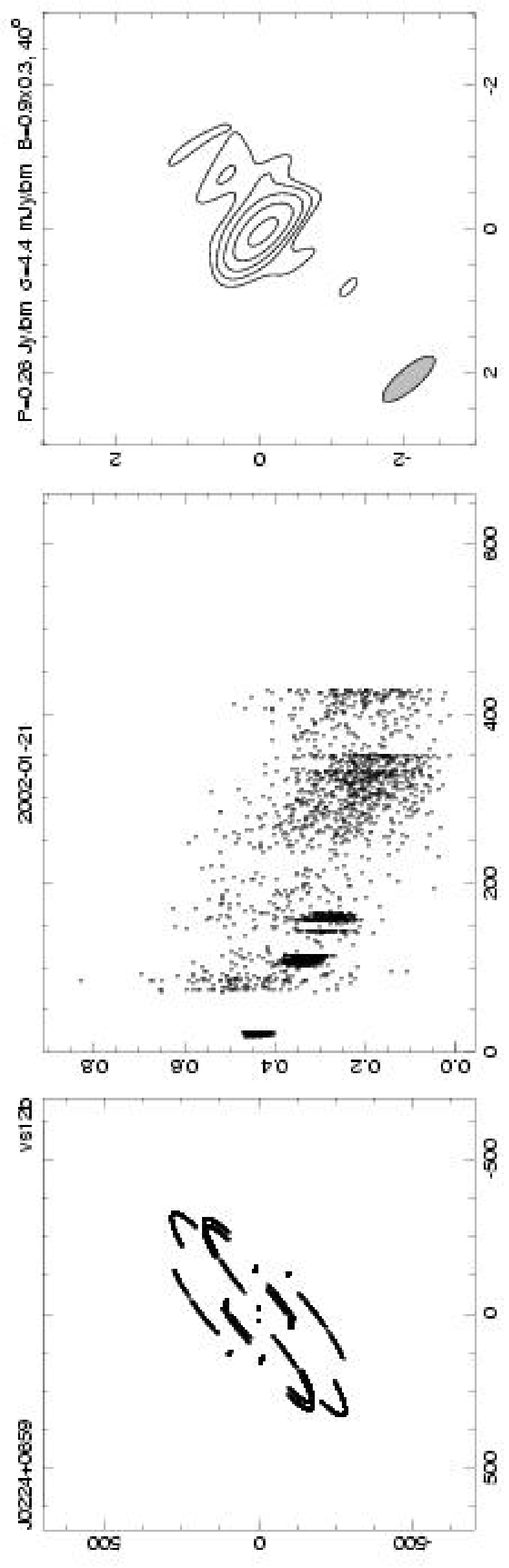}
\spfig{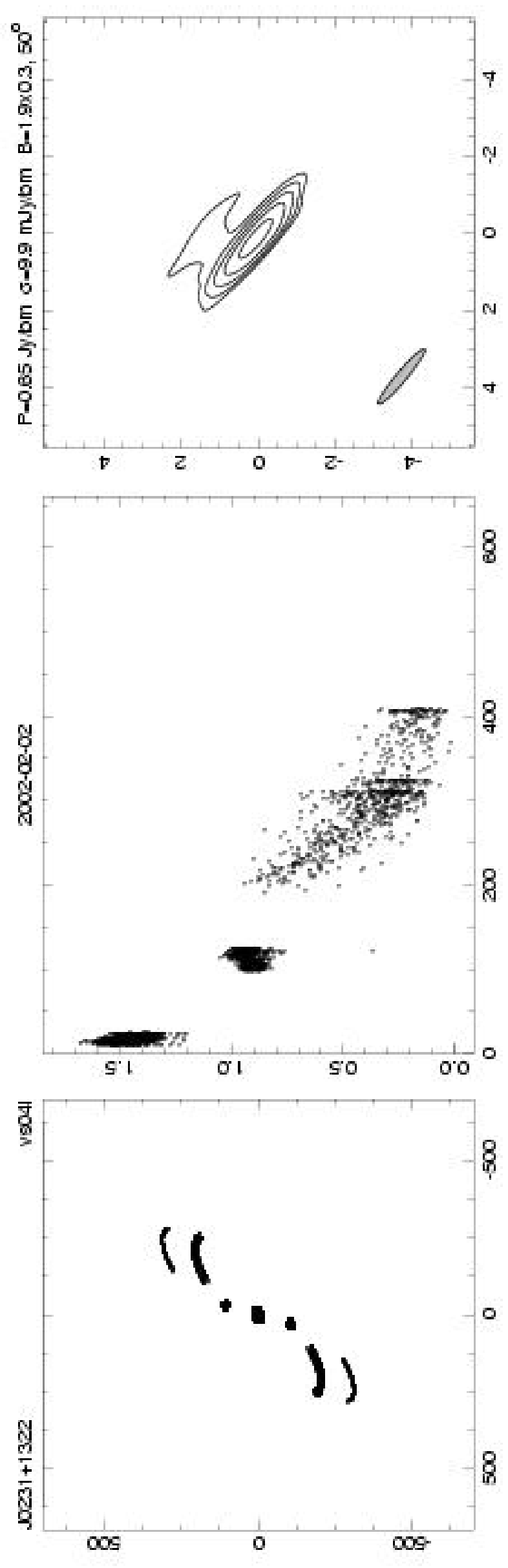}
\spfig{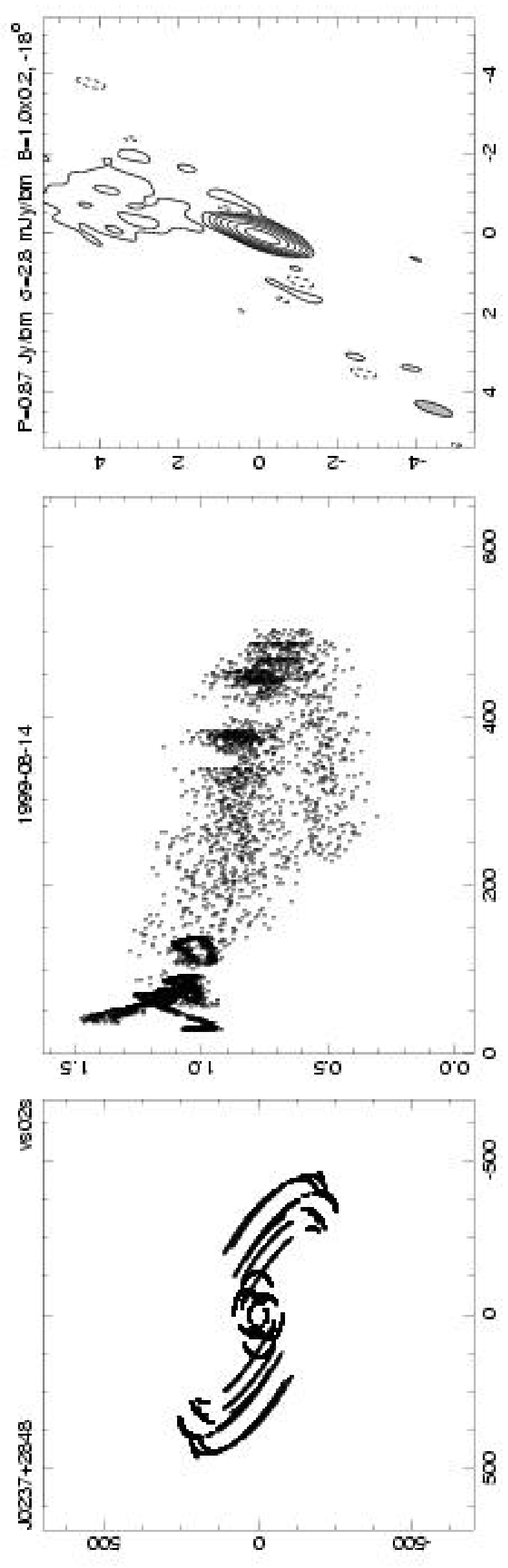}
\spfig{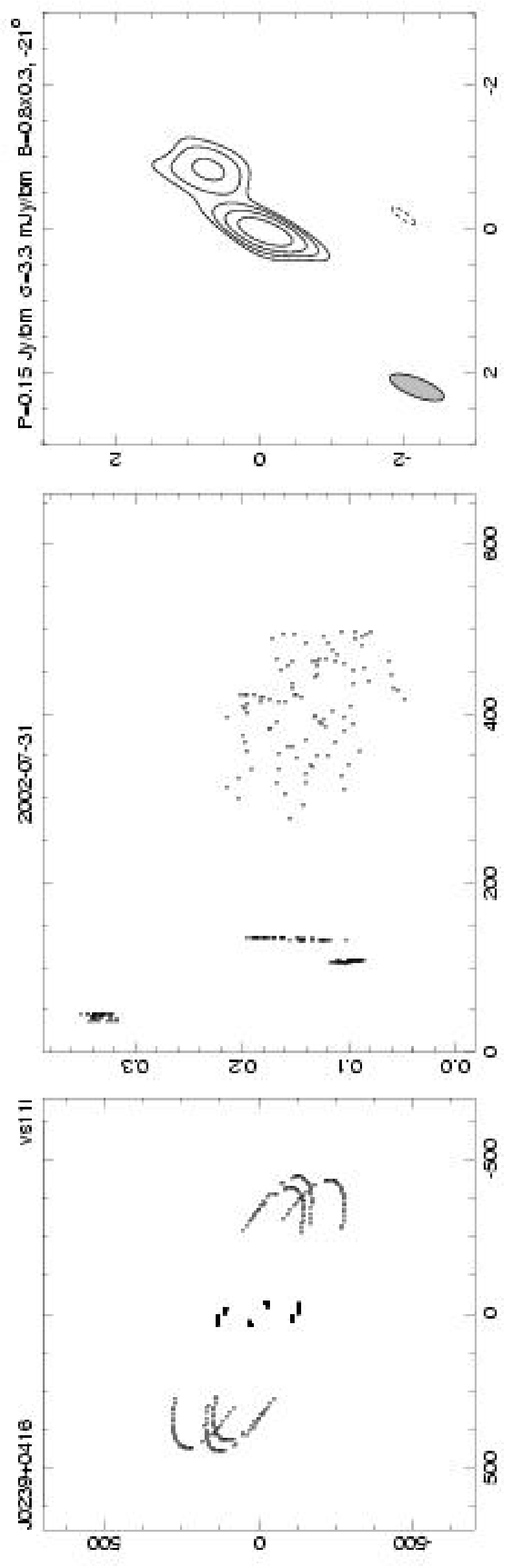}
\spfig{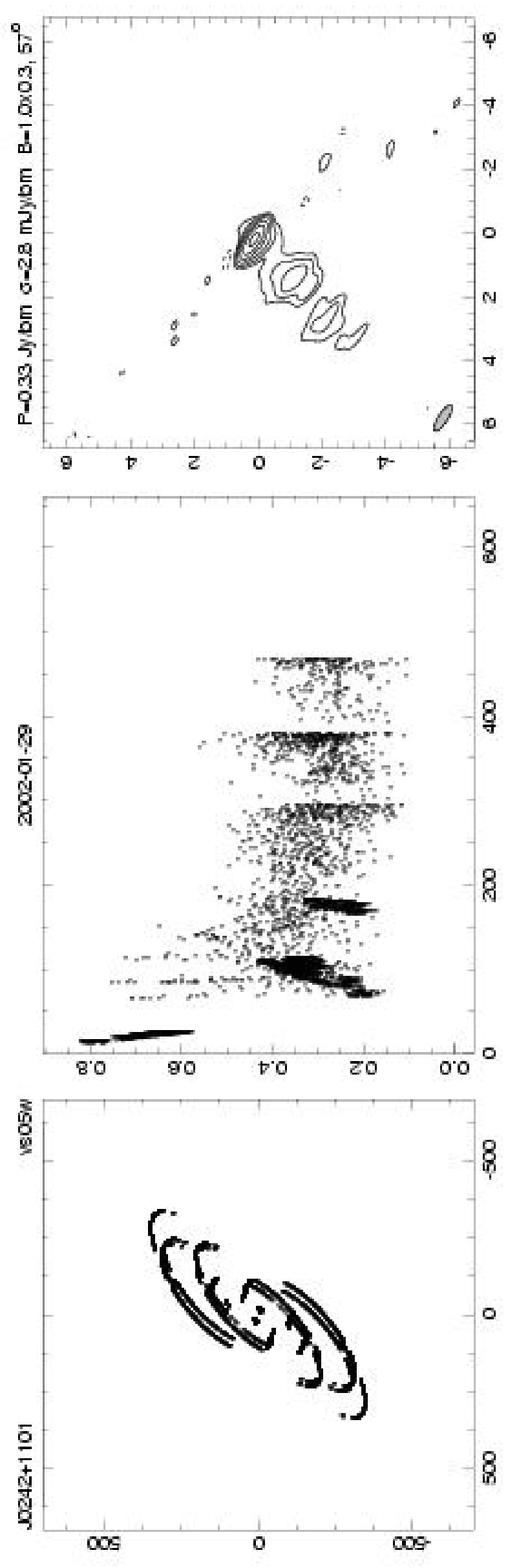}

{Fig. 1. -- {\em continued}}
\end{figure}
\clearpage
\begin{figure}
\spfig{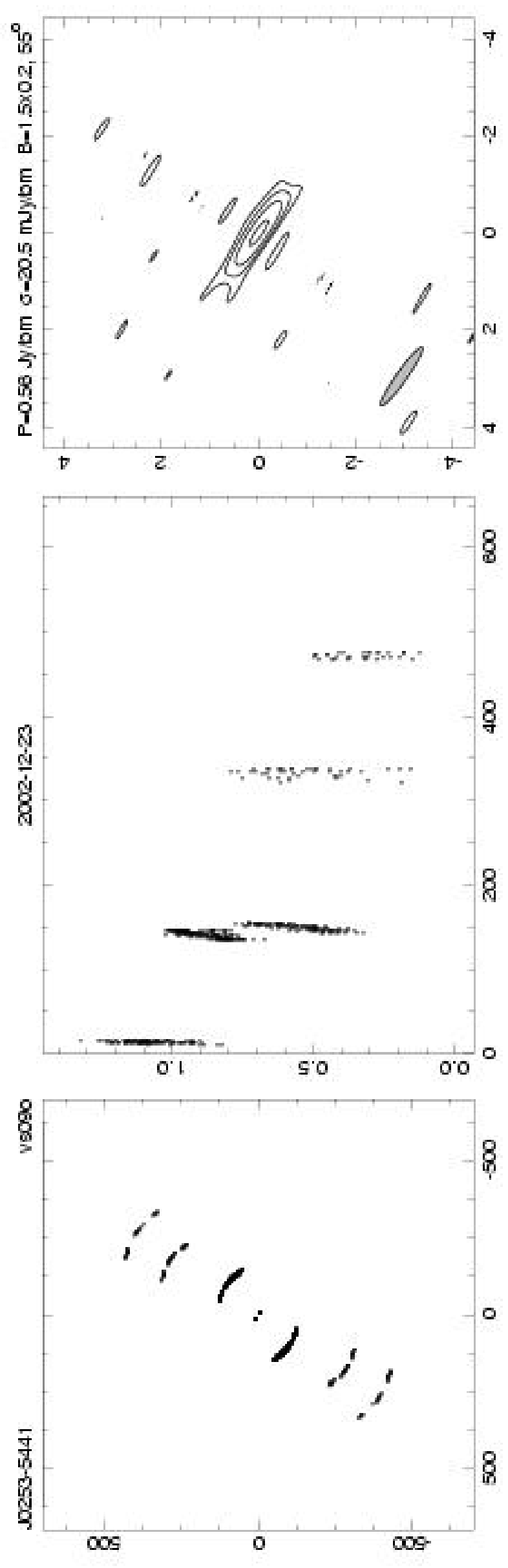}
\spfig{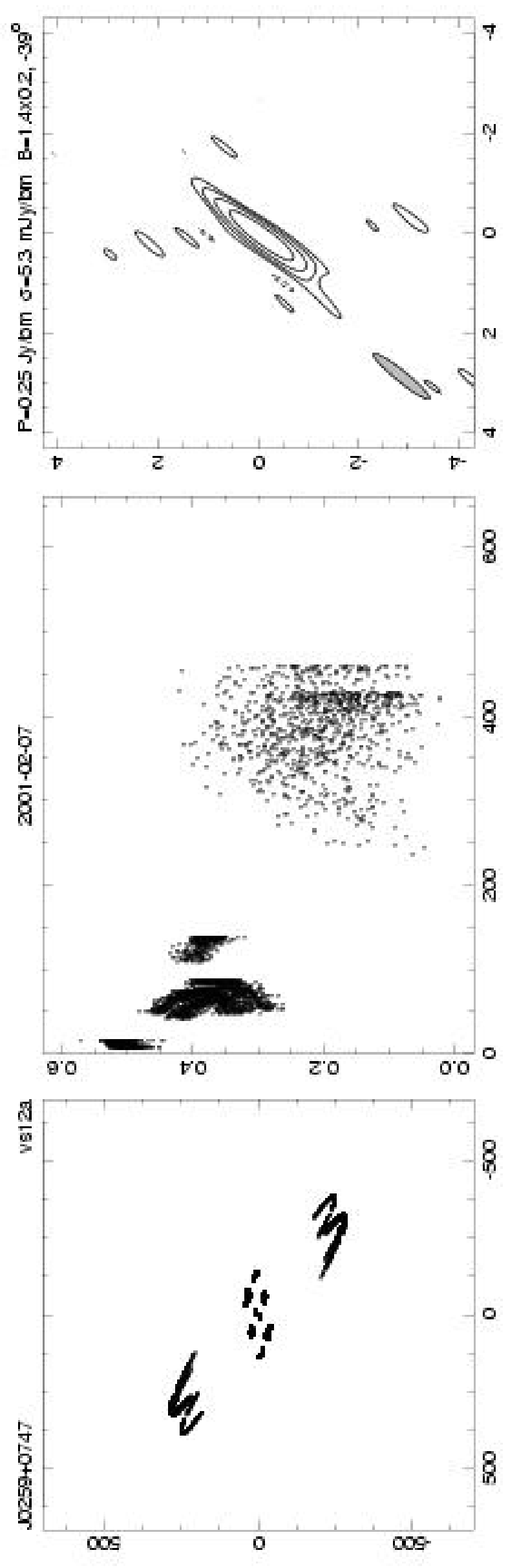}
\spfig{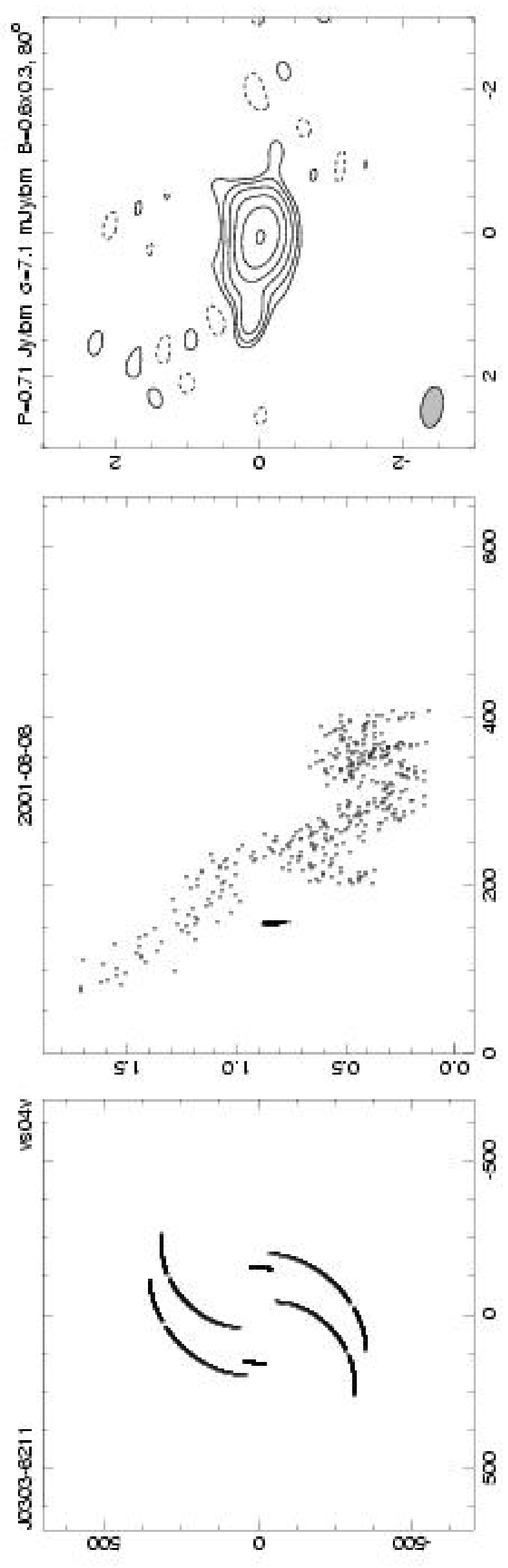}
\spfig{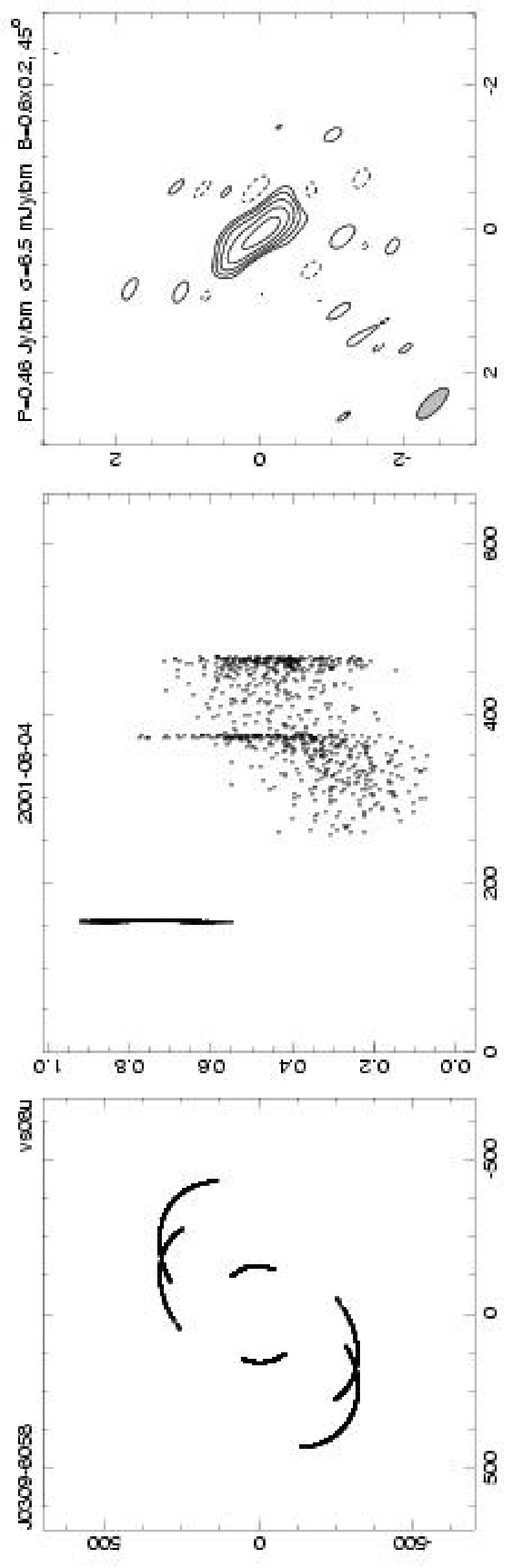} \typeout{no png}
\spfig{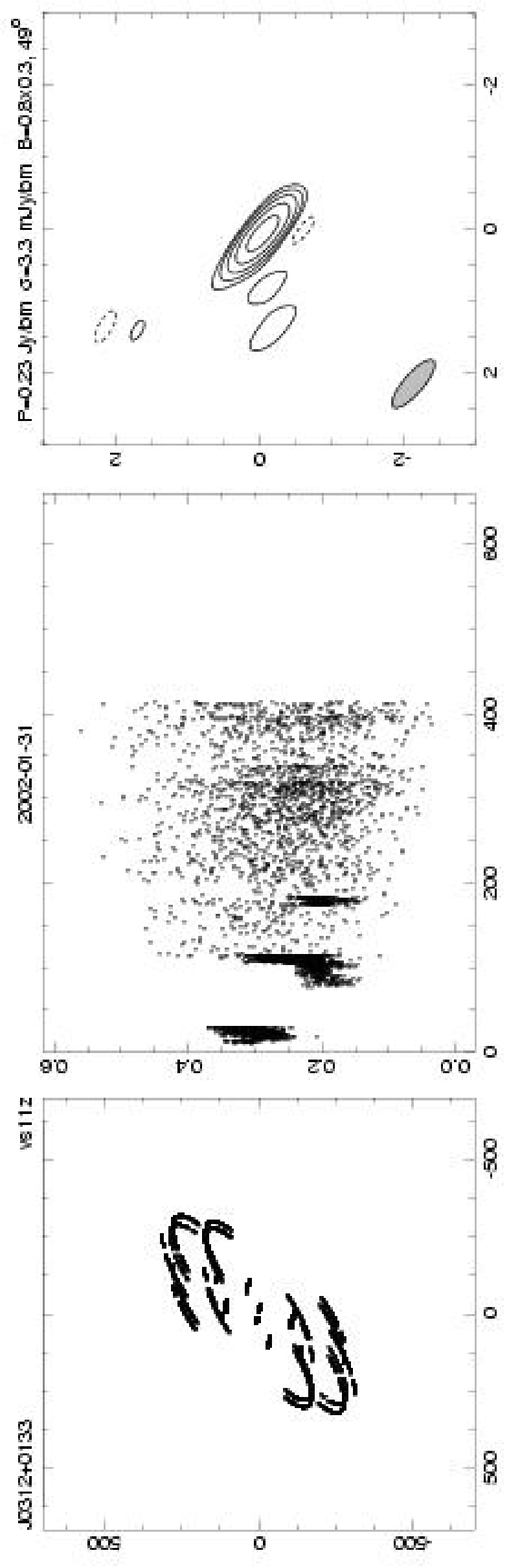}

{Fig. 1. -- {\em continued}}
\end{figure}
\clearpage
\begin{figure}
\spfig{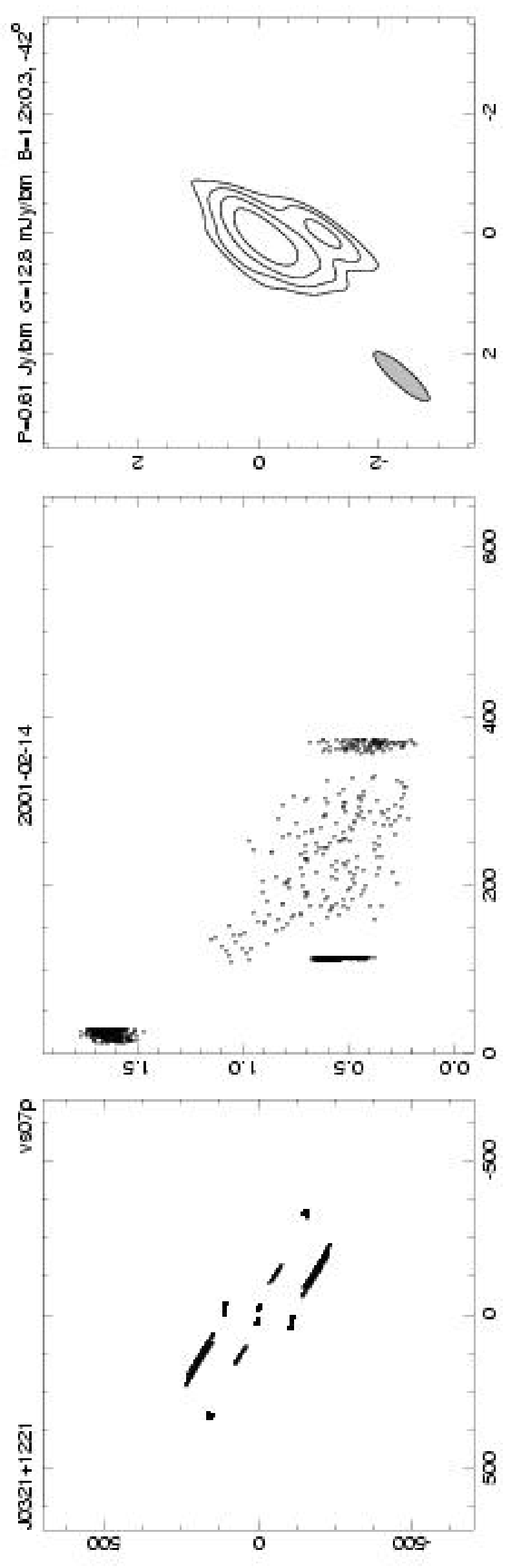}
\spfig{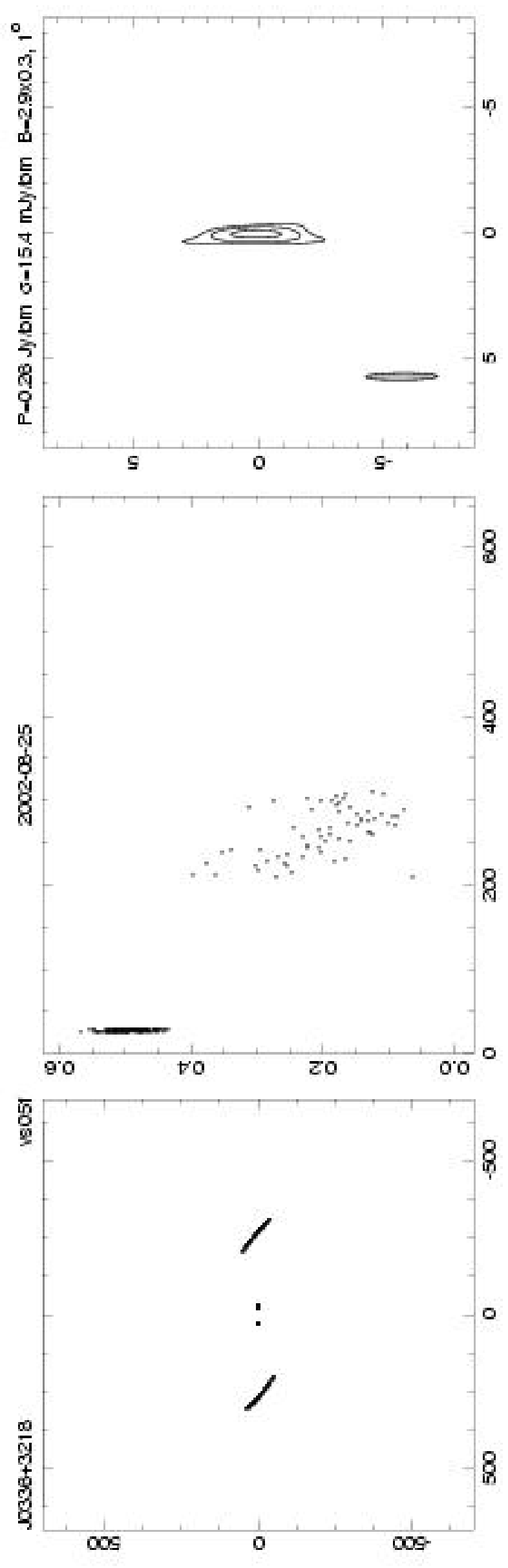}
\spfig{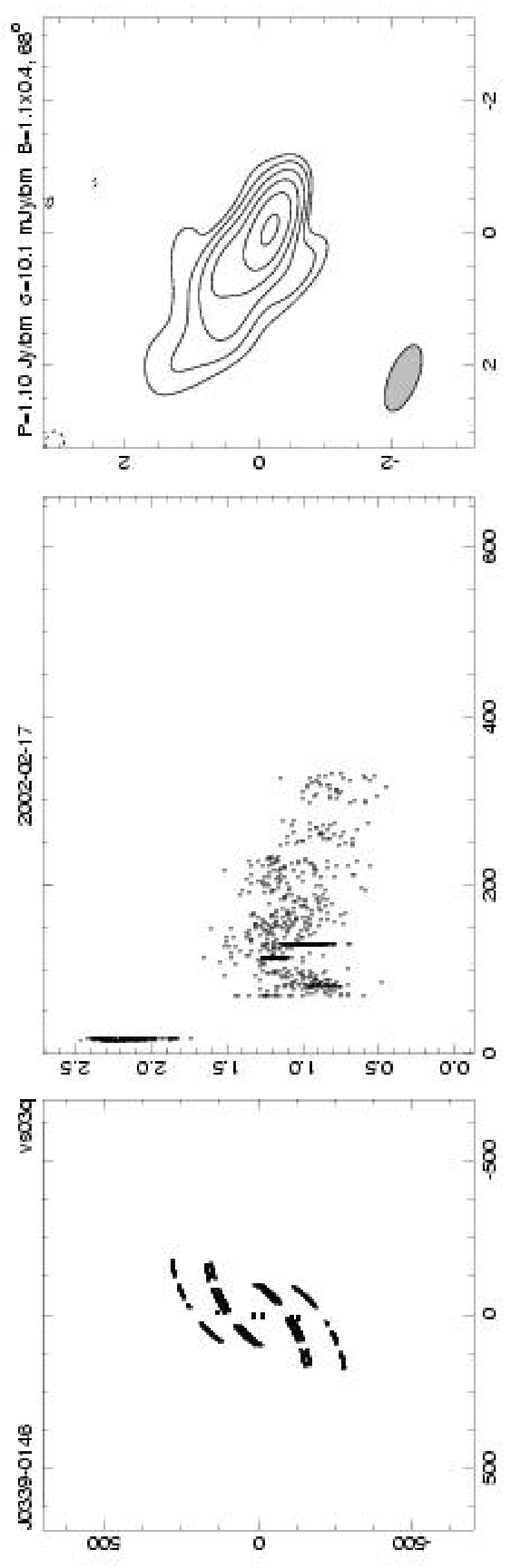}
\spfig{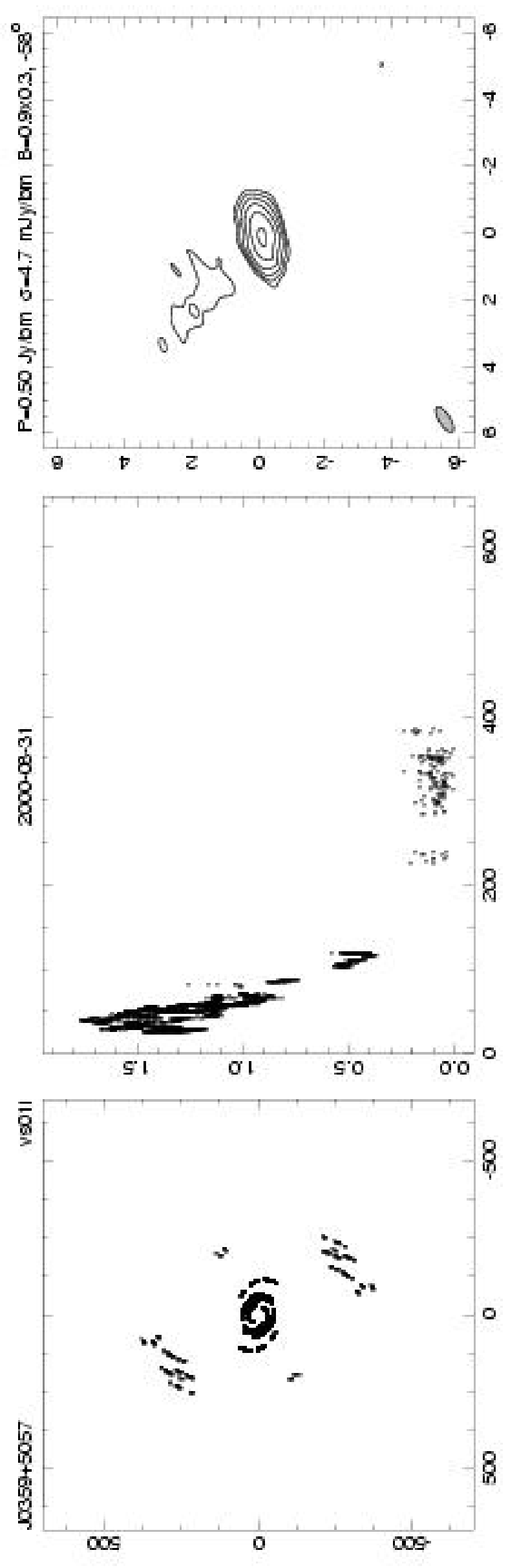} \typeout{no png}
\spfig{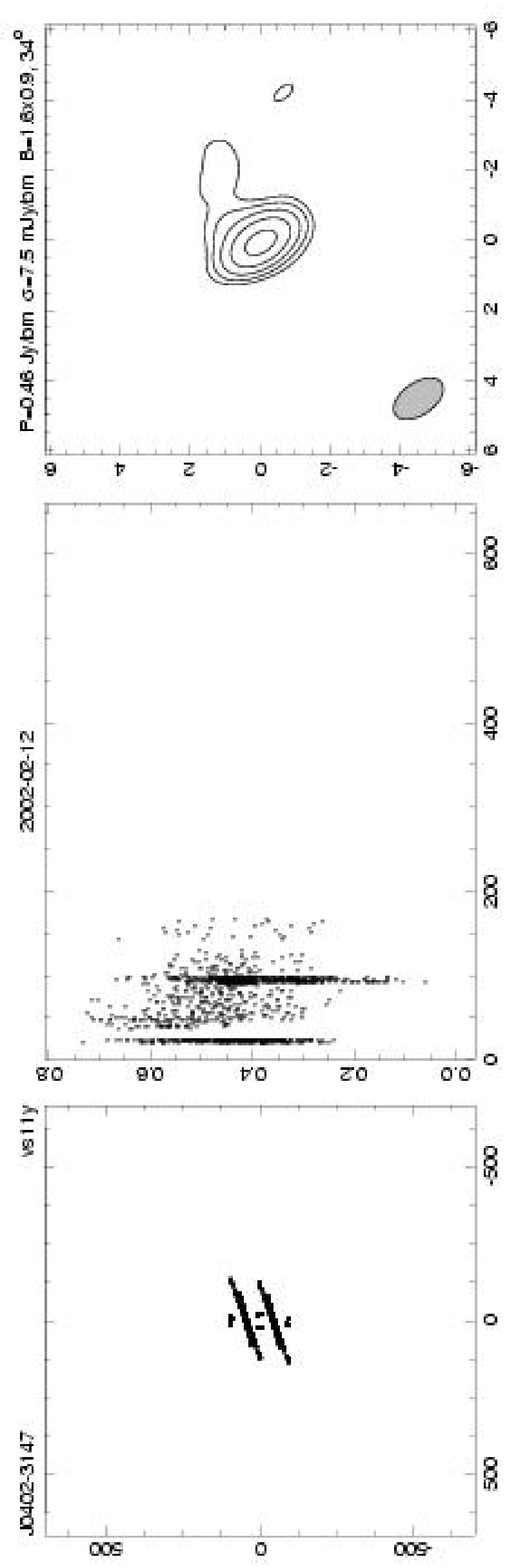}

{Fig. 1. -- {\em continued}}
\end{figure}
\clearpage
\begin{figure}
\spfig{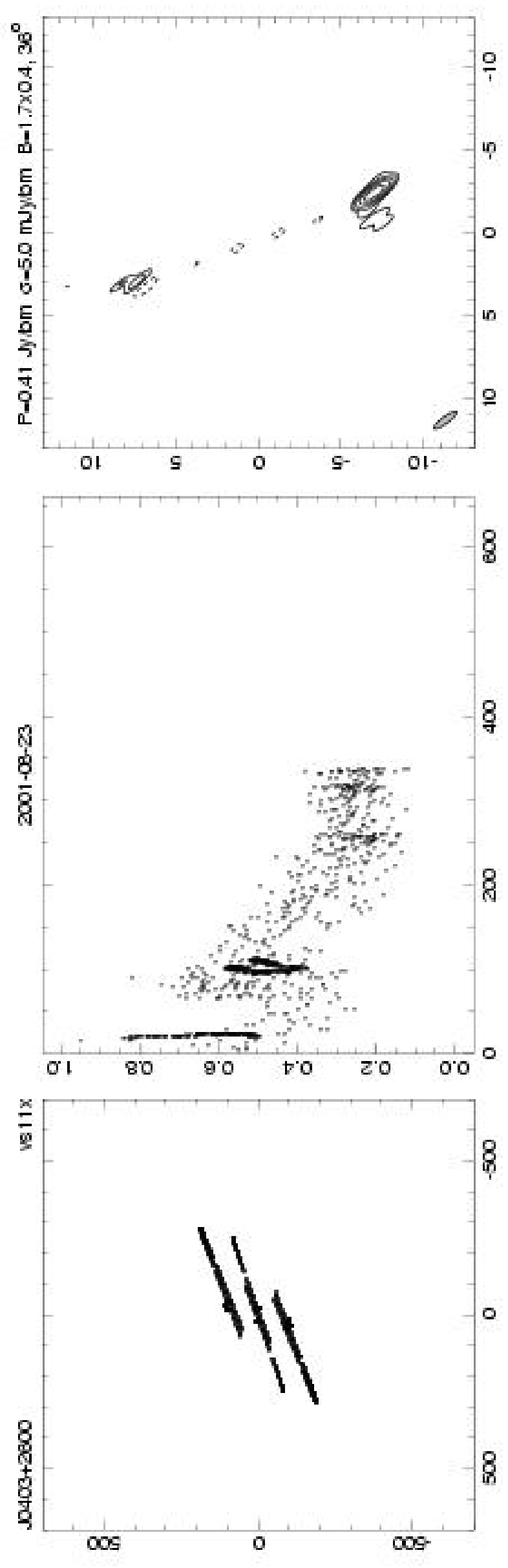} \typeout{needs a shift. Emission at 0,0 & 5,15}
\spfig{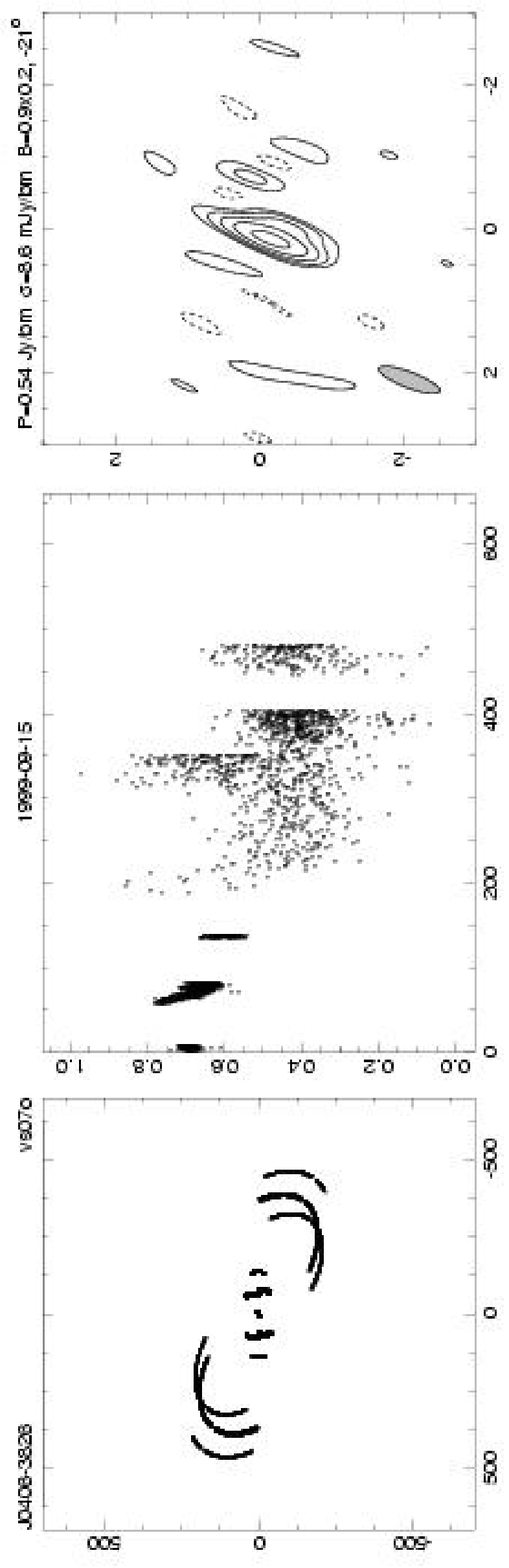}
\spfig{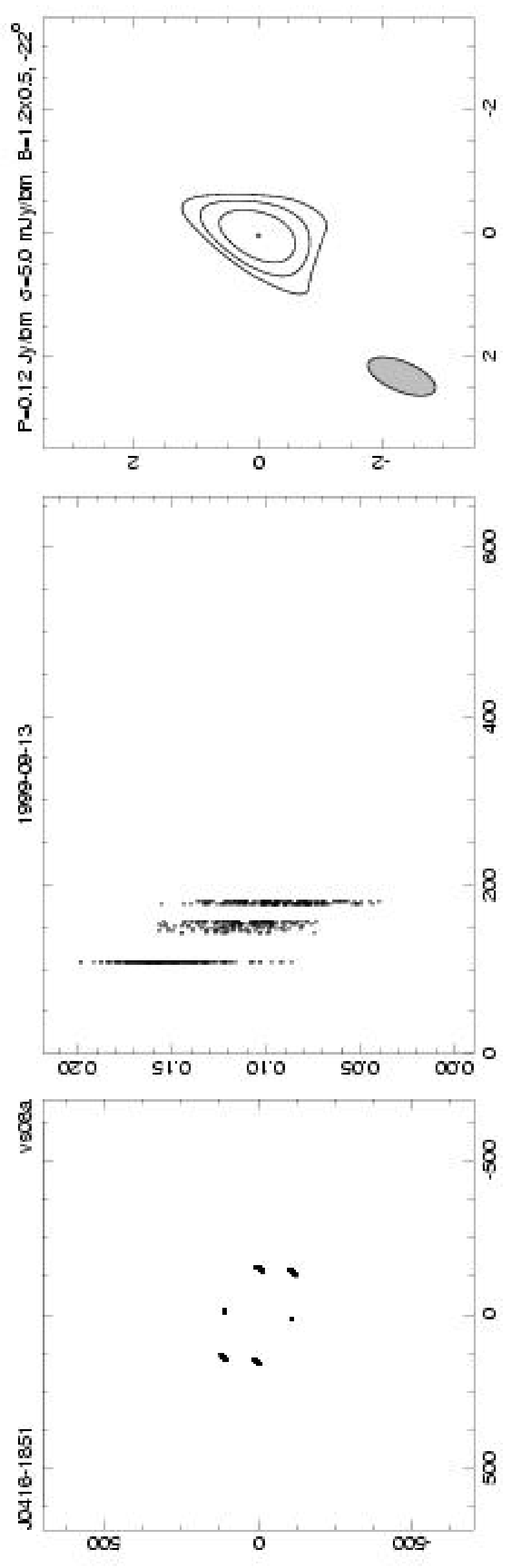}\typeout{no image?}
\spfig{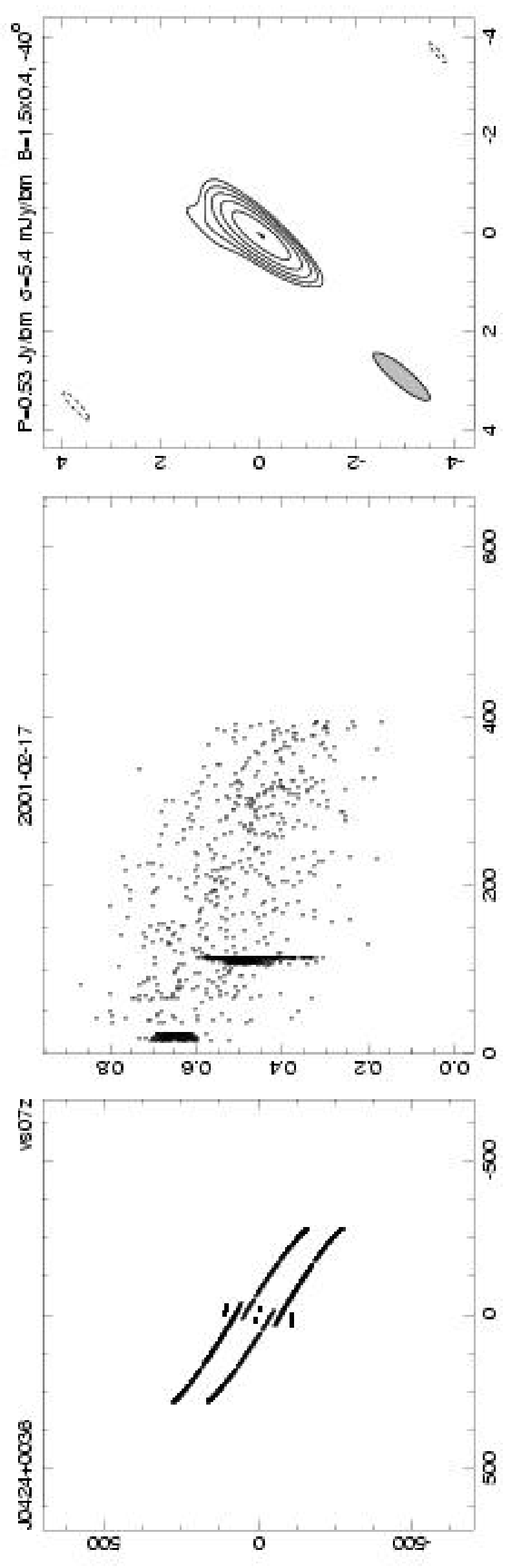}
\spfig{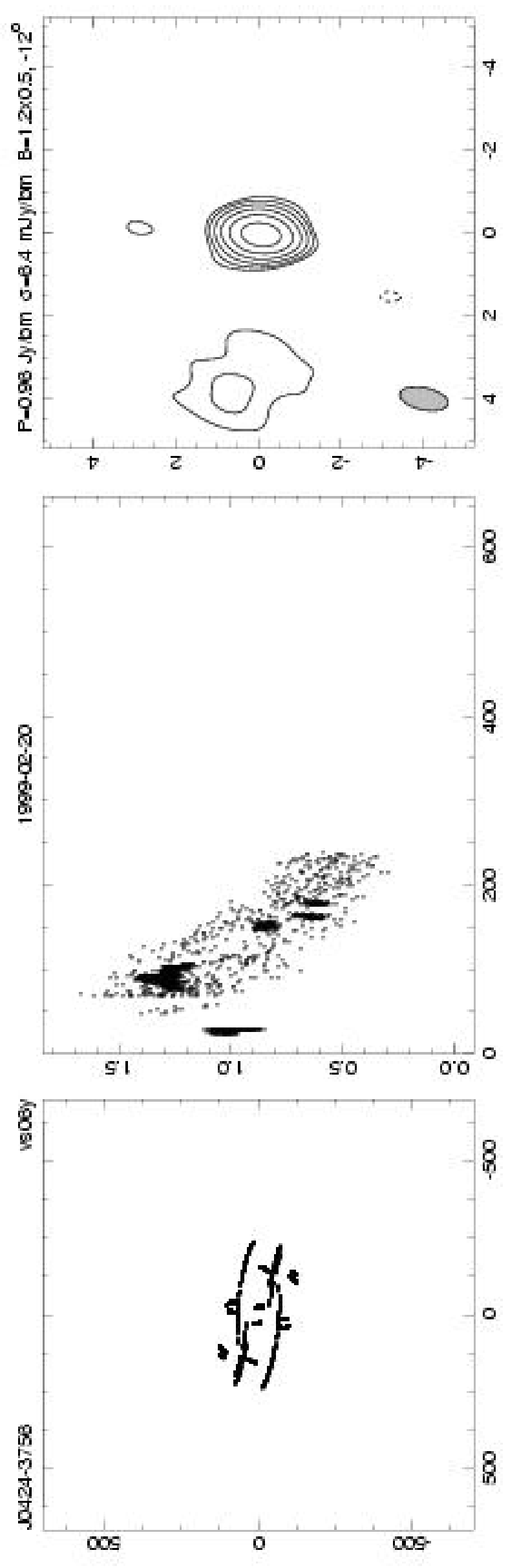}

{Fig. 1. -- {\em continued}}
\end{figure}
\clearpage
\begin{figure}
\spfig{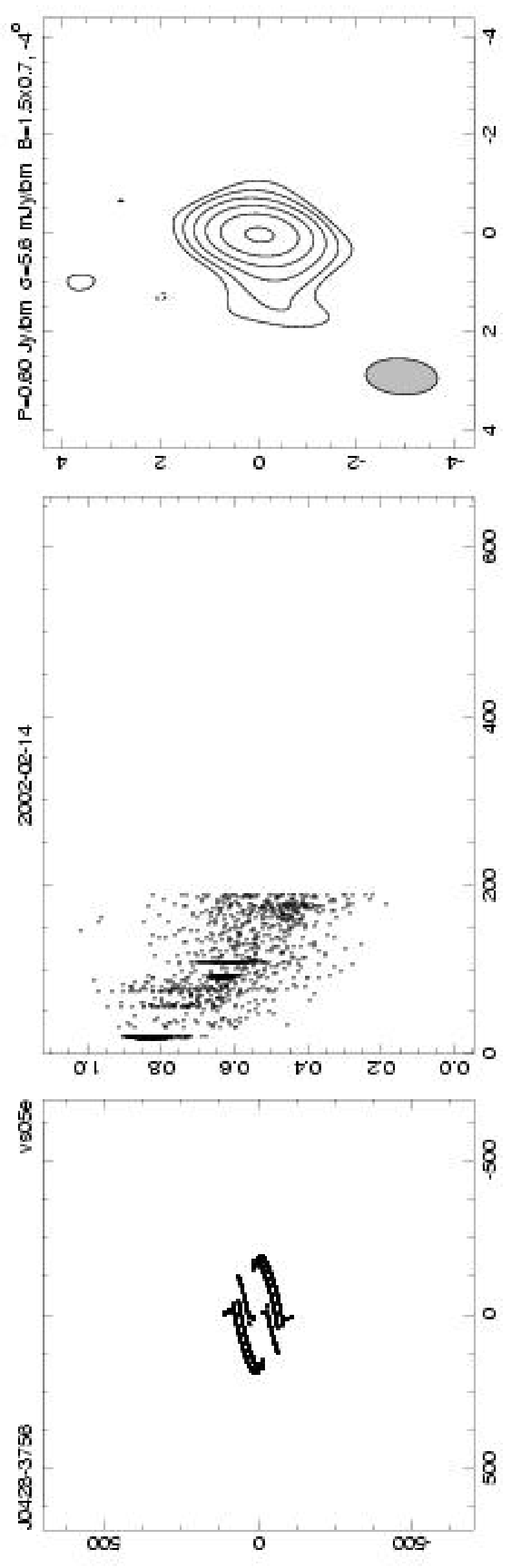}
\spfig{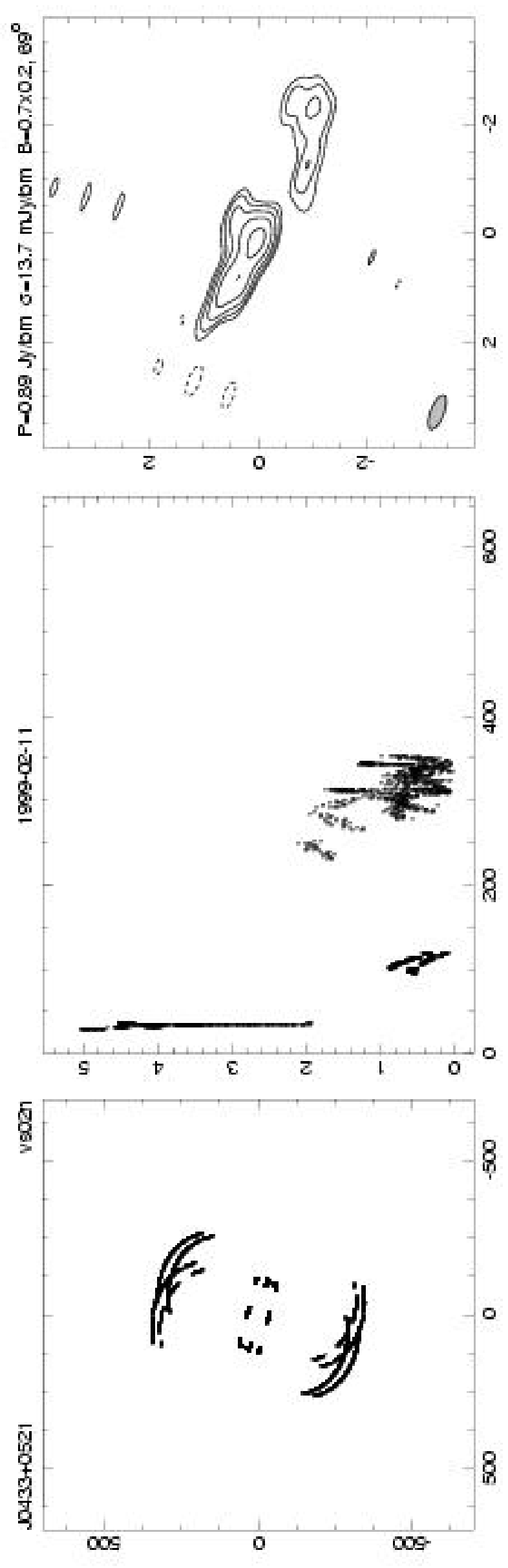}
\spfig{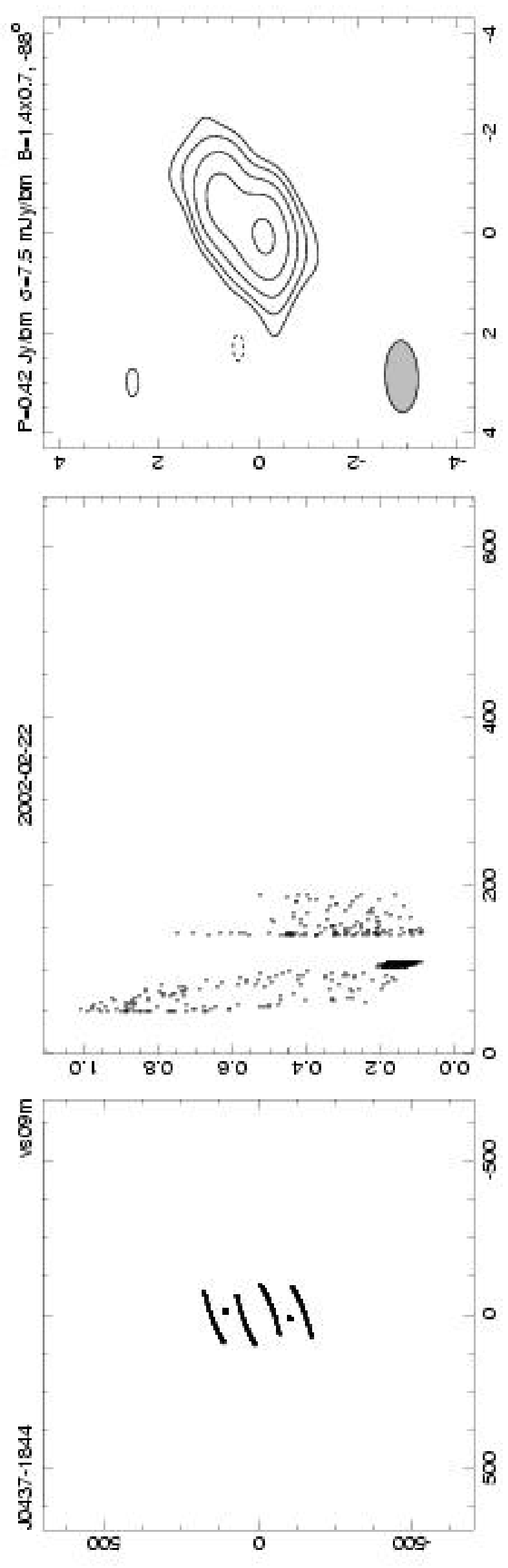}
\spfig{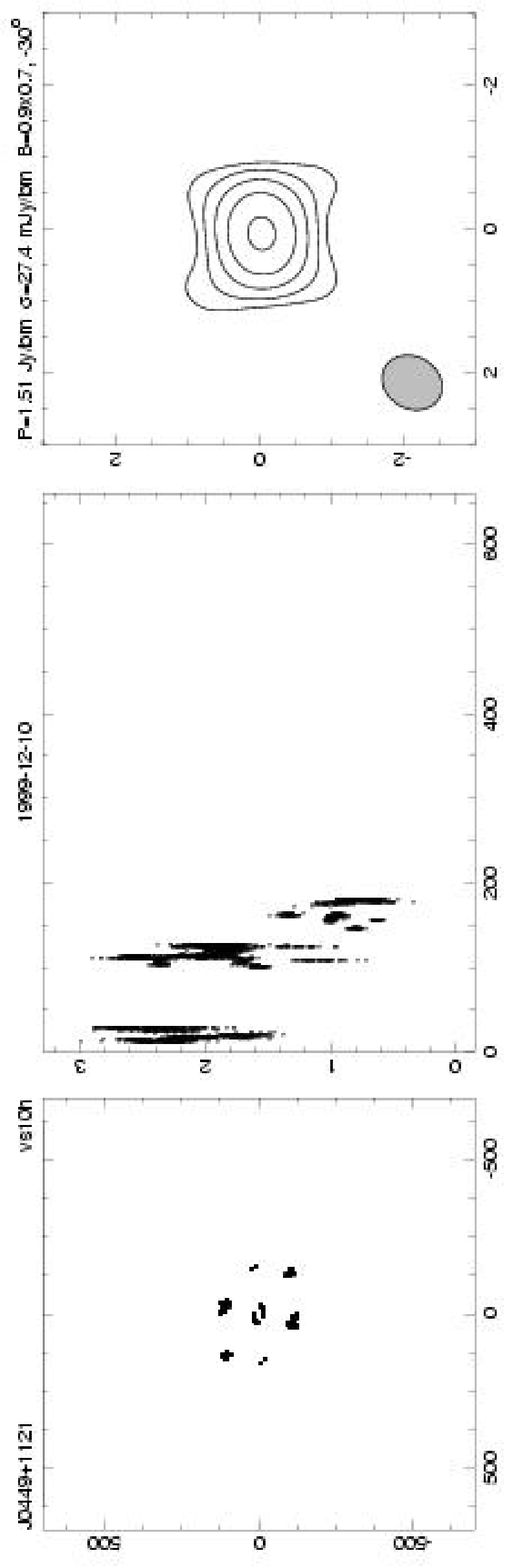}
\spfig{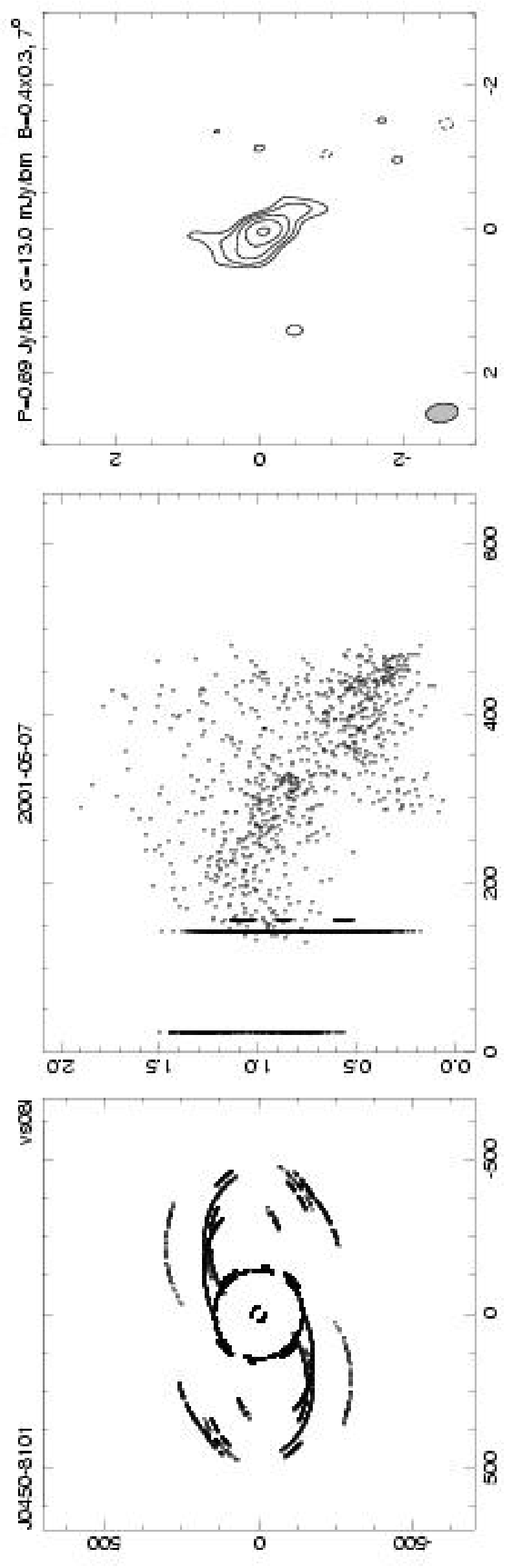}

{Fig. 1. -- {\em continued}}
\end{figure}
\clearpage
\begin{figure}
\spfig{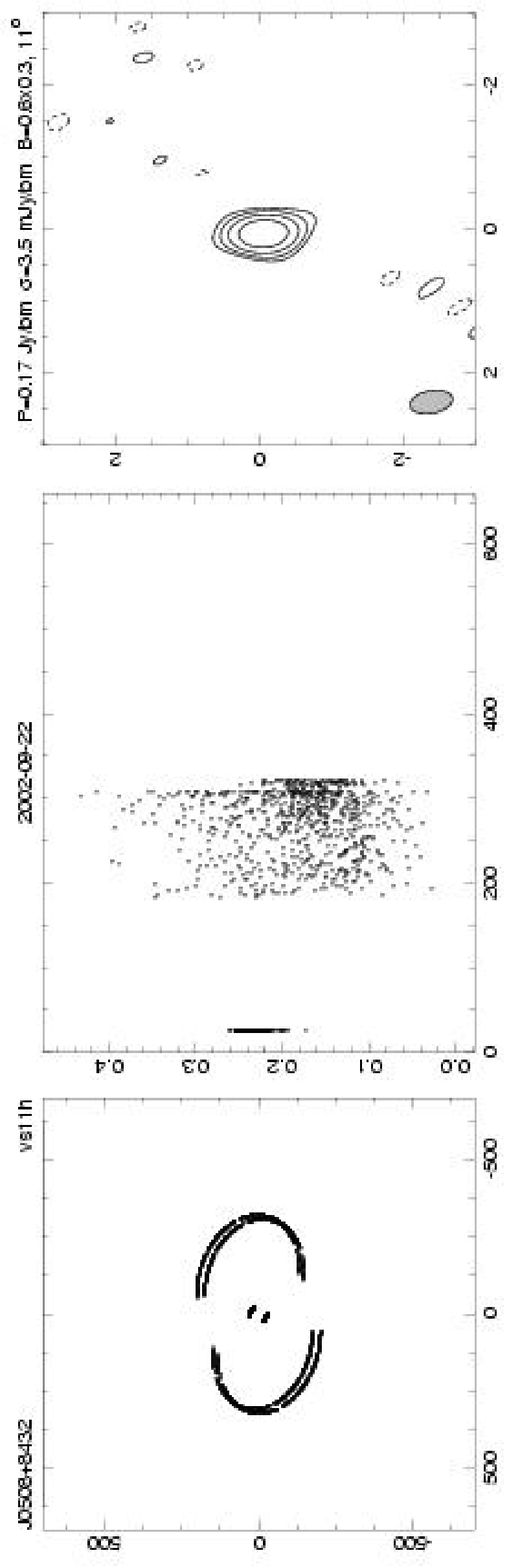}
\spfig{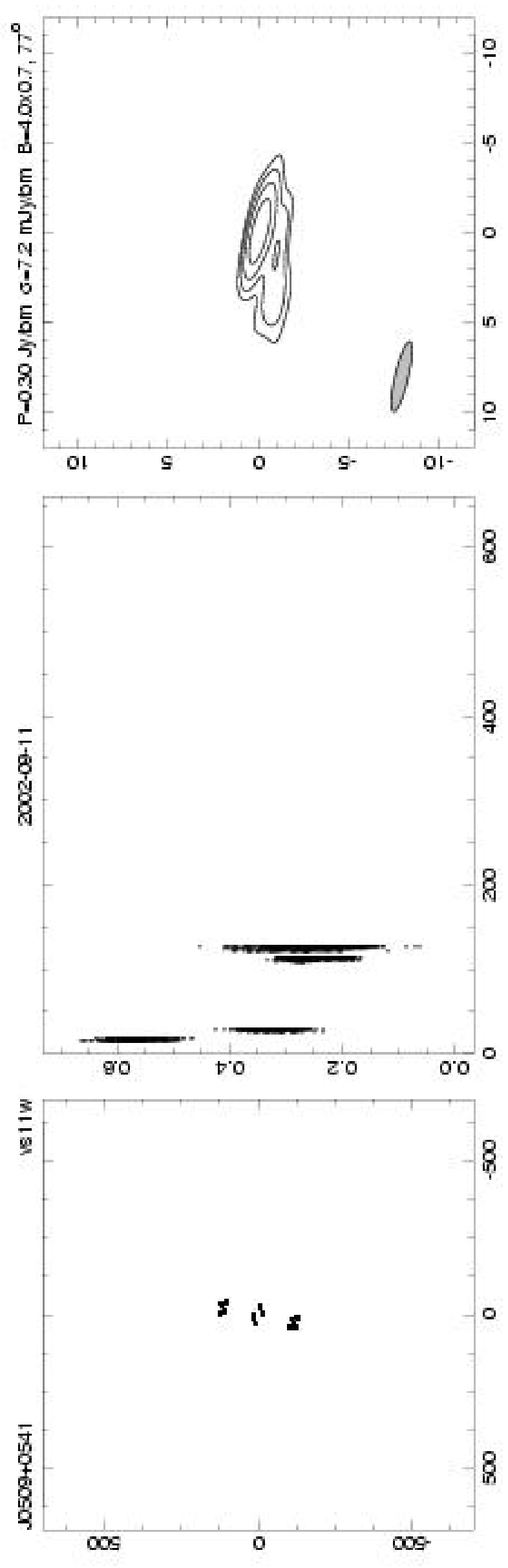}
\spfig{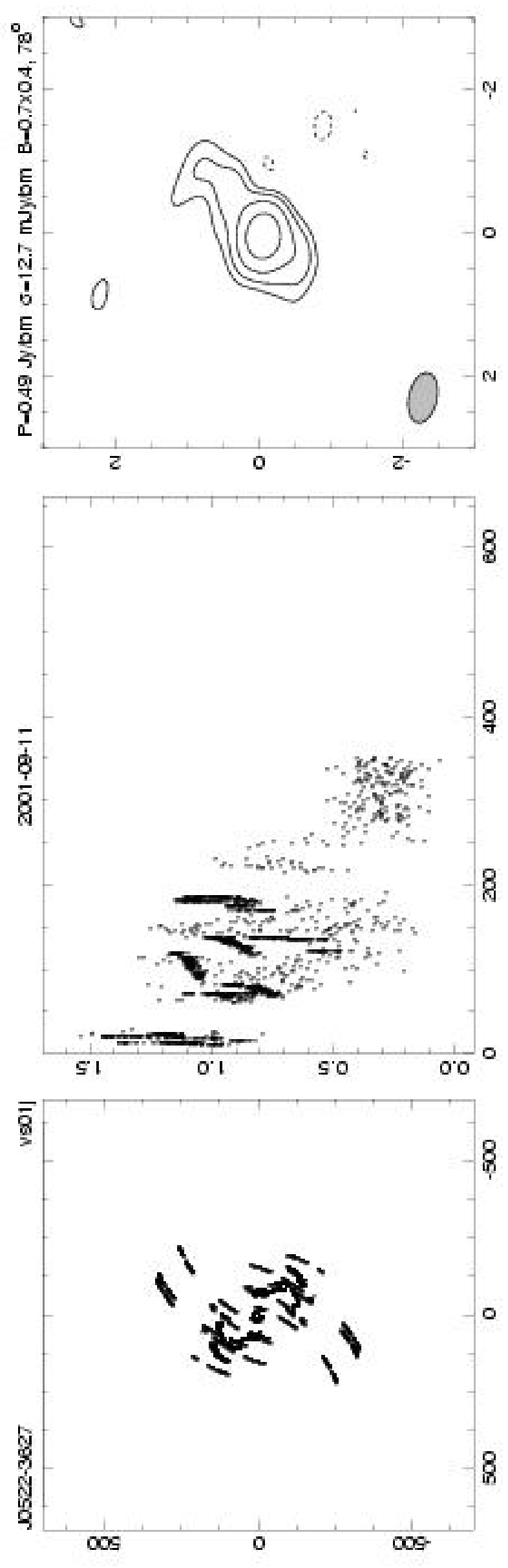}
\spfig{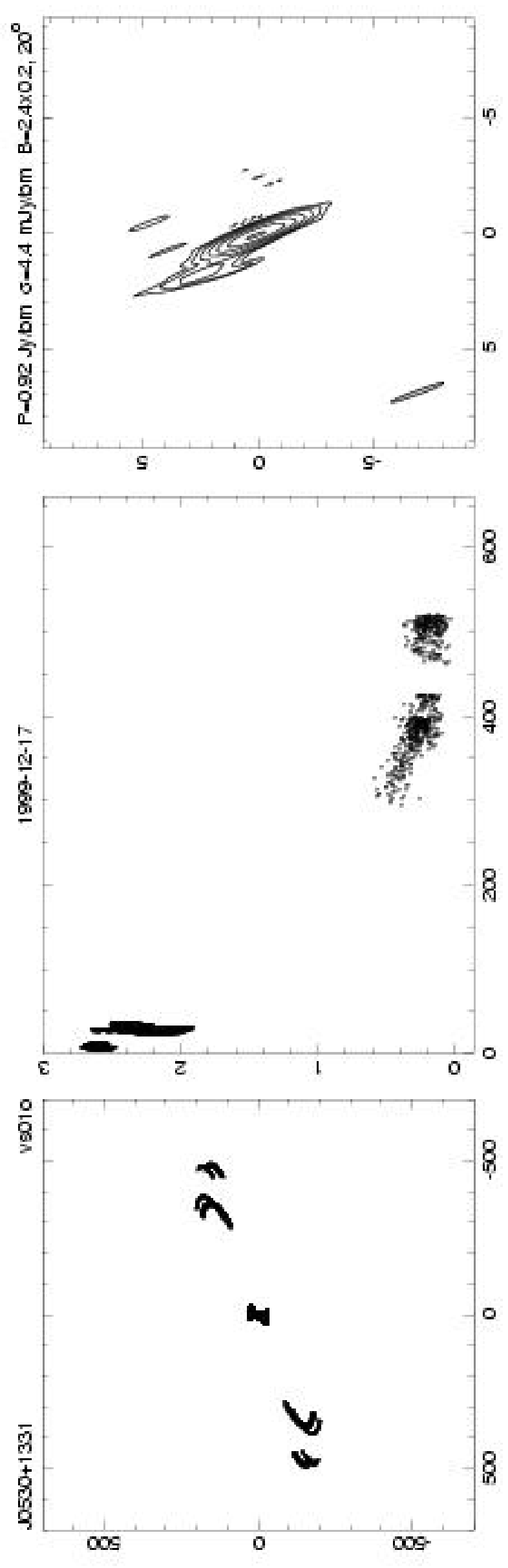} \typeout{different png ps. plot flux is too high}
\spfig{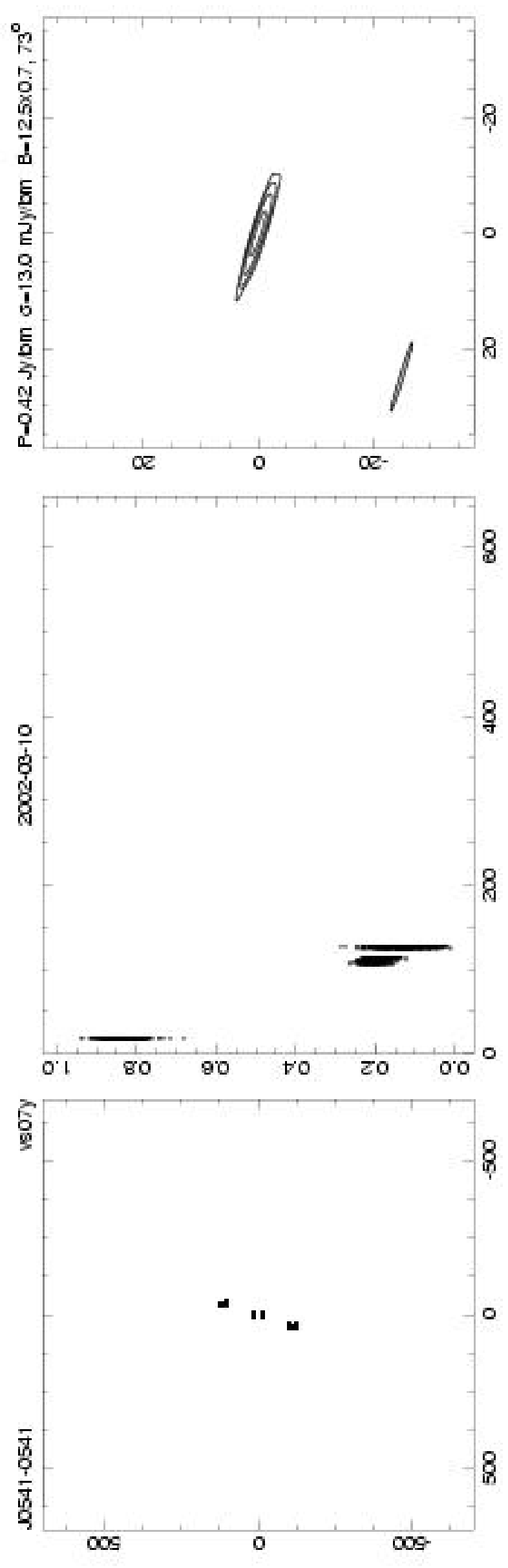} \typeout{no image?}

{Fig. 1. -- {\em continued}}
\end{figure}
\clearpage
\begin{figure}
\spfig{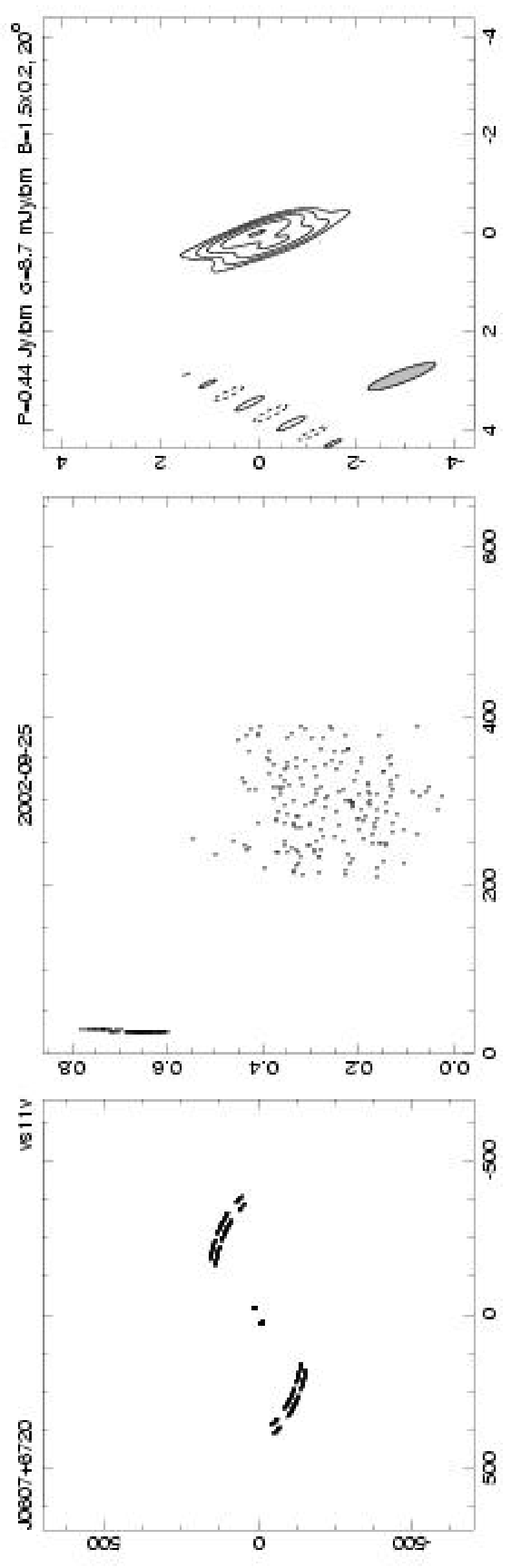}
\spfig{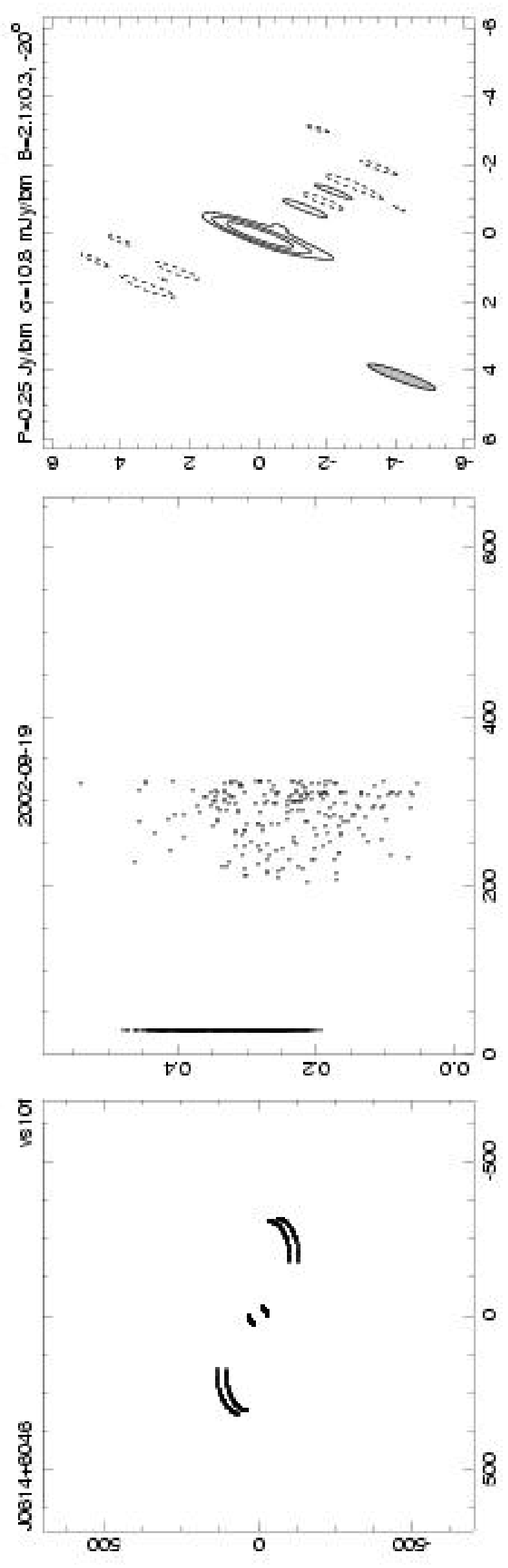}
\spfig{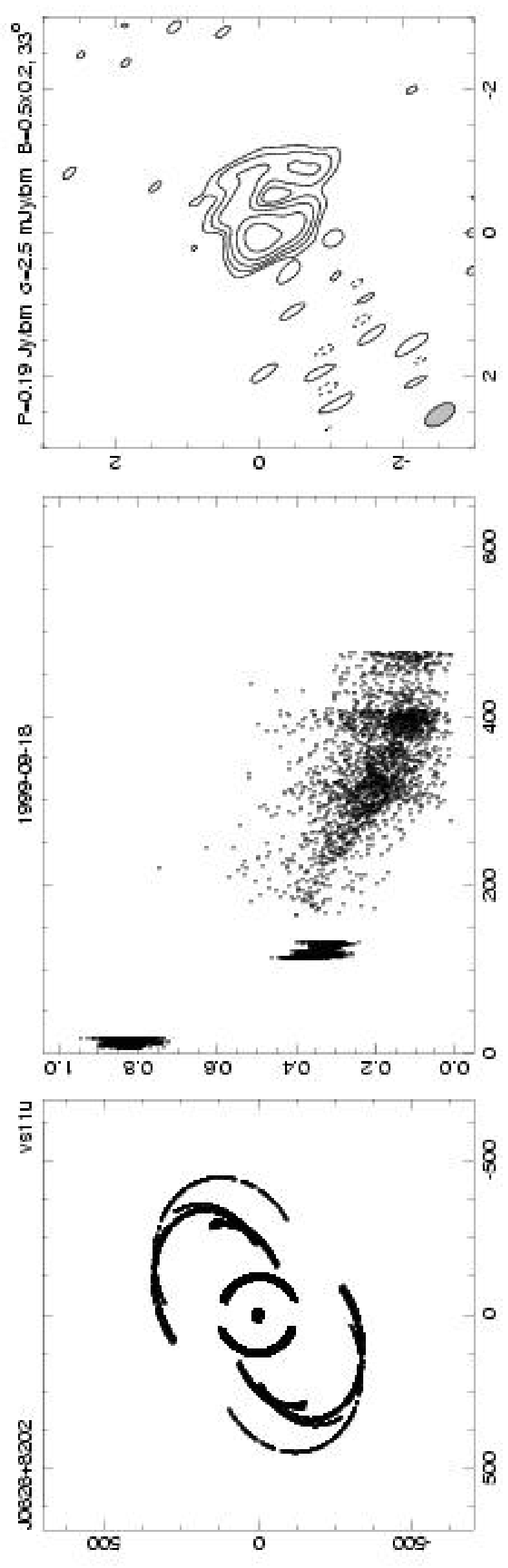}
\spfig{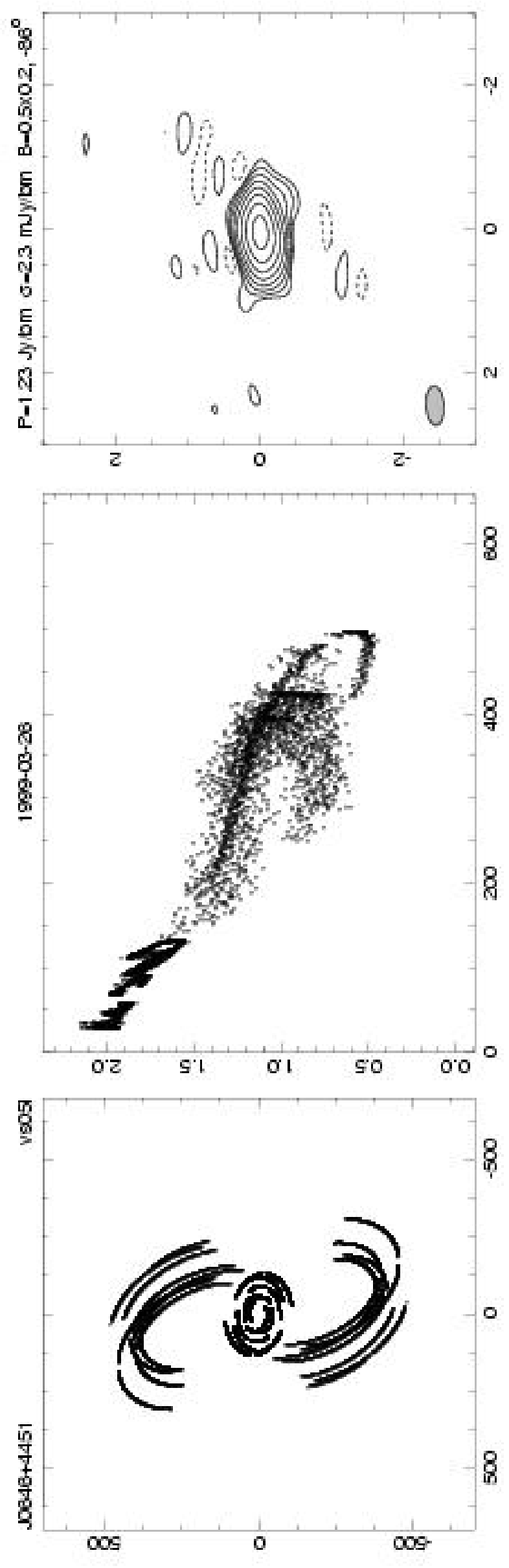}
\spfig{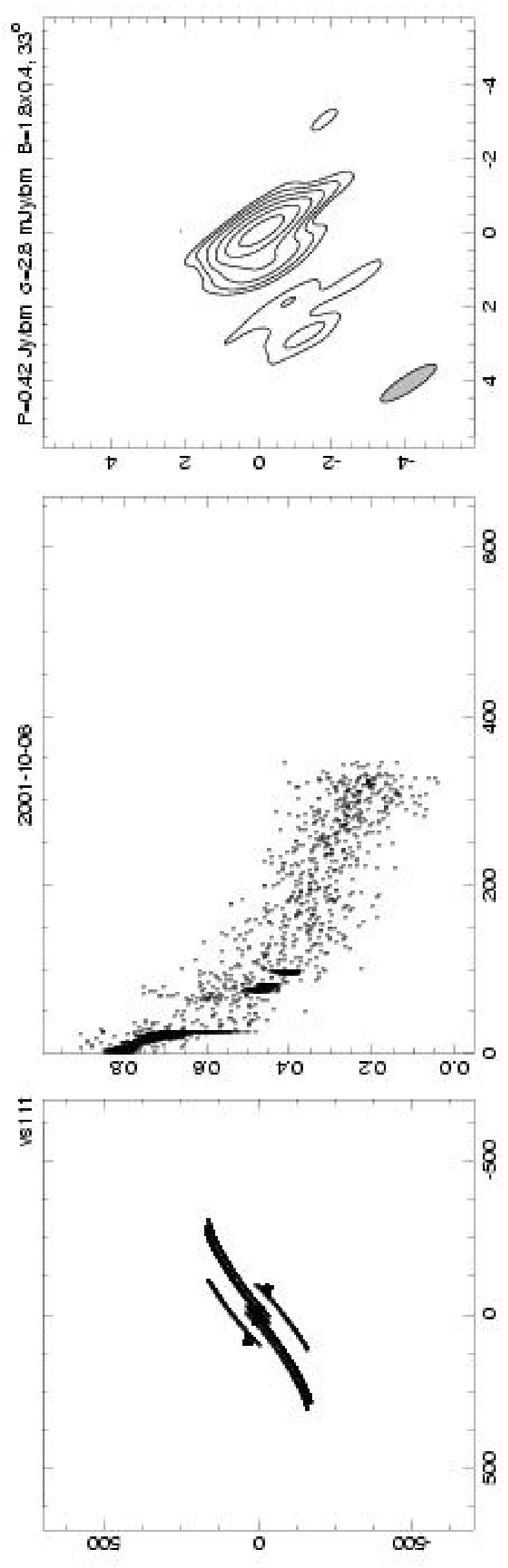}

{Fig. 1. -- {\em continued}}
\end{figure}
\clearpage
\begin{figure}
\spfig{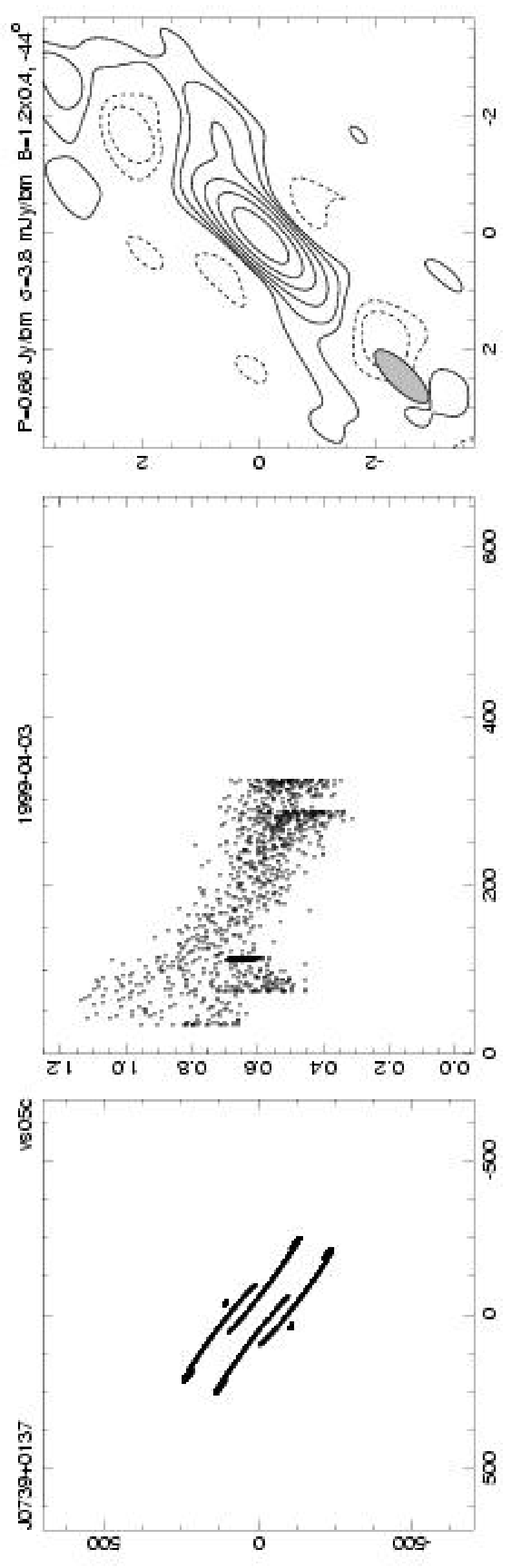}
\spfig{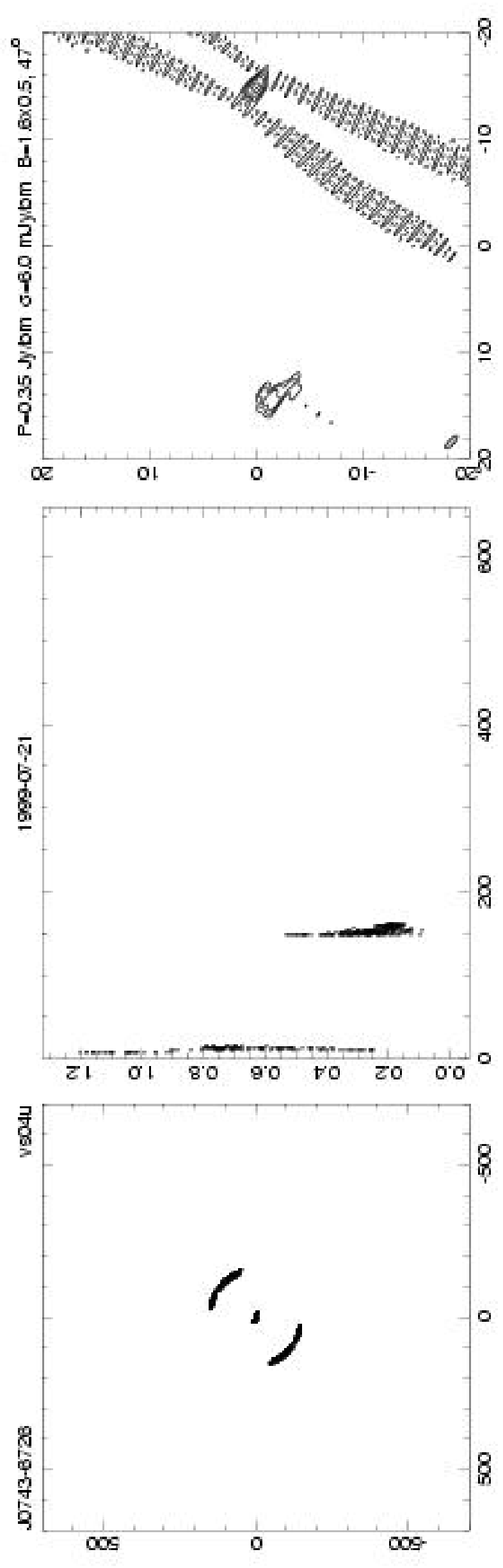} \typeout{different png ps. levels too low}
\spfig{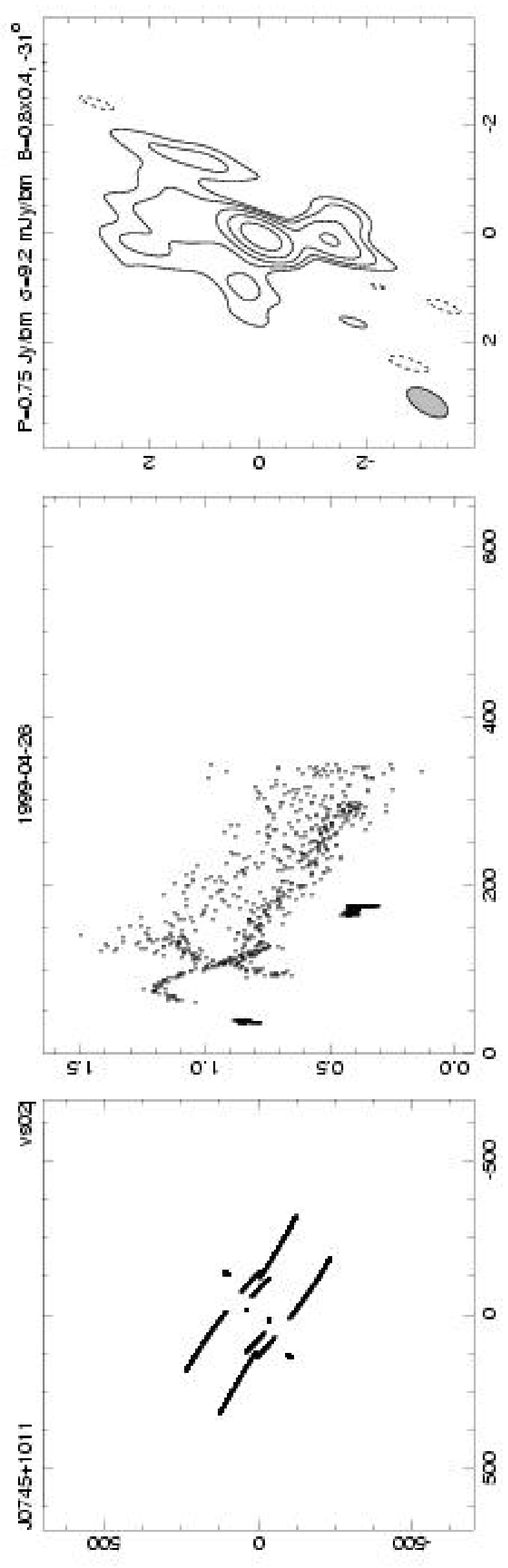}
\spfig{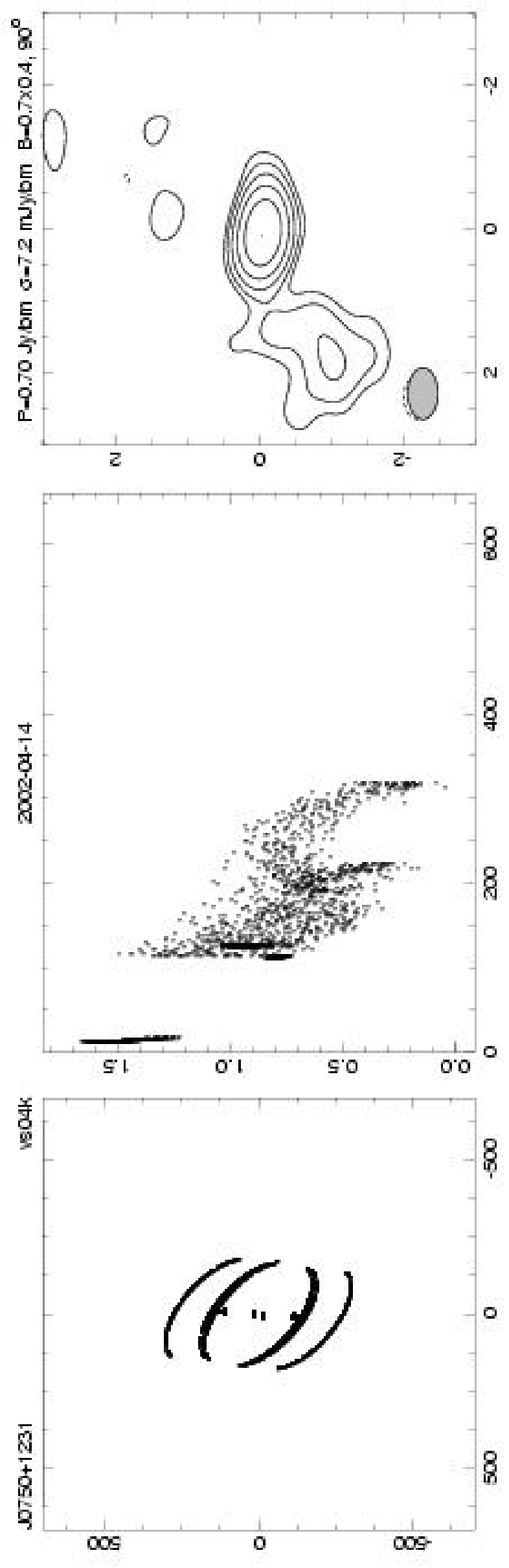}
\spfig{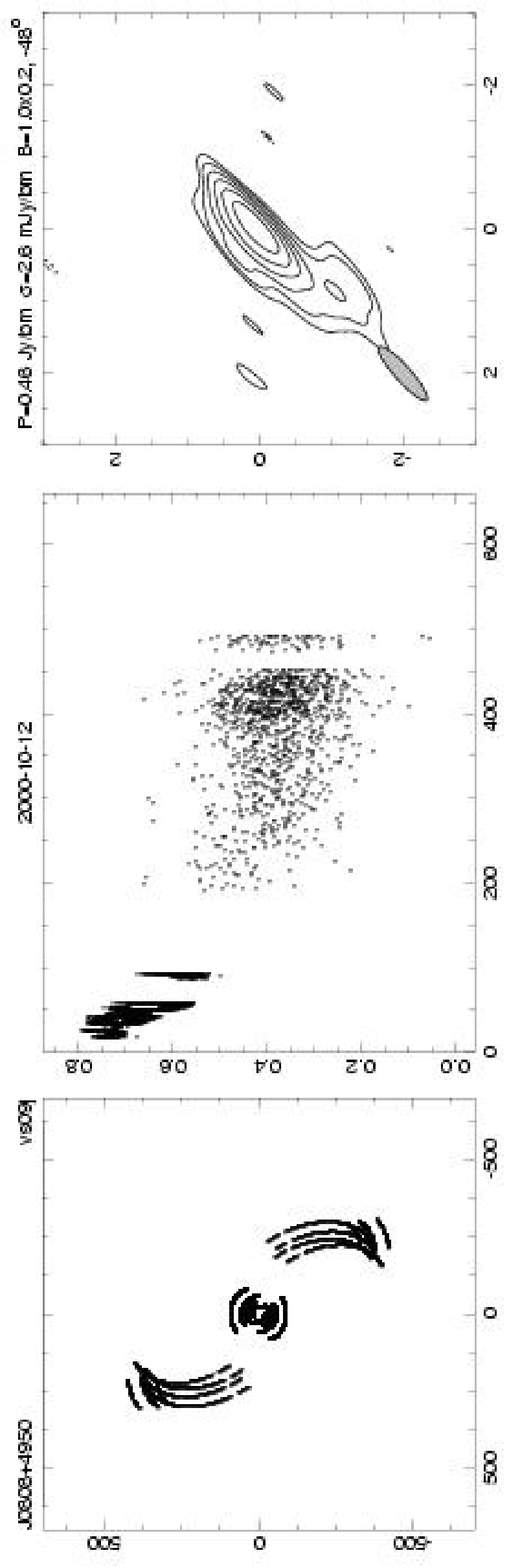}

{Fig. 1. -- {\em continued}}
\end{figure}
\clearpage
\begin{figure}
\spfig{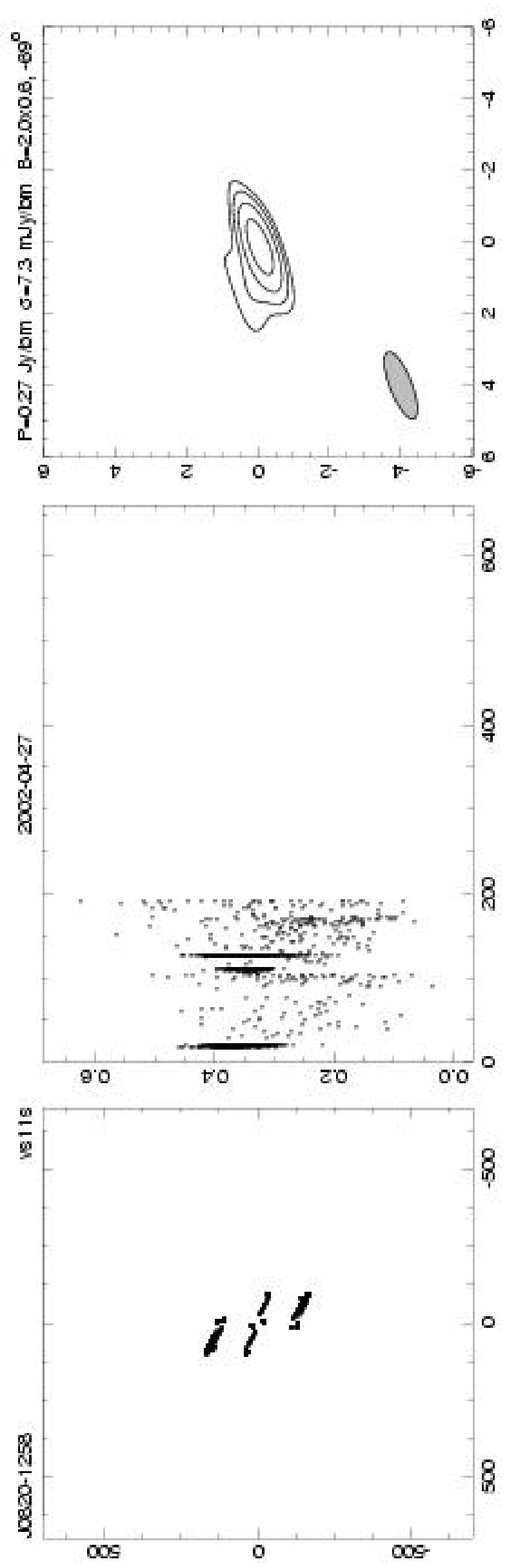}
\spfig{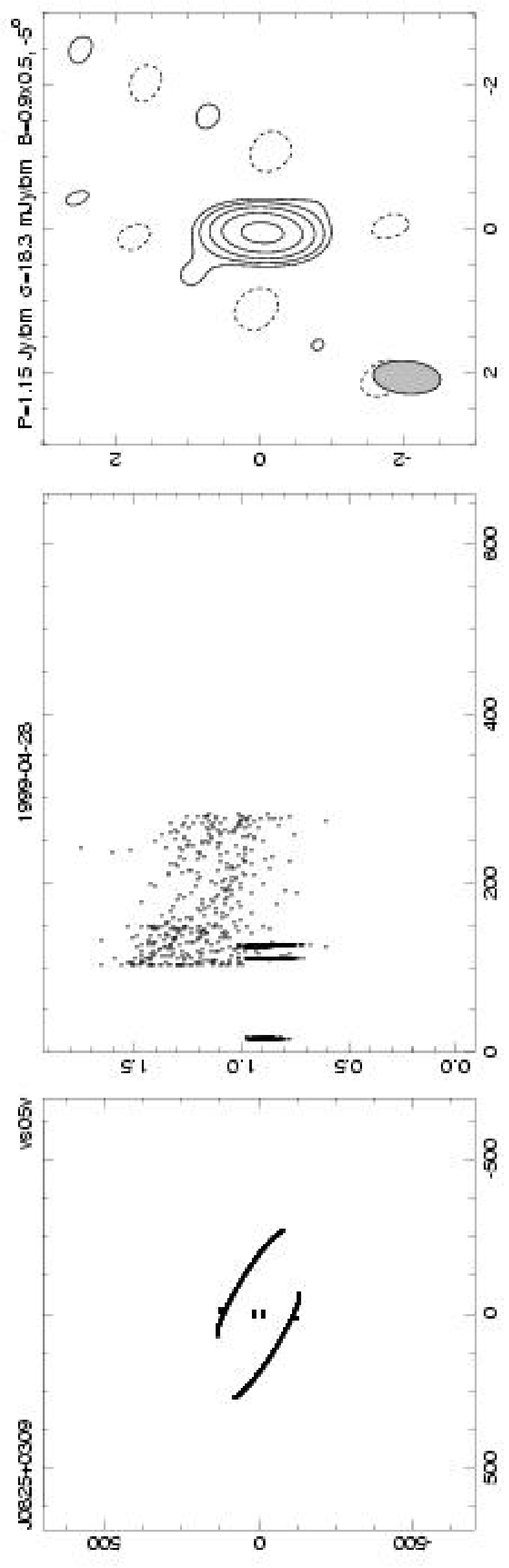}
\spfig{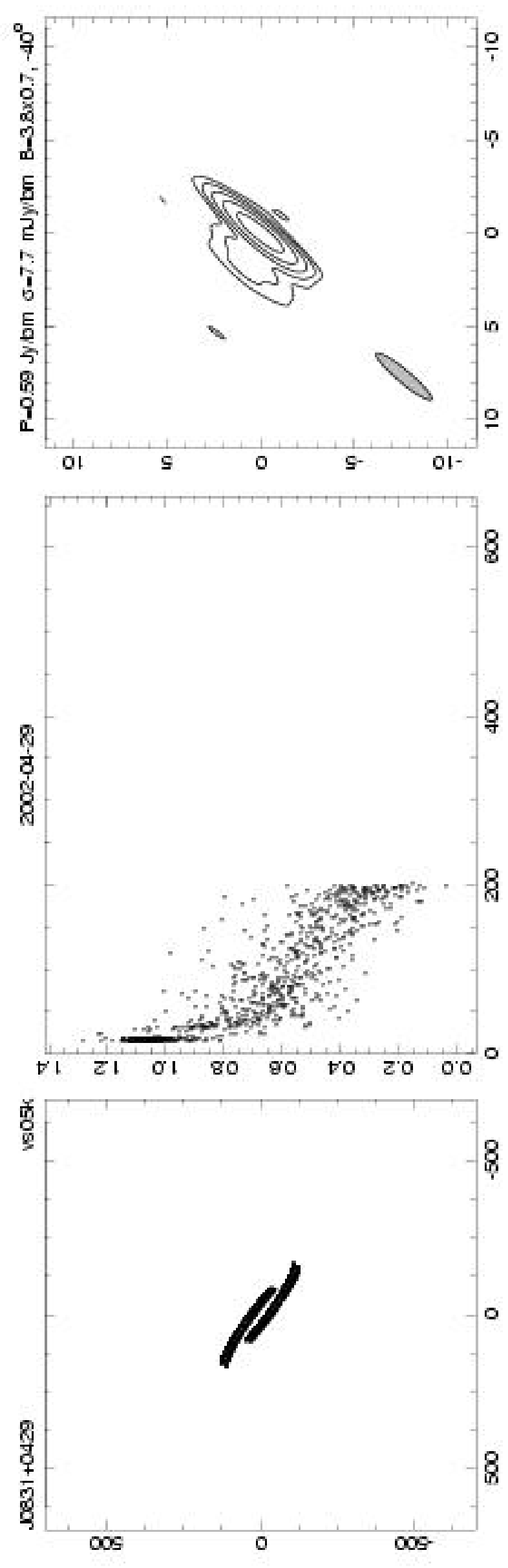}
\spfig{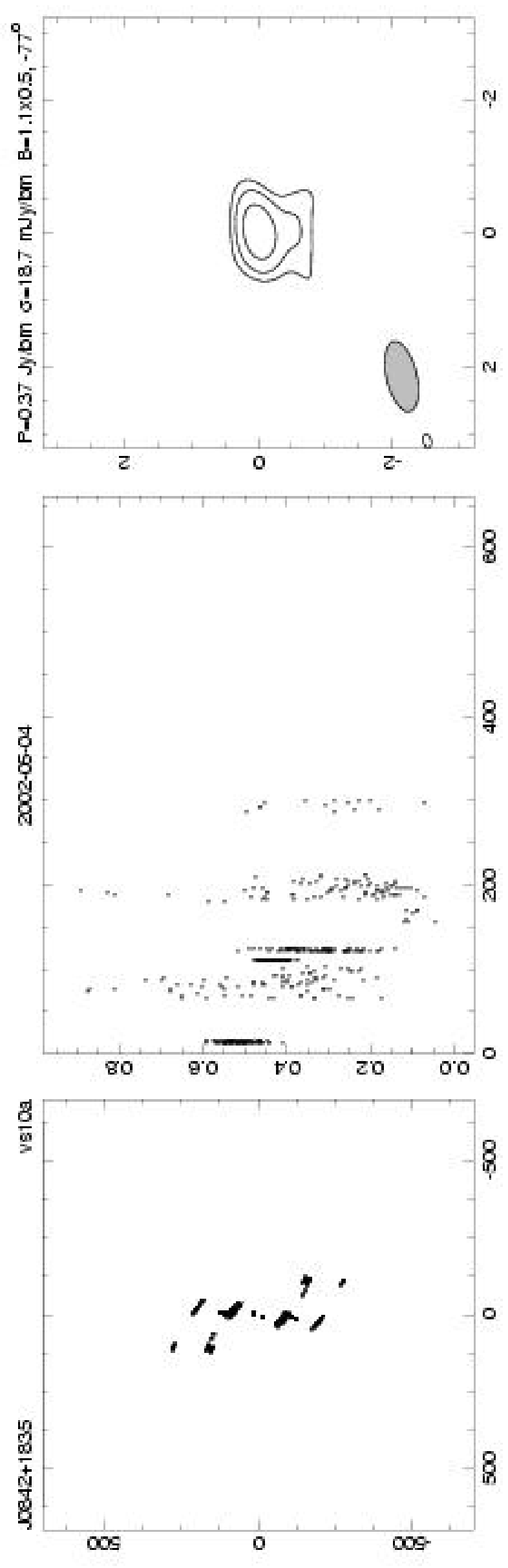}
\spfig{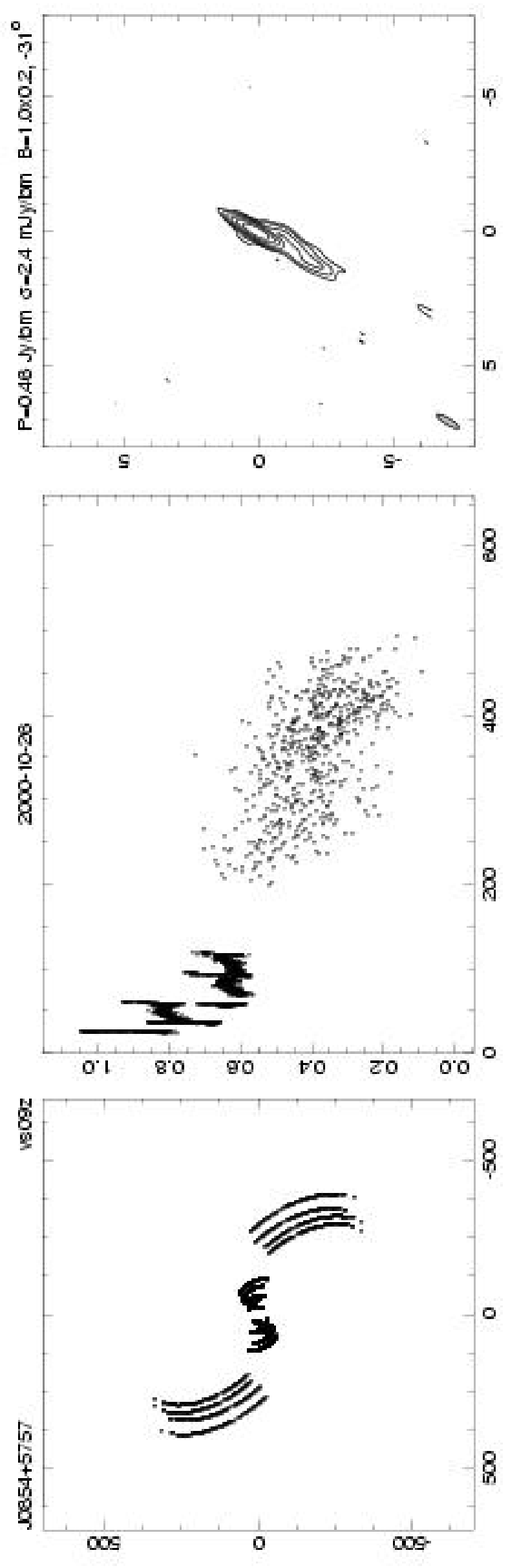}

{Fig. 1. -- {\em continued}}
\end{figure}
\clearpage
\begin{figure}
\spfig{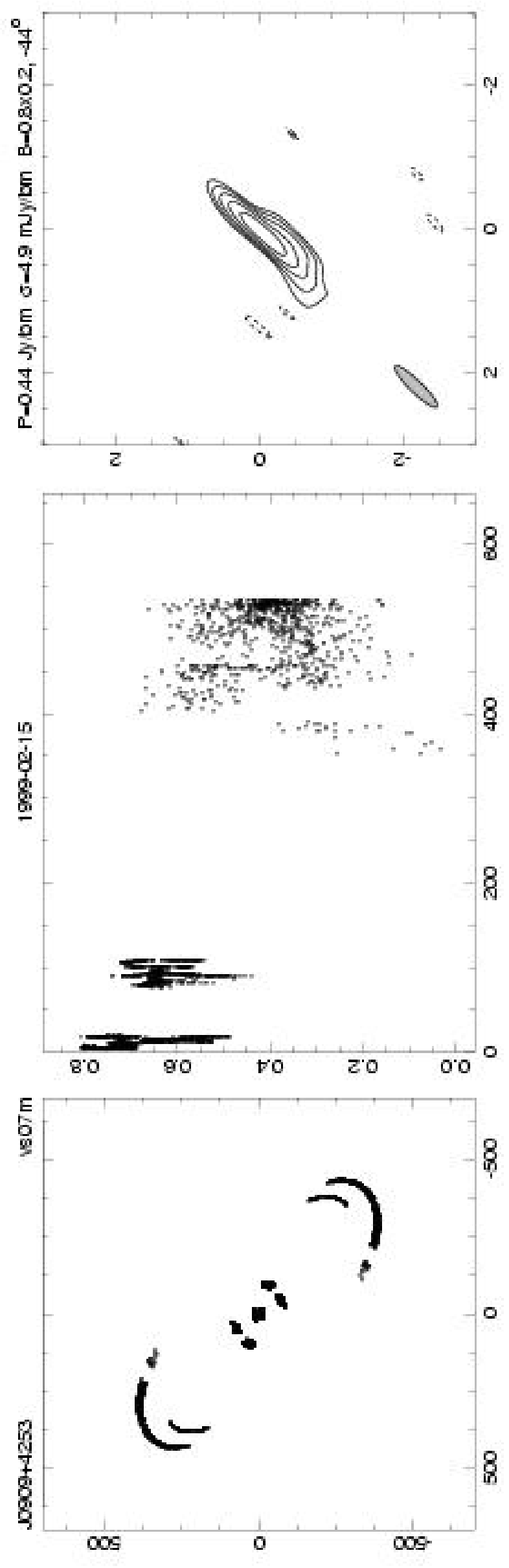}
\spfig{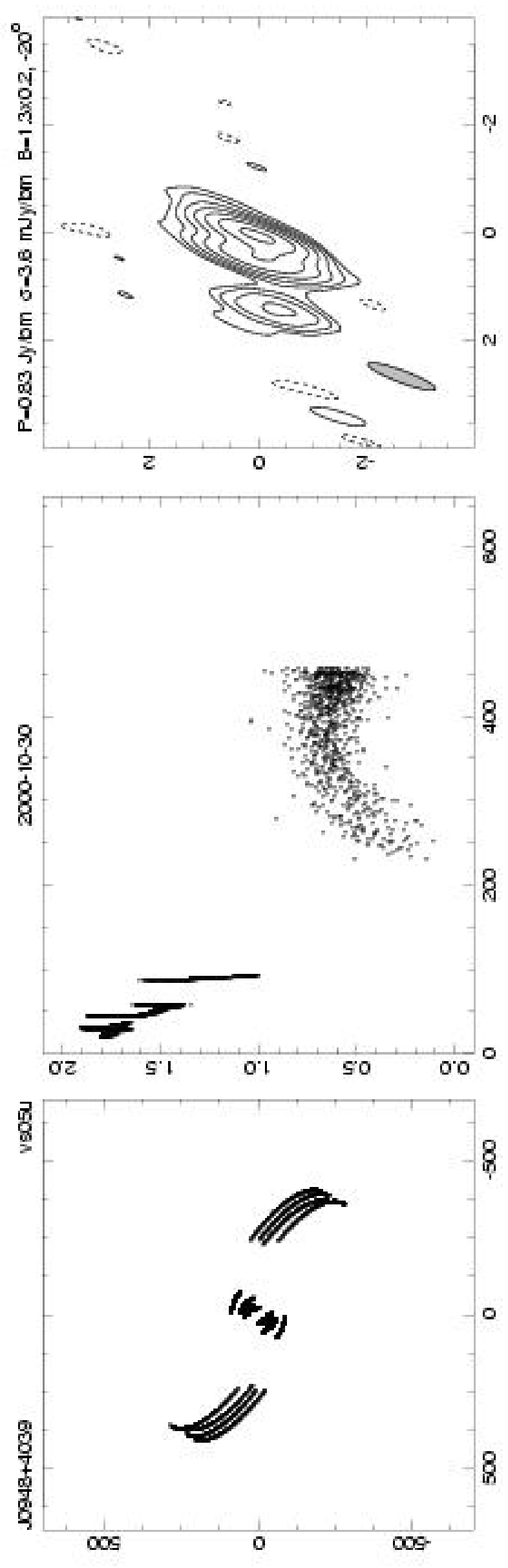}
\spfig{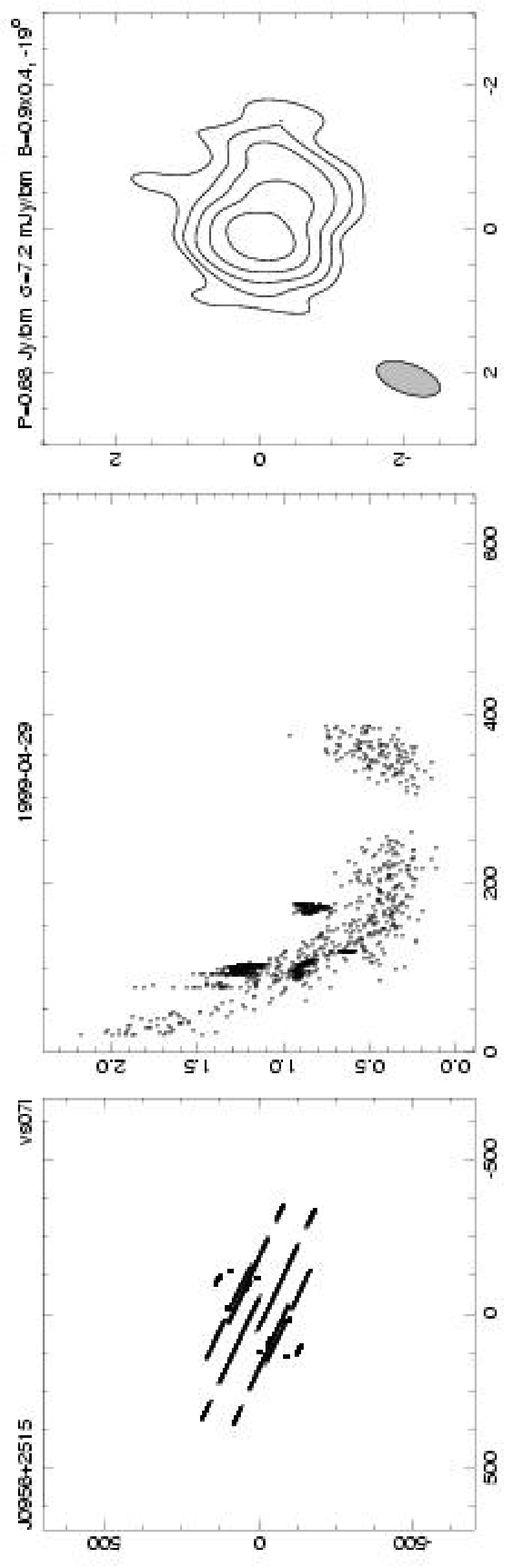}
\spfig{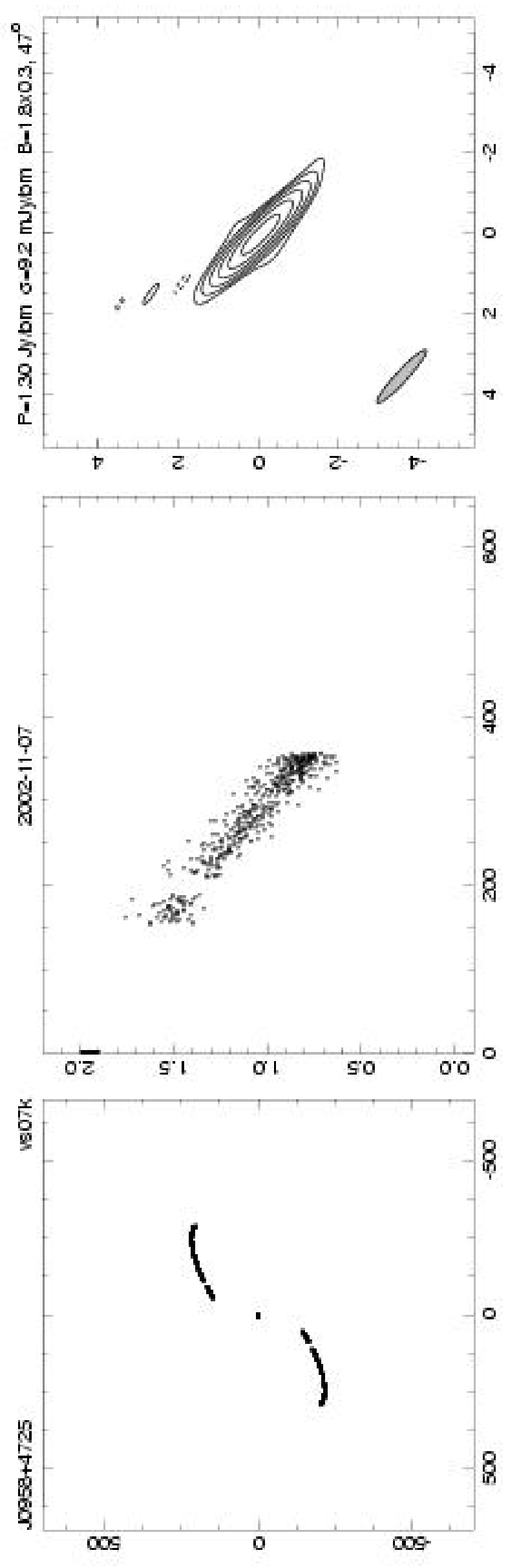}
\spfig{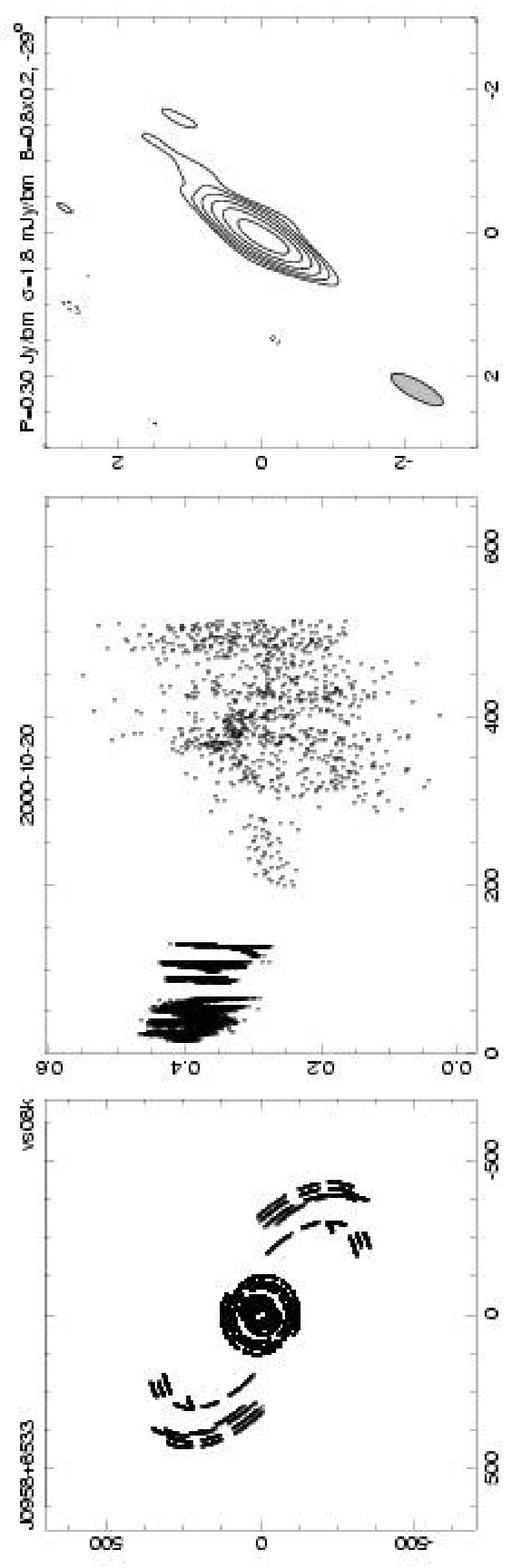}

{Fig. 1. -- {\em continued}}
\end{figure}
\clearpage
\begin{figure}
\spfig{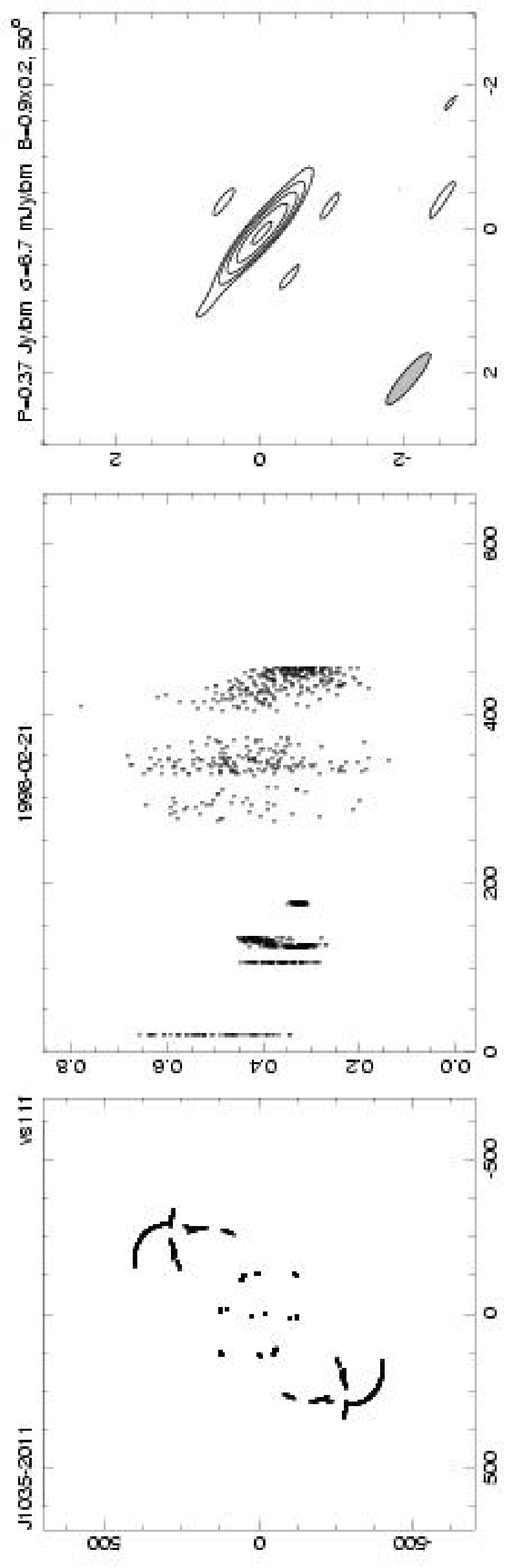}
\spfig{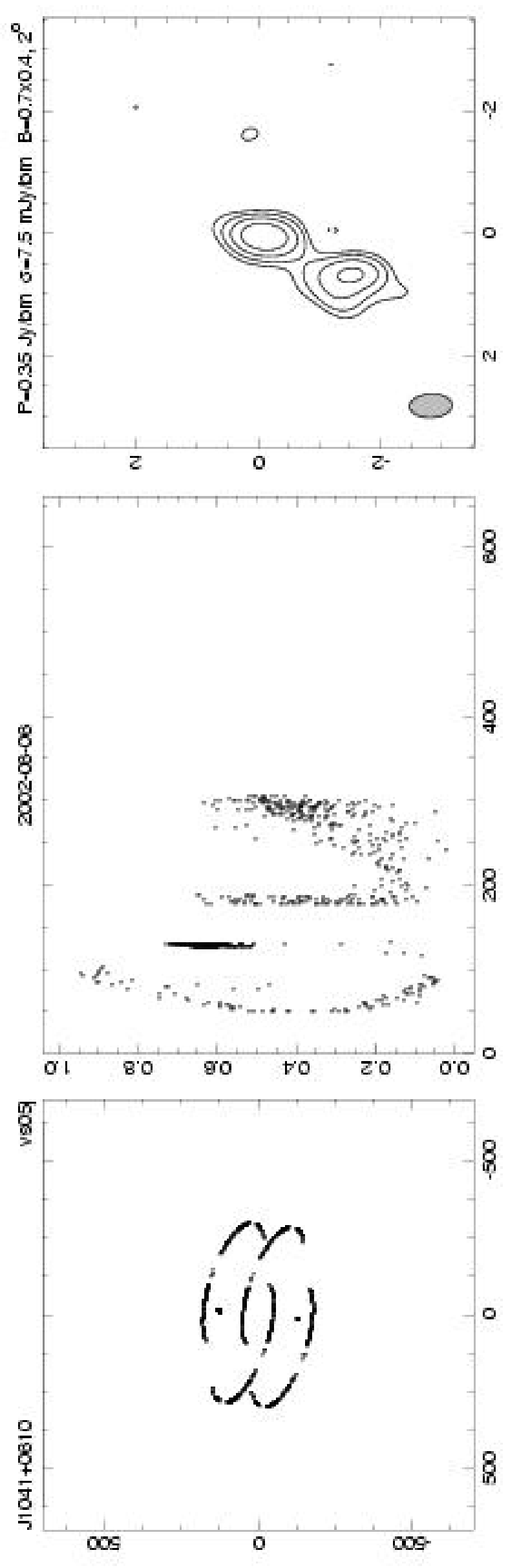}
\spfig{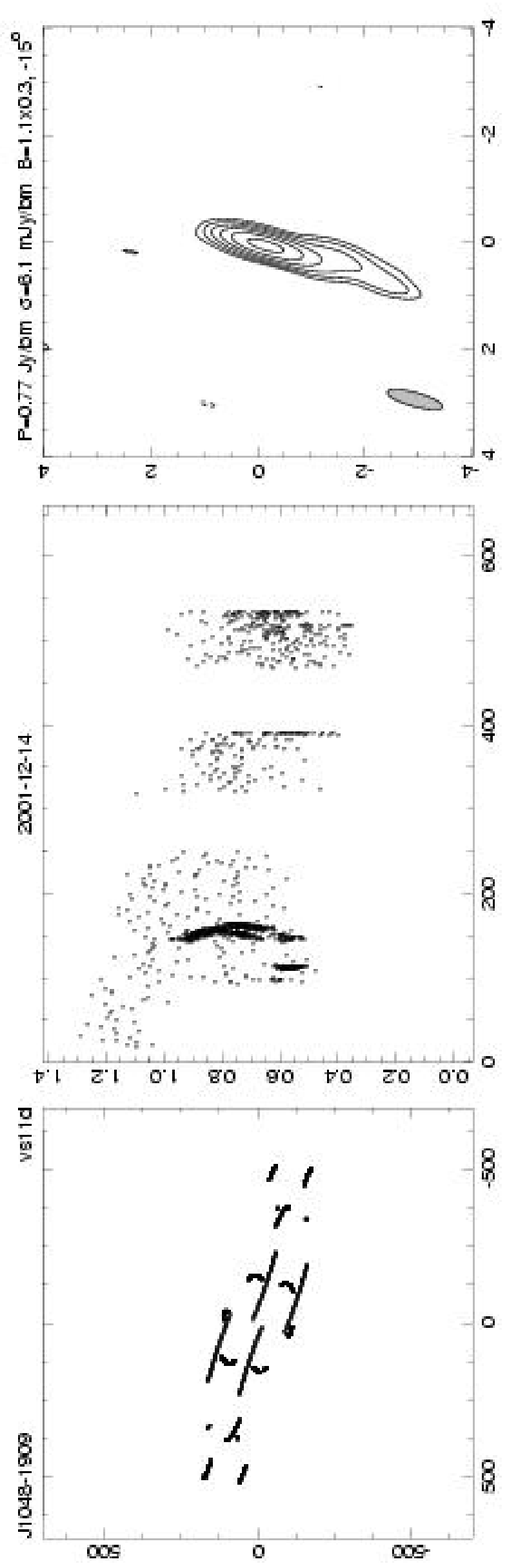}
\spfig{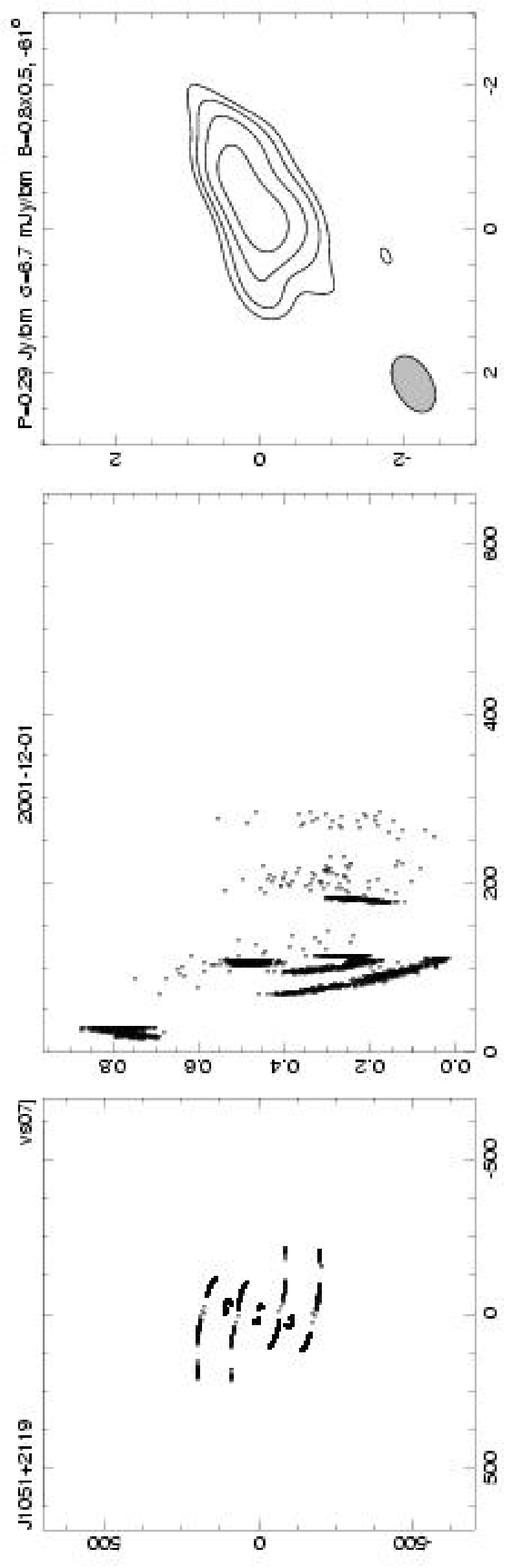}
\spfig{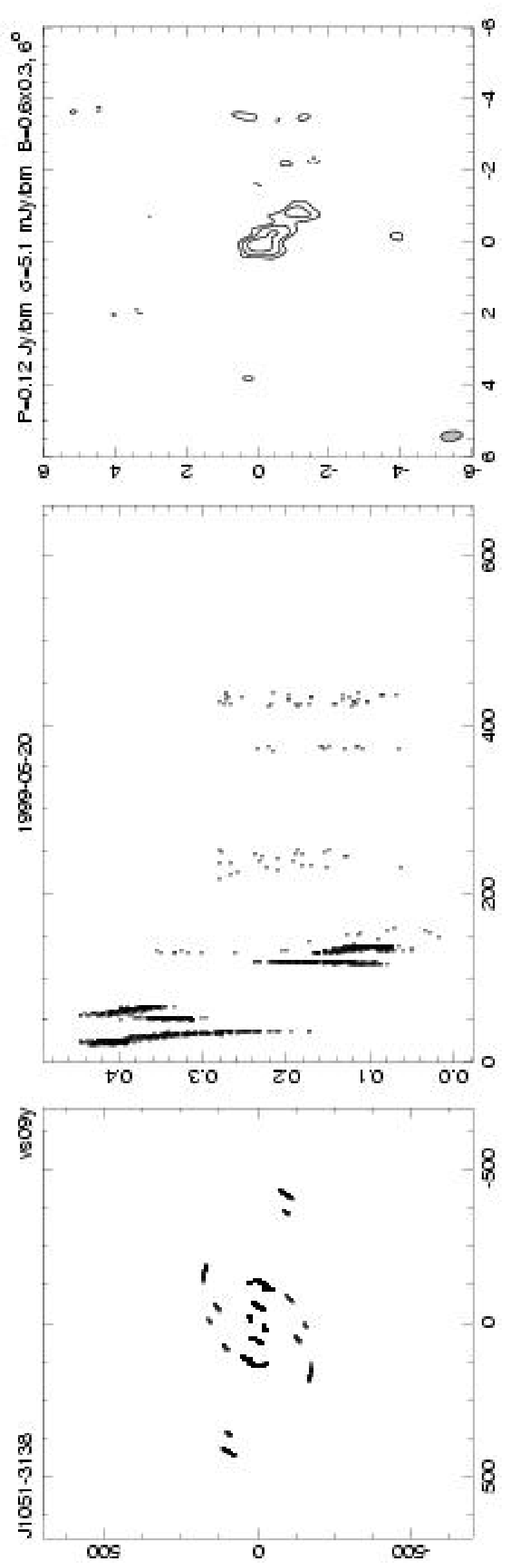}

{Fig. 1. -- {\em continued}}
\end{figure}
\clearpage
\begin{figure}
\spfig{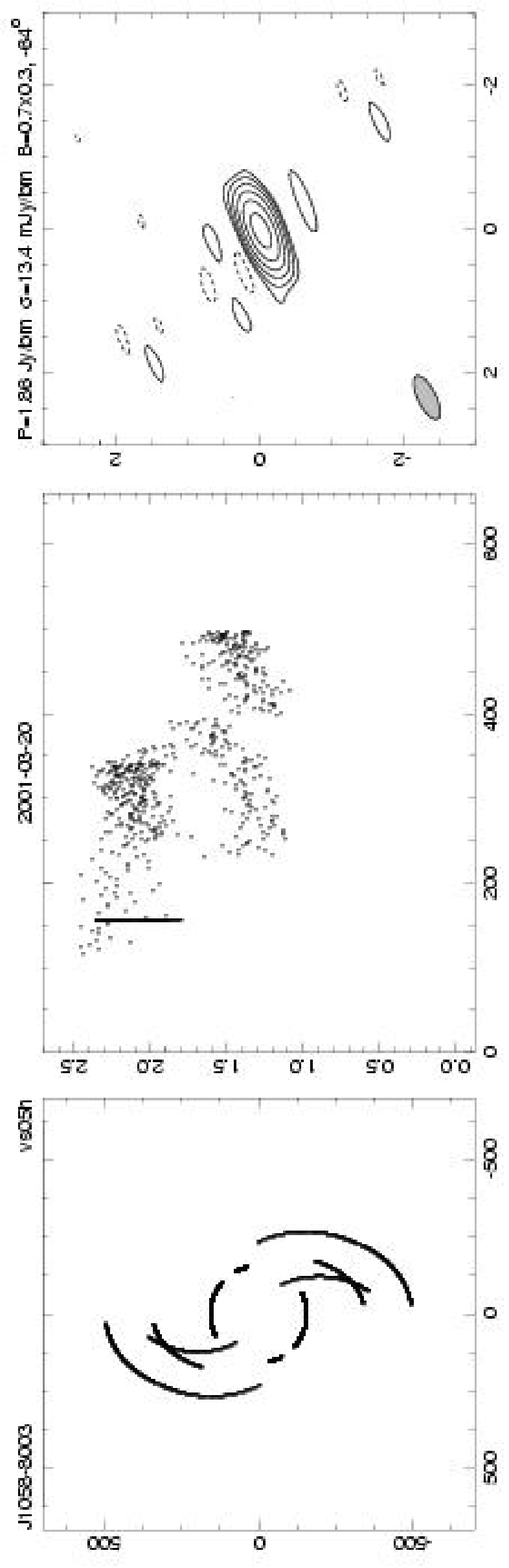}
\spfig{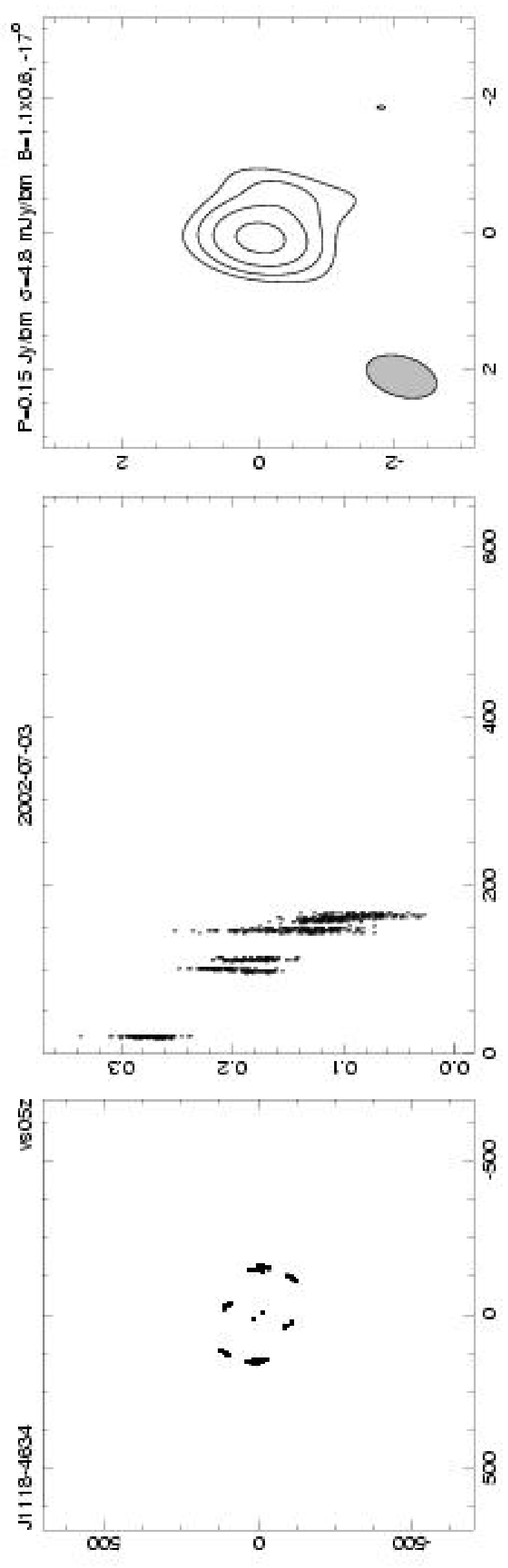}
\spfig{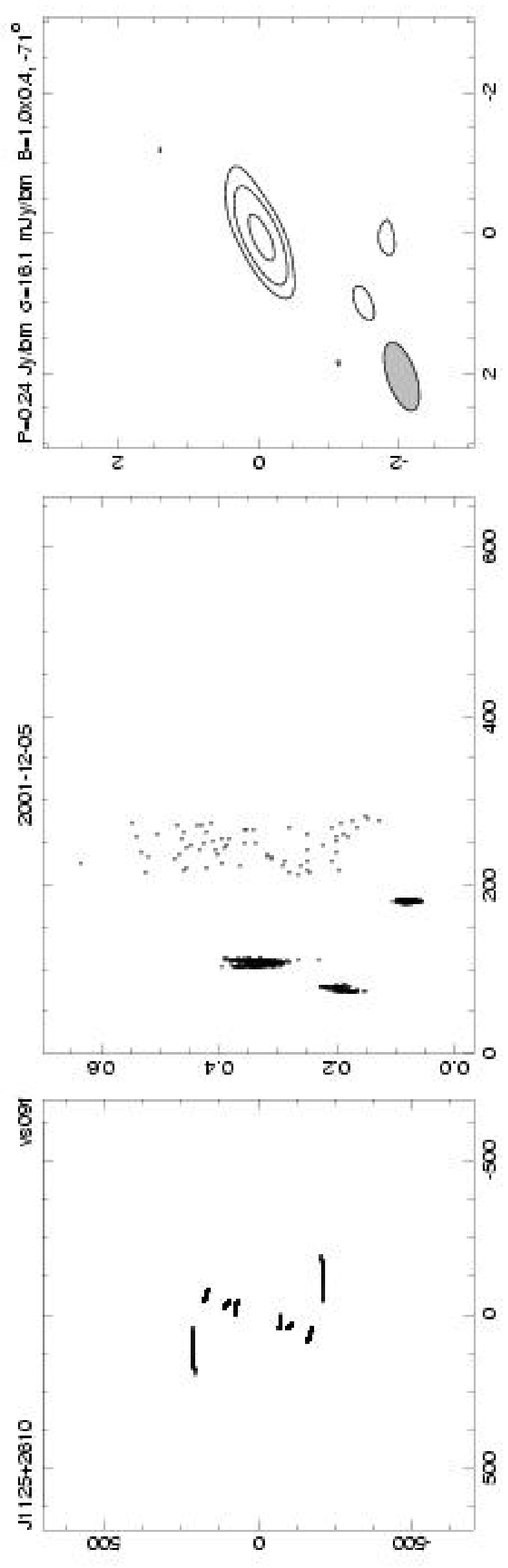}
\spfig{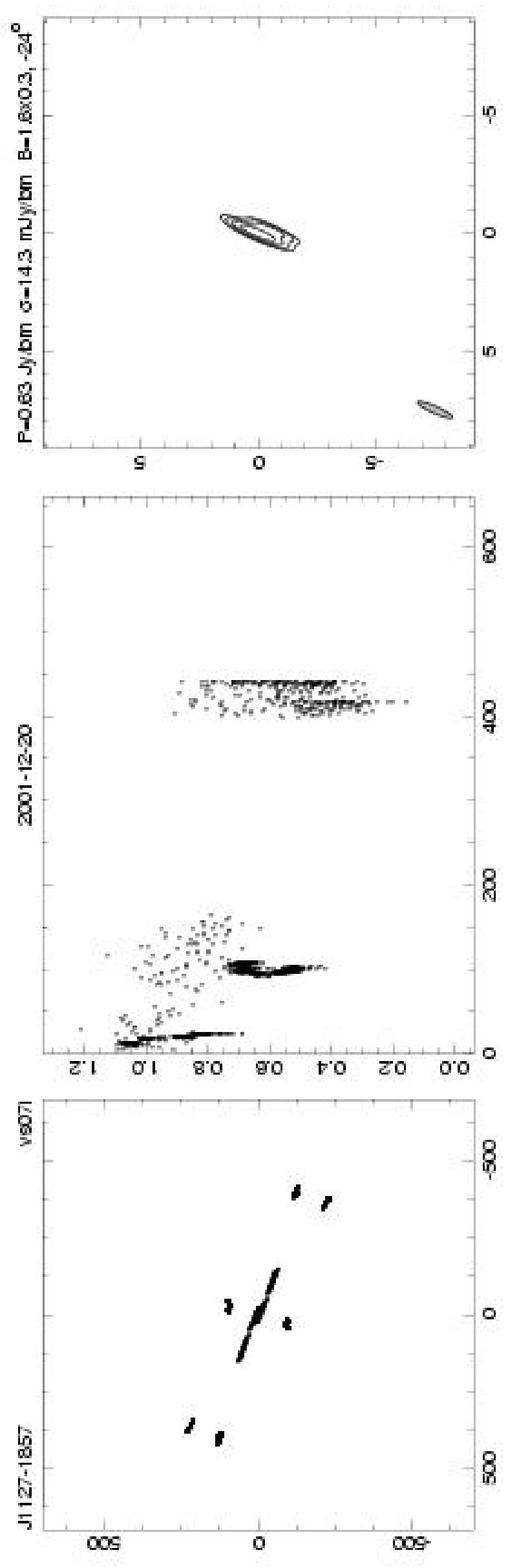}
\spfig{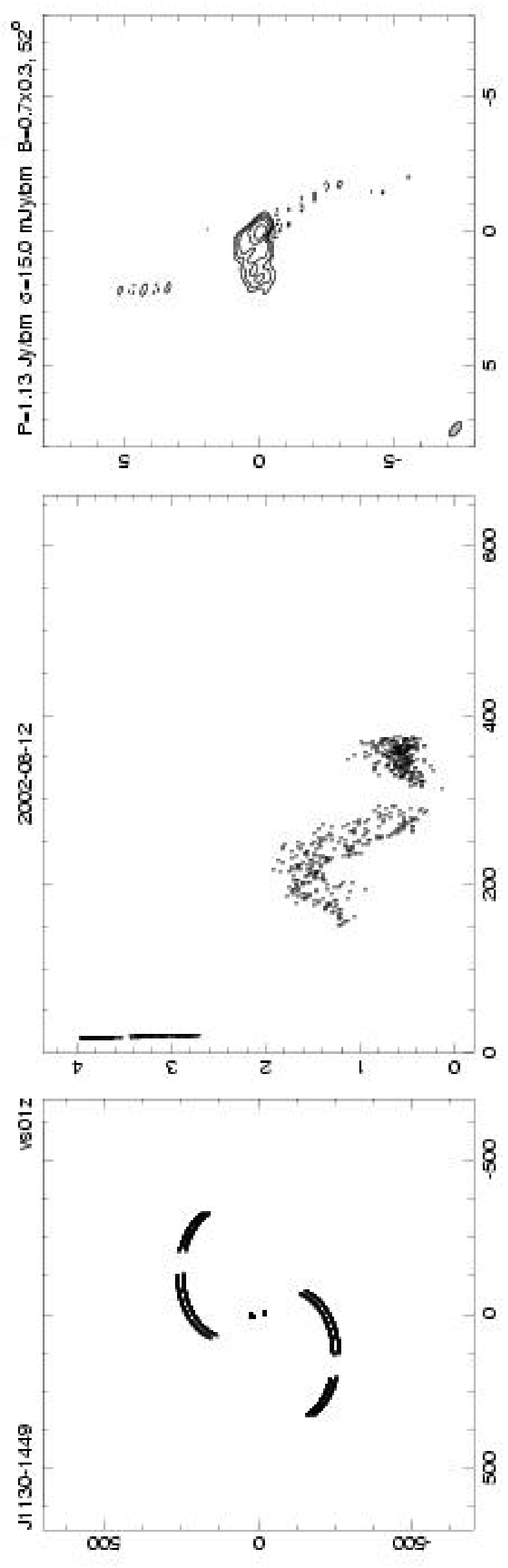} 

{Fig. 1. -- {\em continued}}
\end{figure}
\clearpage
\begin{figure}
\spfig{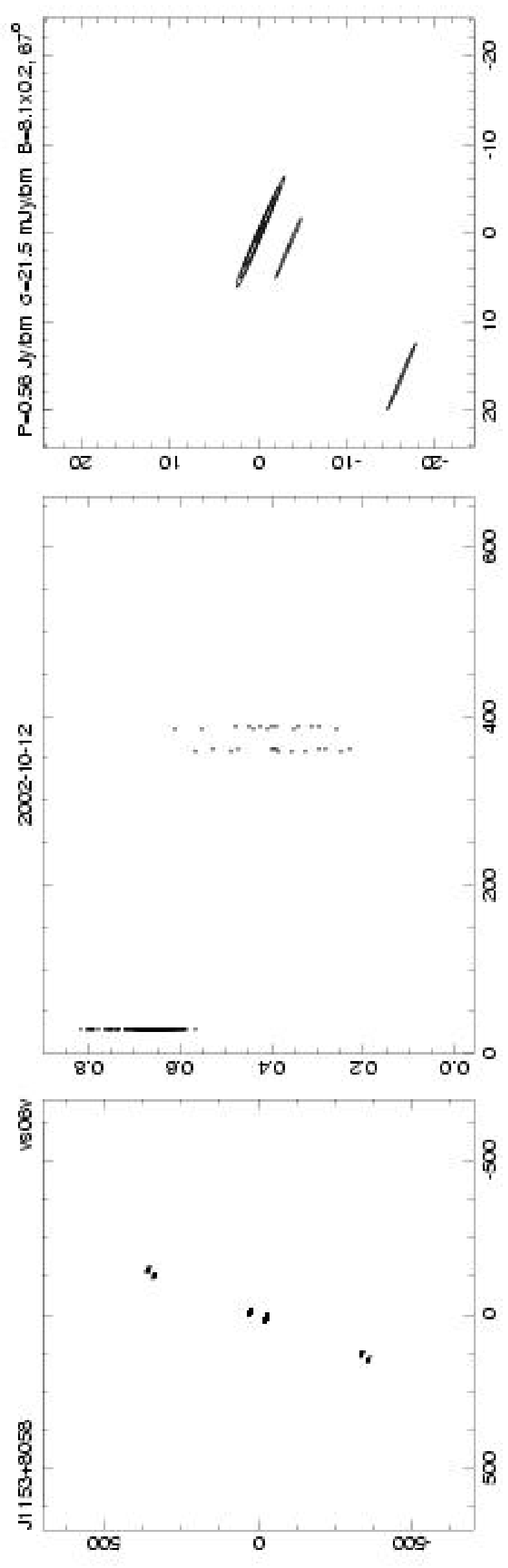}
\spfig{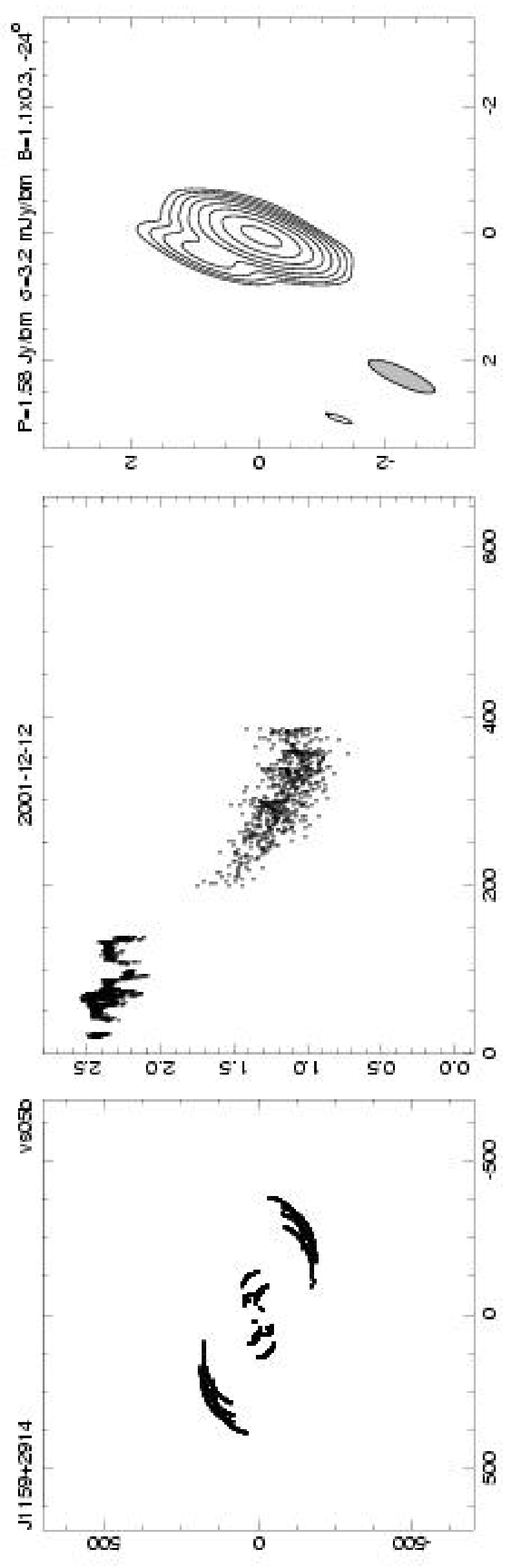}
\spfig{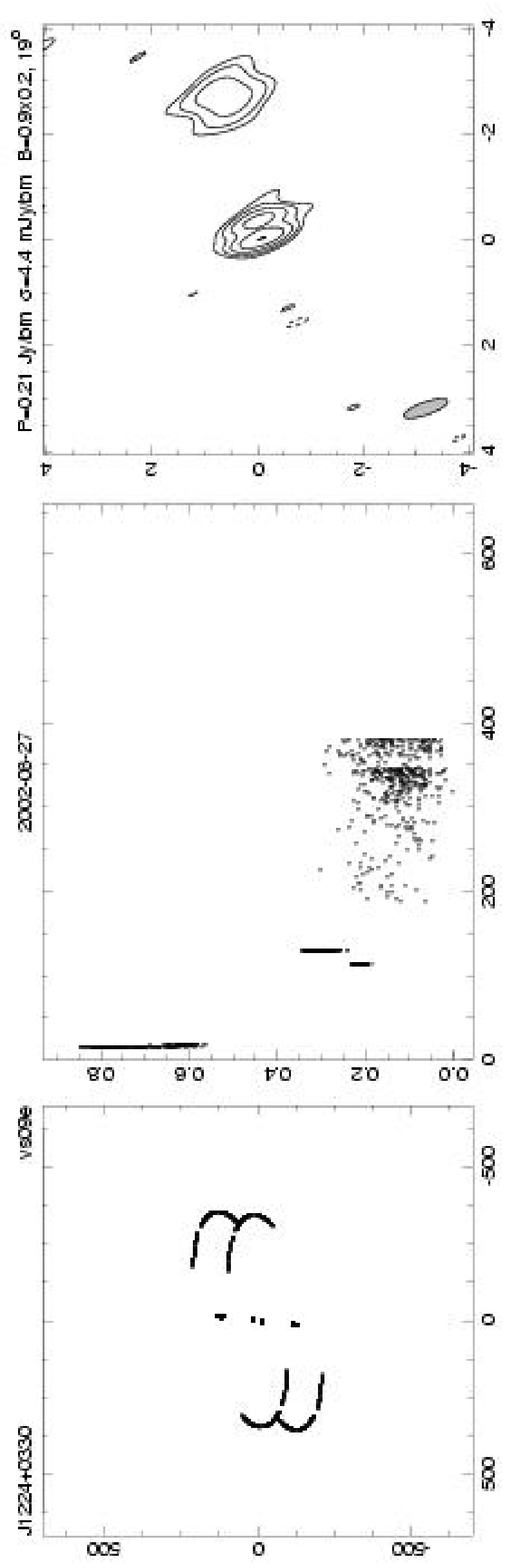}
\spfig{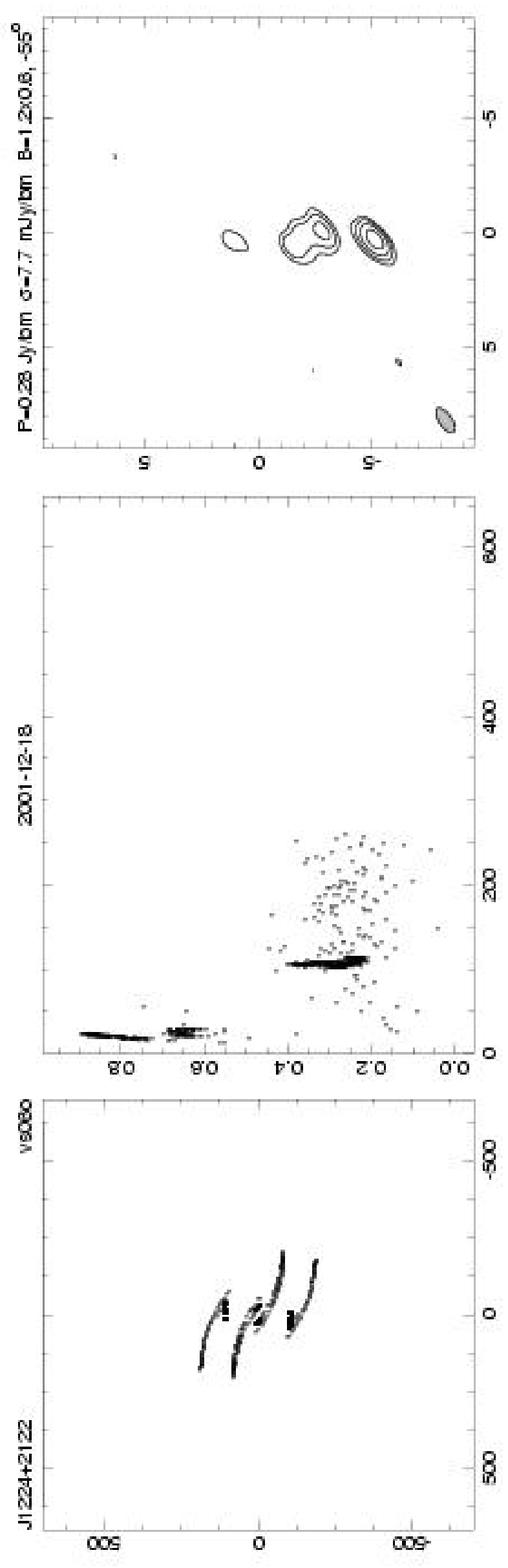}
\spfig{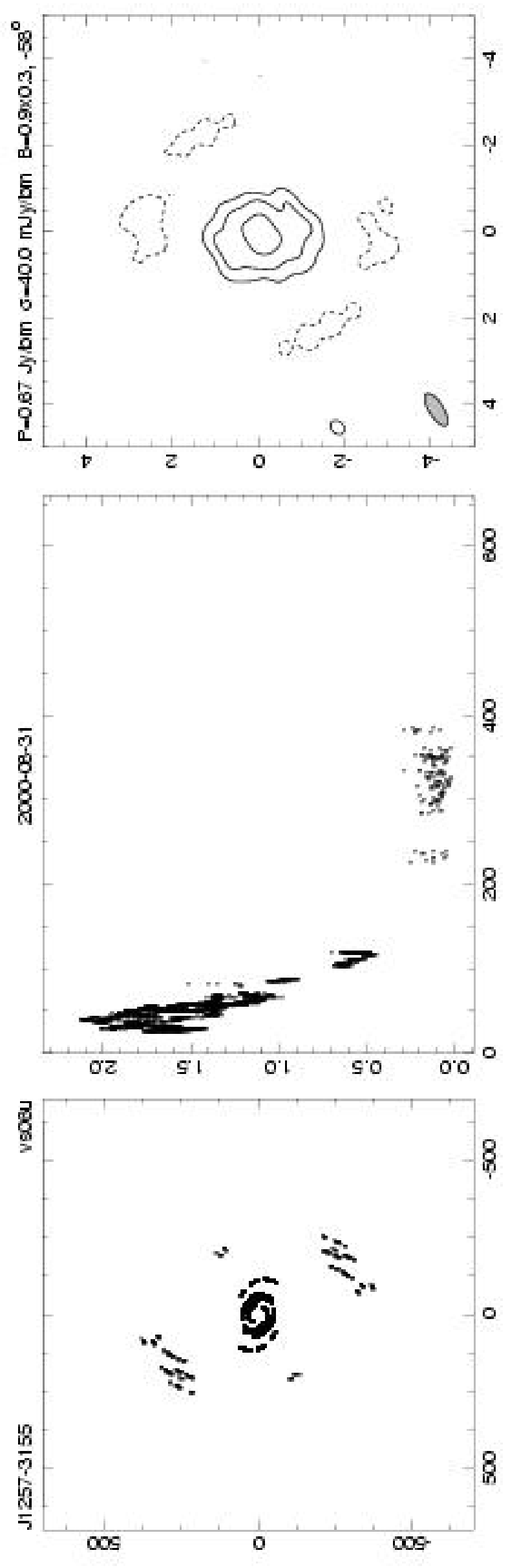}

{Fig. 1. -- {\em continued}}
\end{figure}
\clearpage
\begin{figure}
\spfig{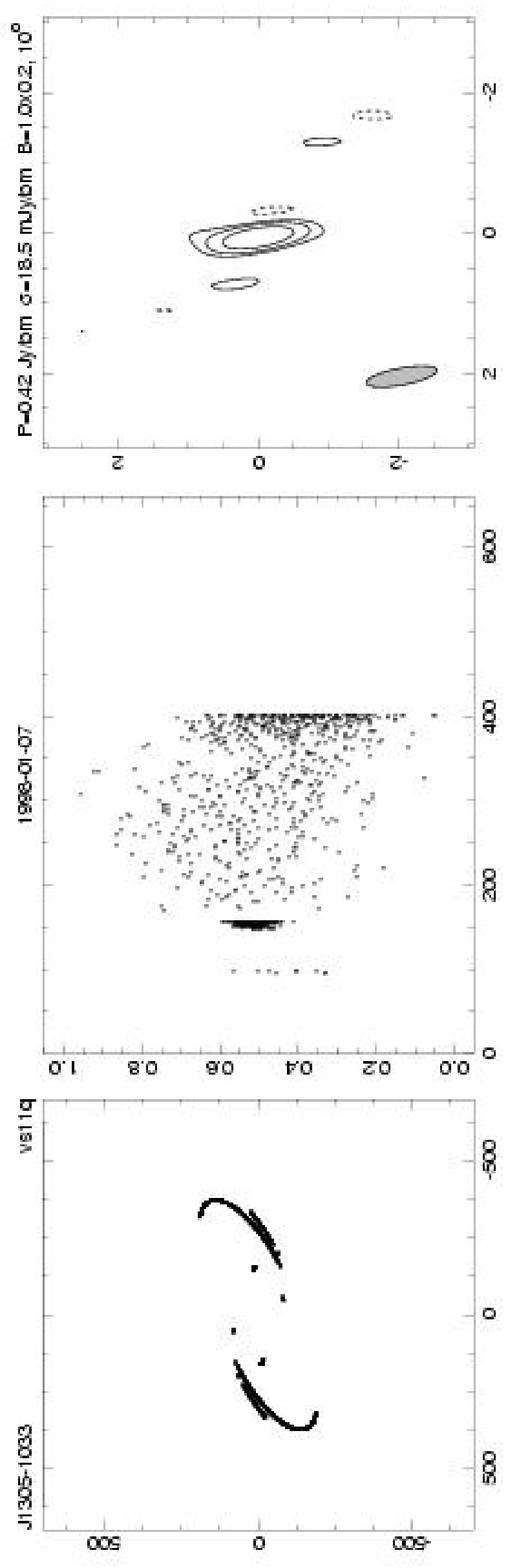}
\spfig{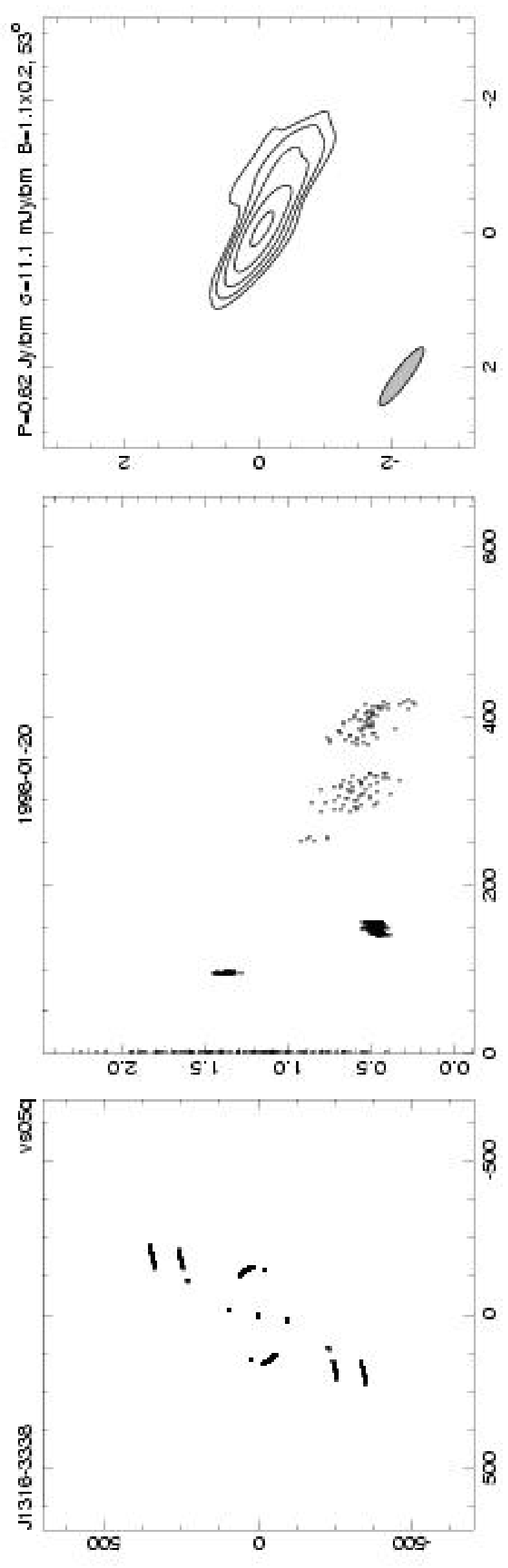} \typeout{different from vsop_difmap -- still one missing IF if one flagged?}
\spfig{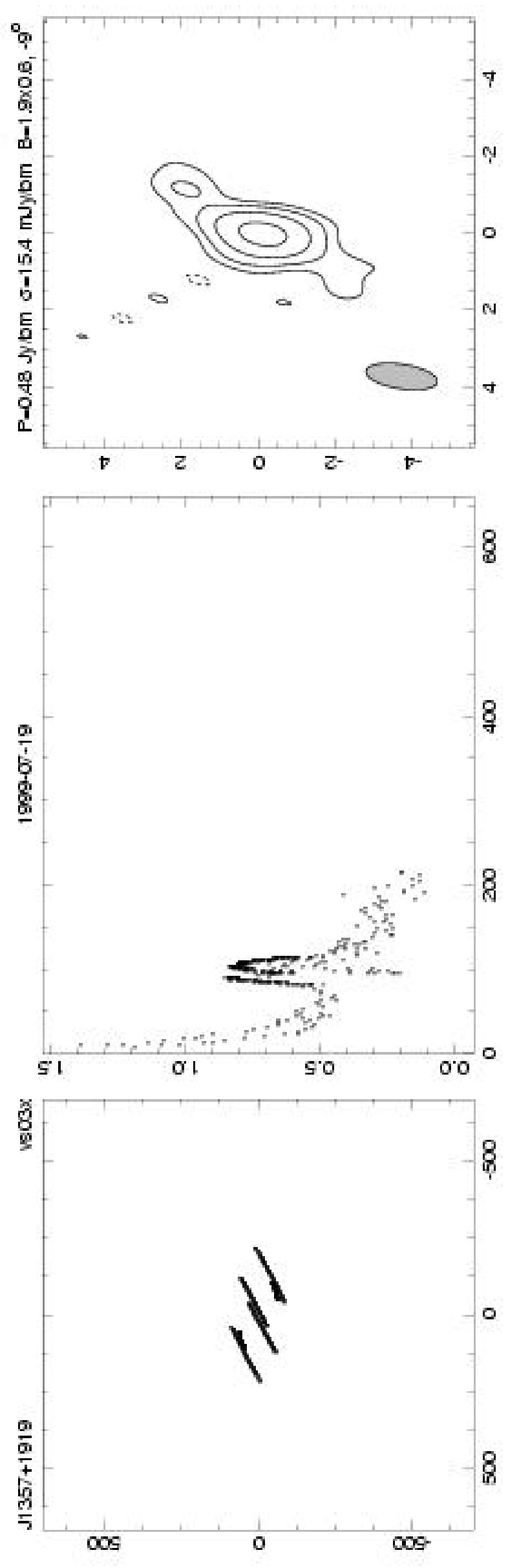}
\spfig{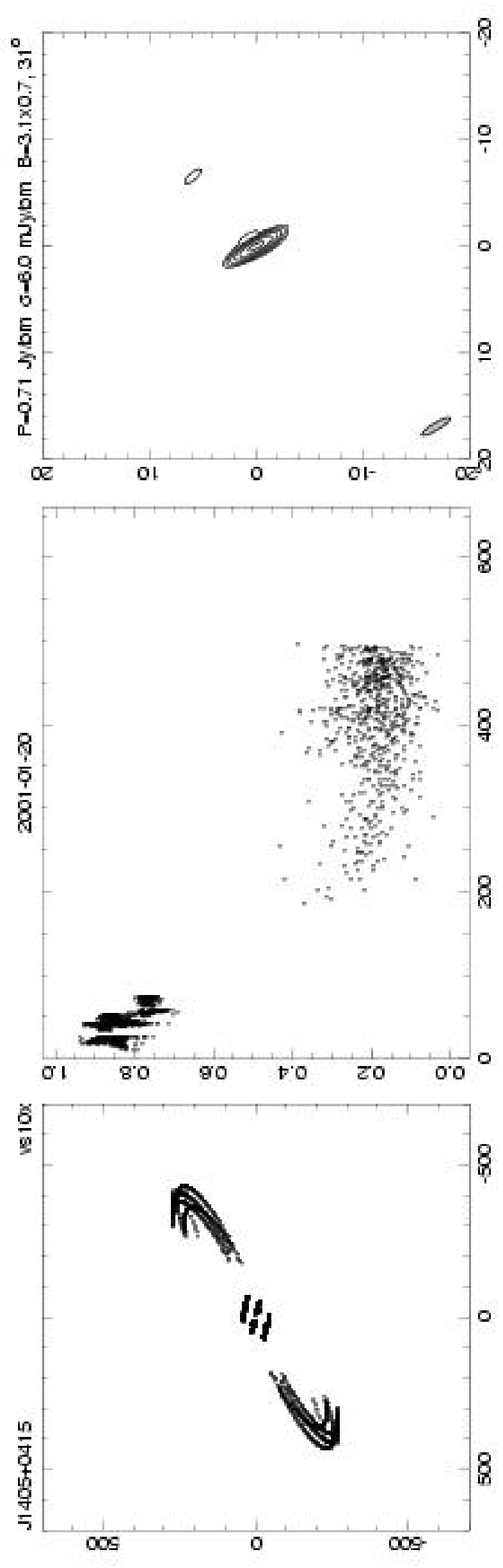} 
\spfig{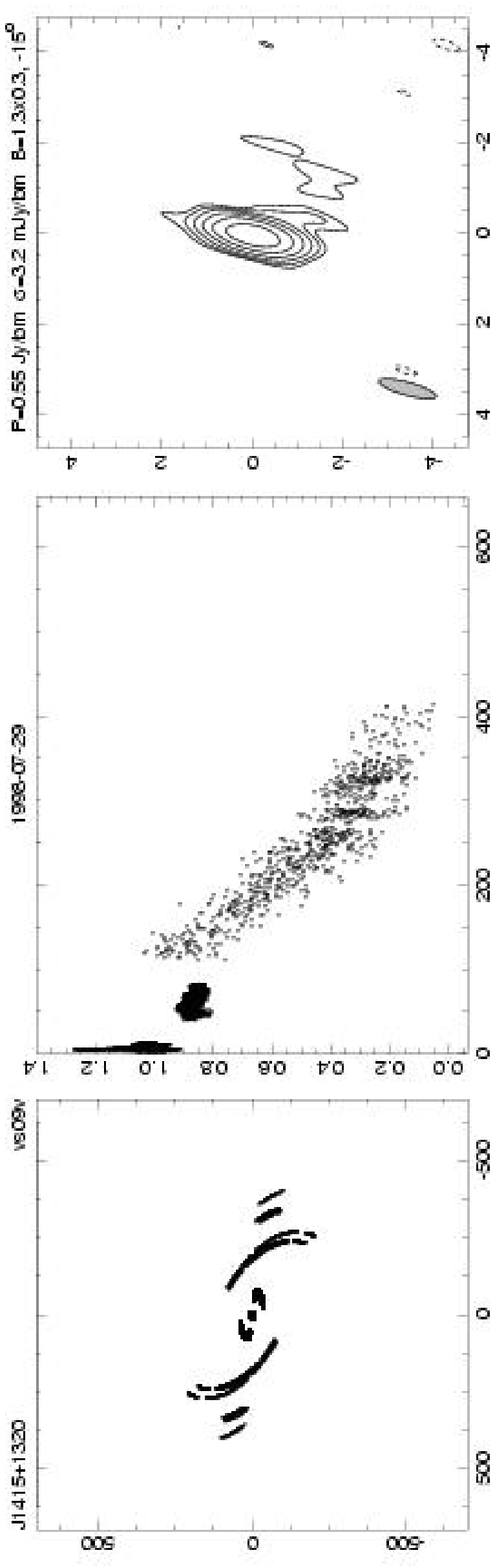}

{Fig. 1. -- {\em continued}}
\end{figure}
\clearpage
\begin{figure}
\spfig{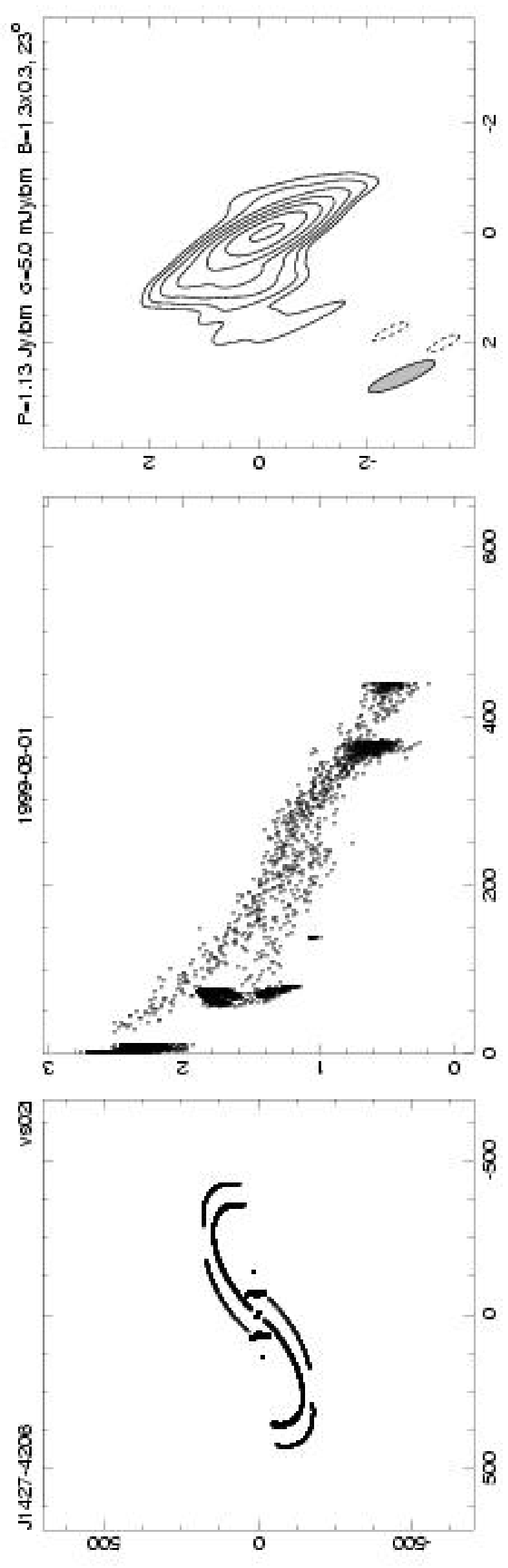}
\spfig{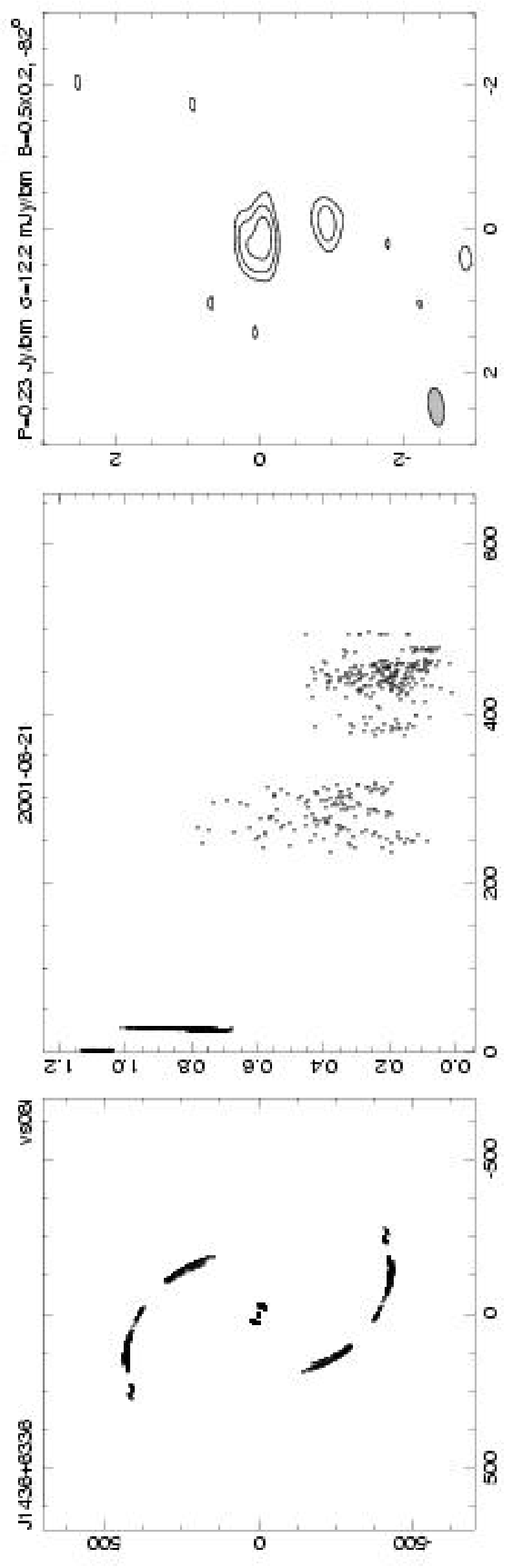} 
\spfig{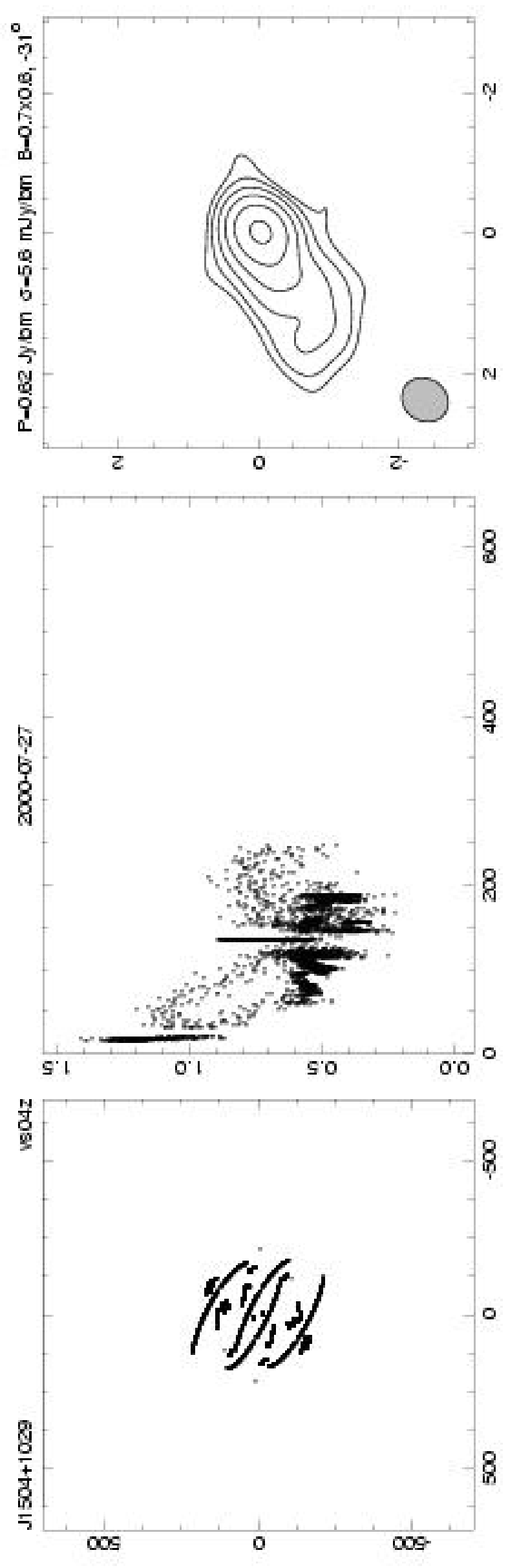}
\spfig{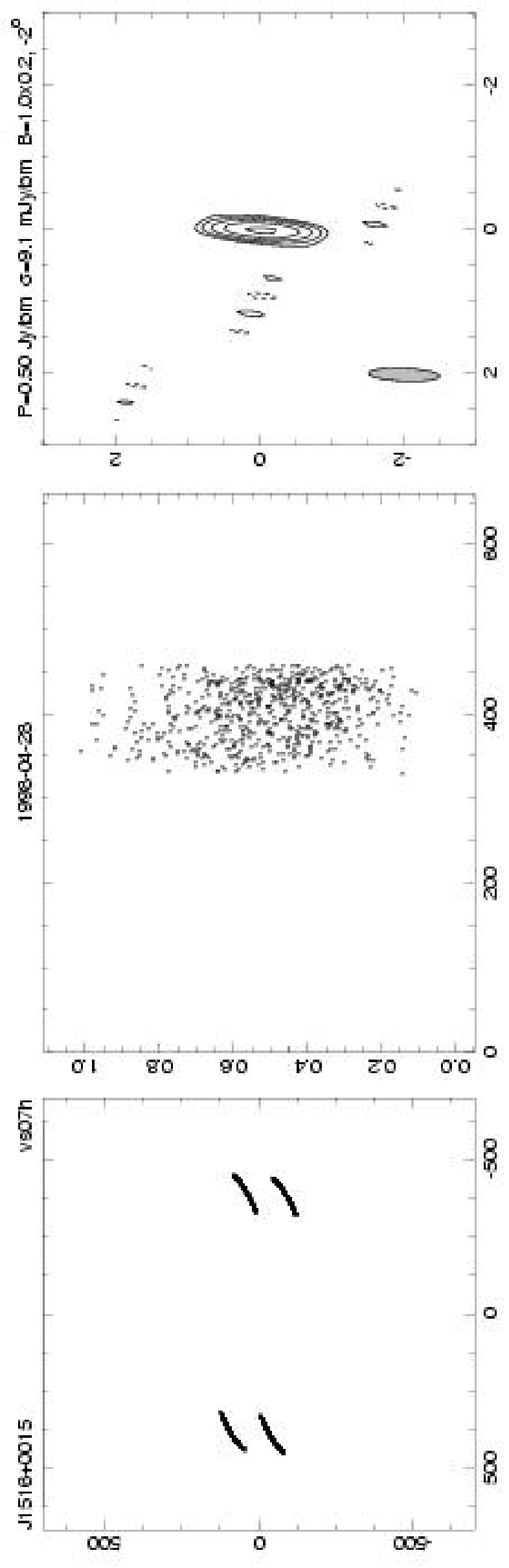}
\spfig{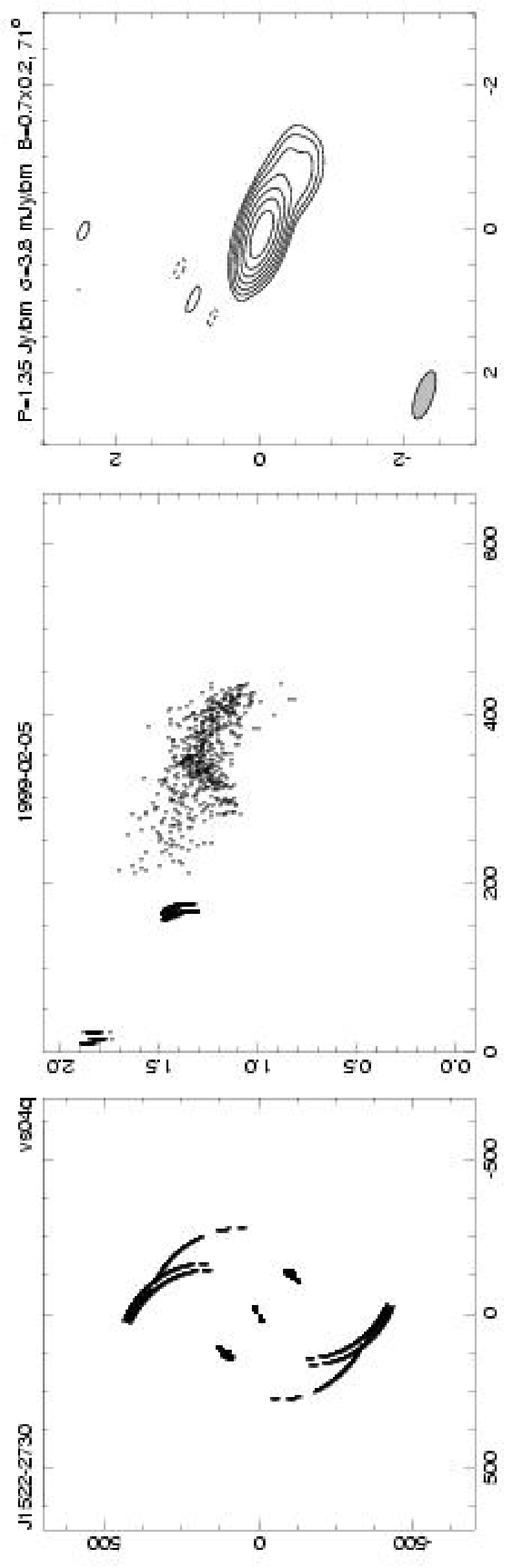}

{Fig. 1. -- {\em continued}}
\end{figure}
\clearpage
\begin{figure}
\spfig{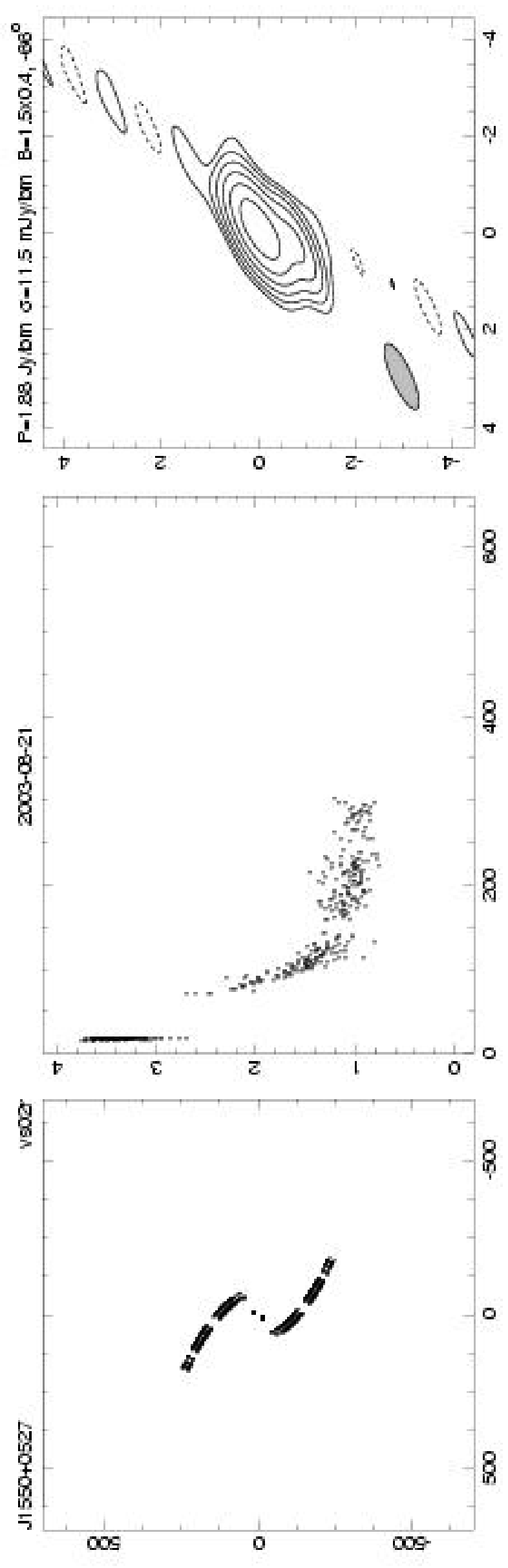}
\spfig{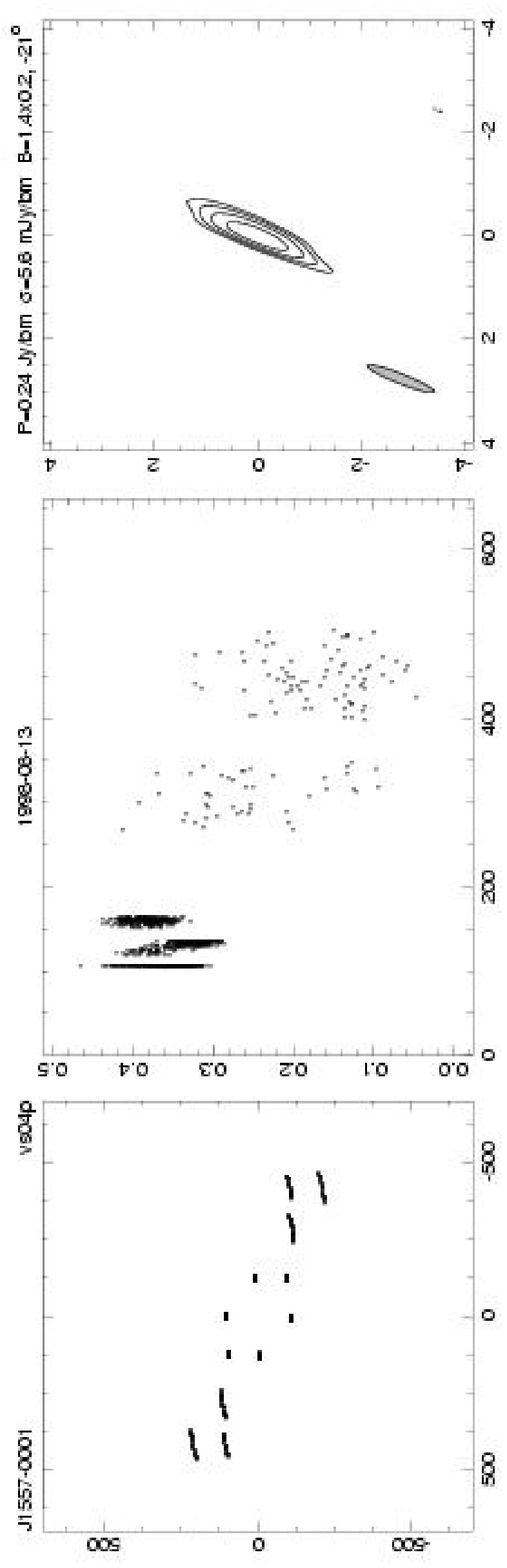}
\spfig{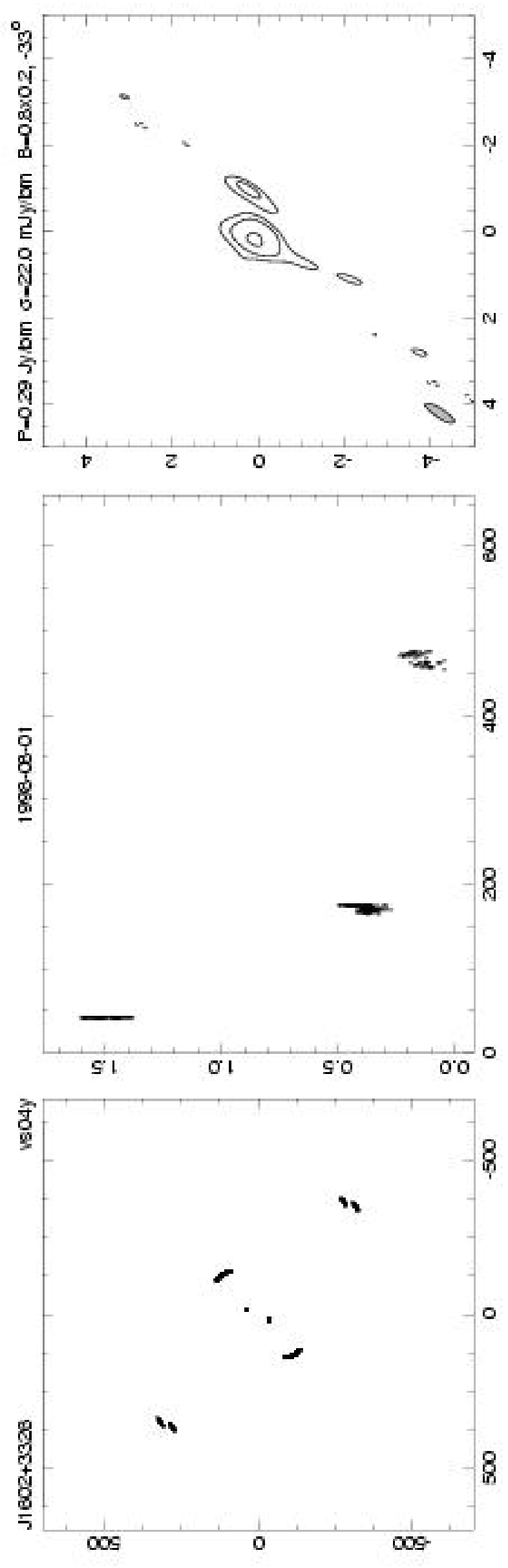} \typeout{different from vsop_difmap -- still one missing IF if one flagged?}
\spfig{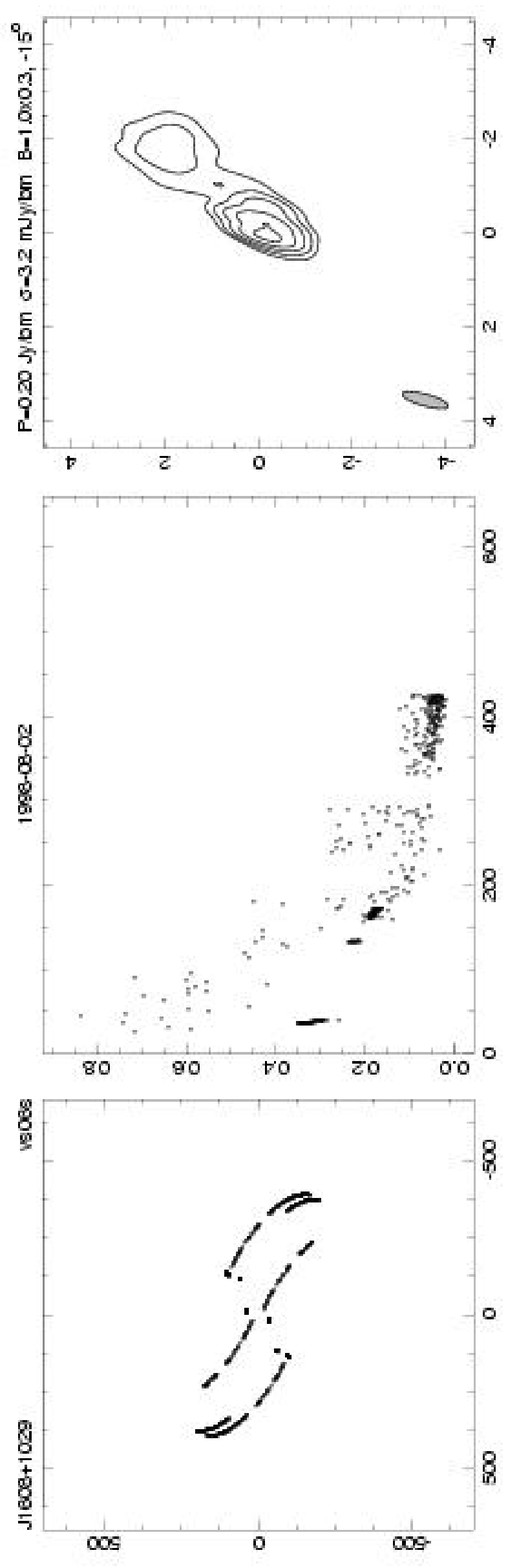}
\spfig{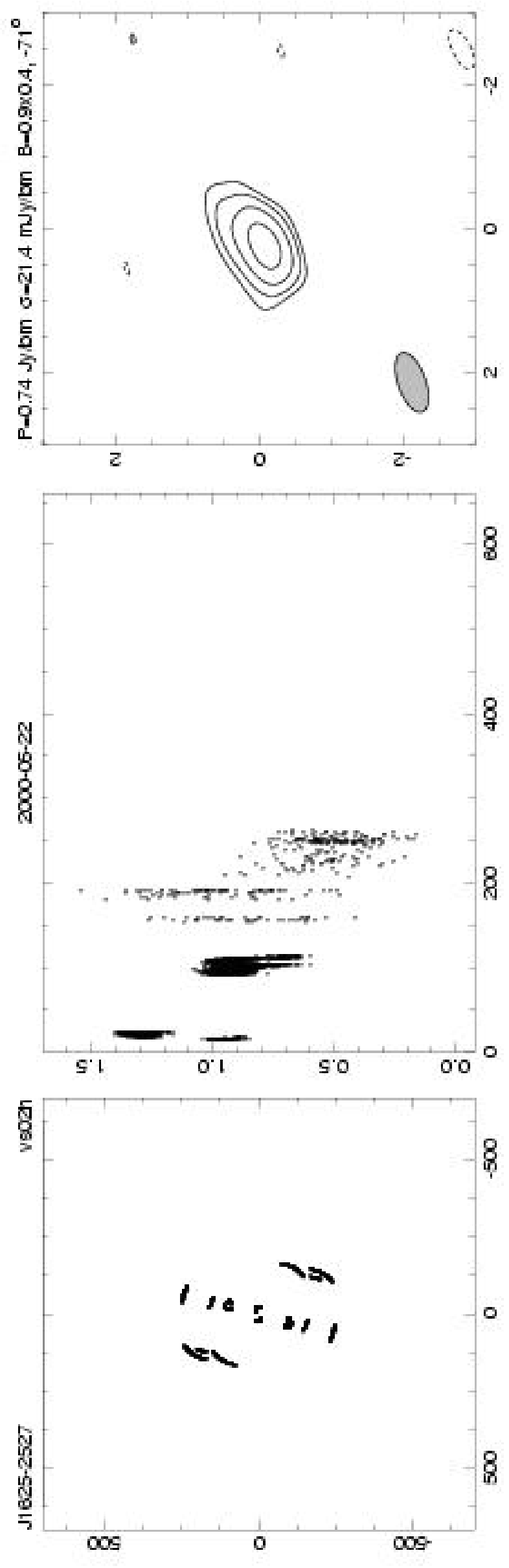}

{Fig. 1. -- {\em continued}}
\end{figure}
\clearpage
\begin{figure}
\spfig{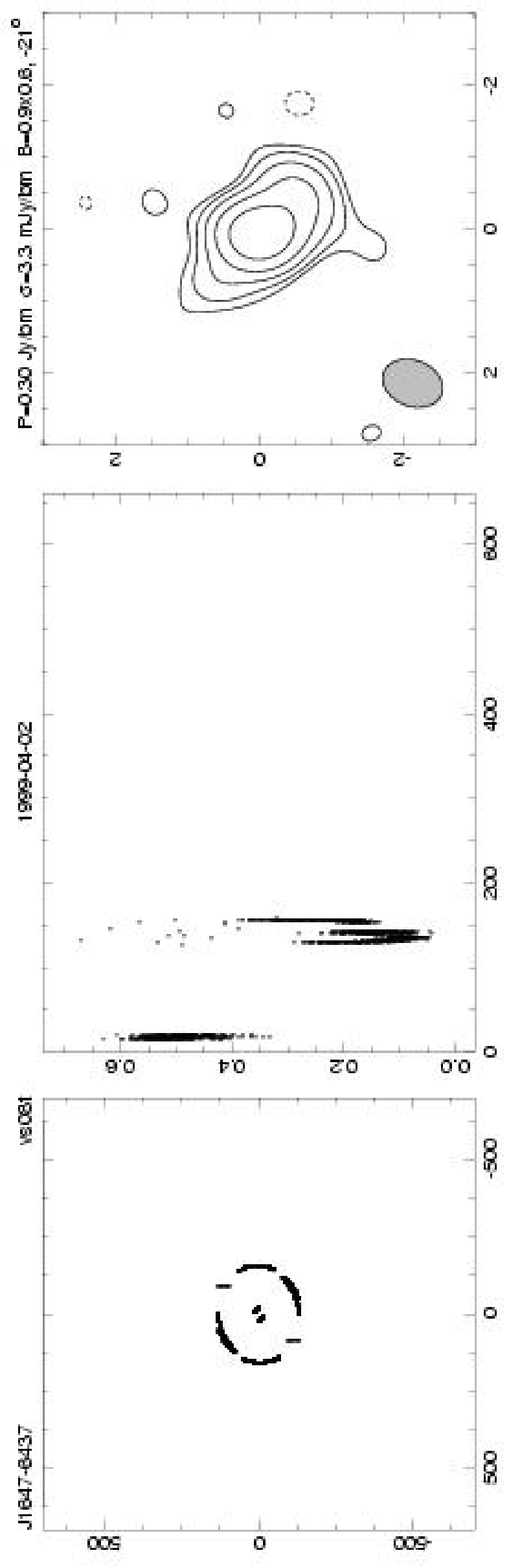}
\spfig{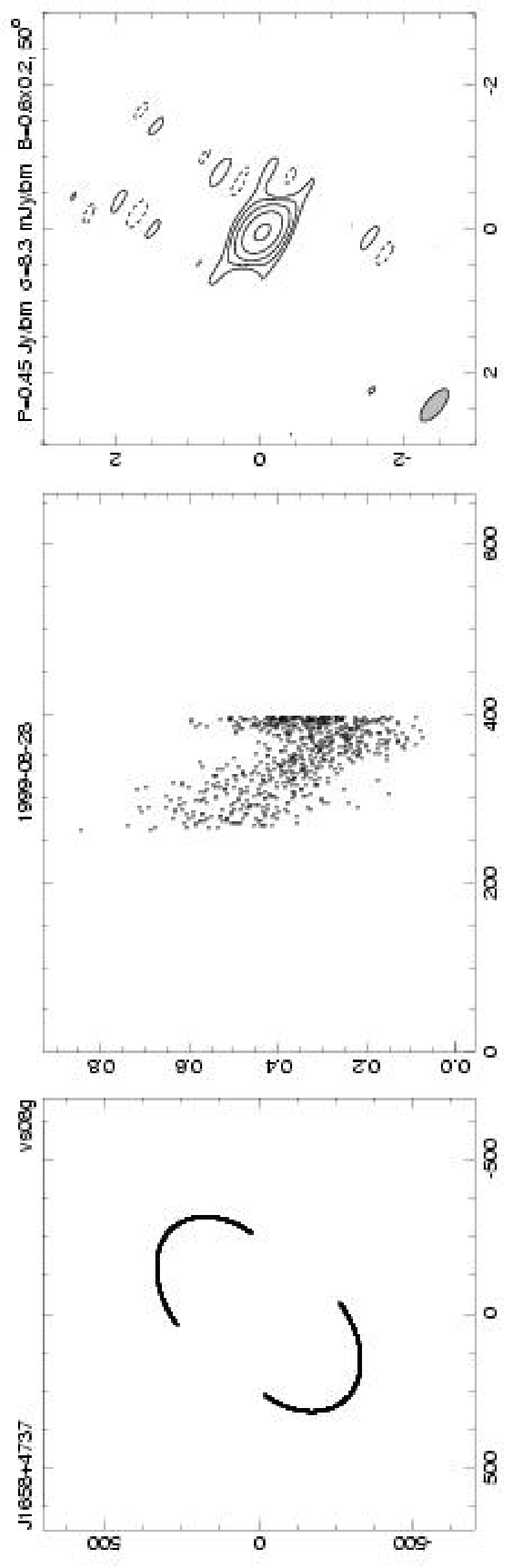}
\spfig{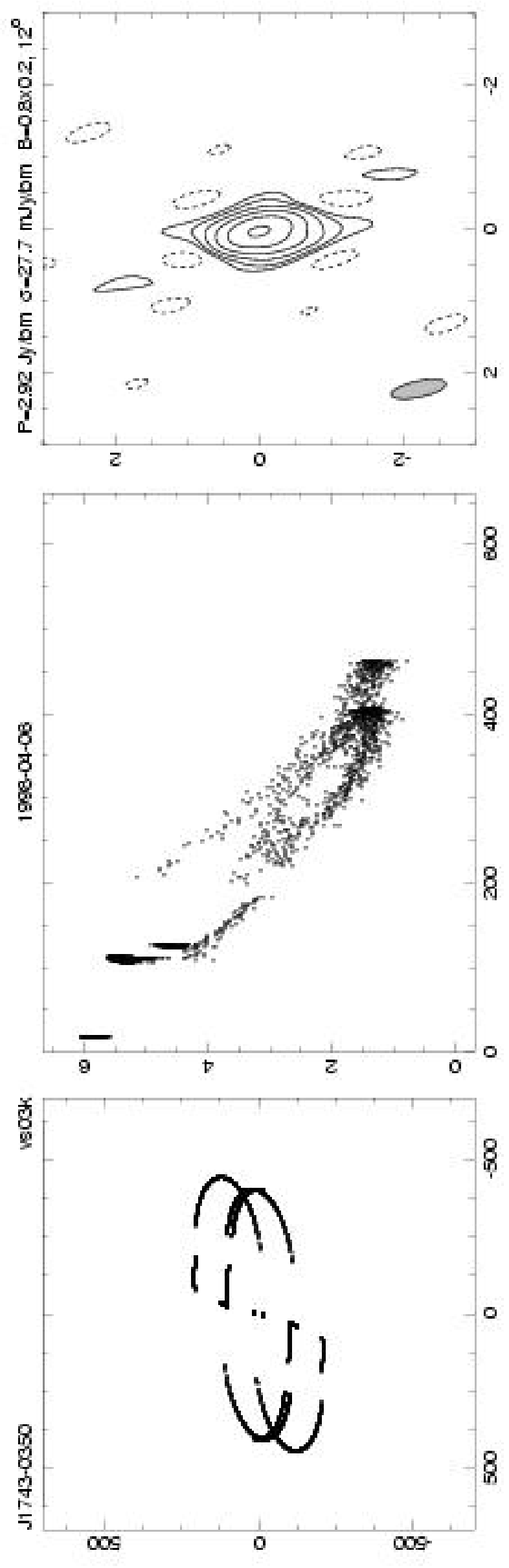}
\spfig{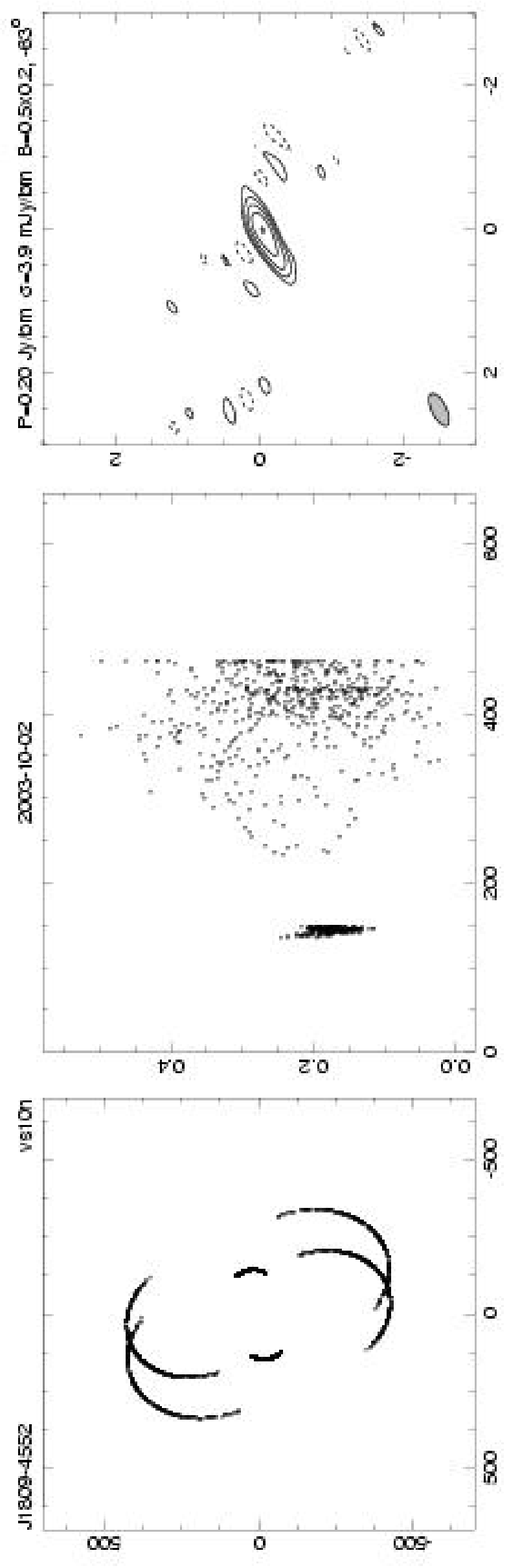}
\spfig{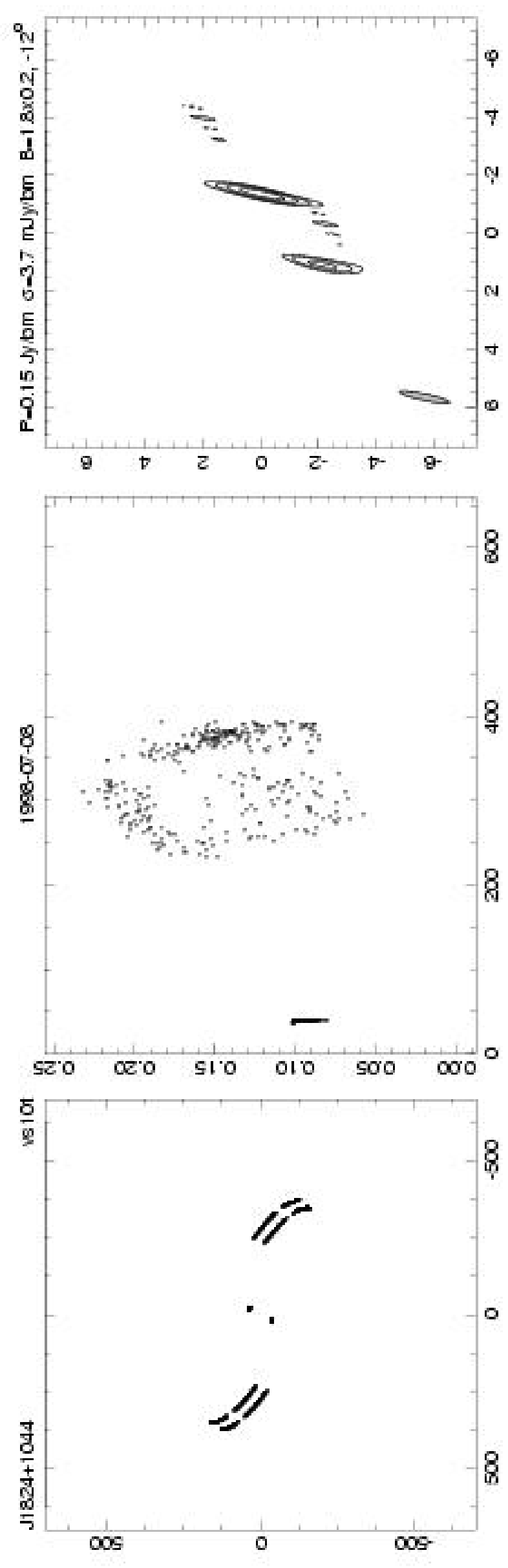} \typeout{different png ps (ps better - shift?) -- vsop_difmap image double -- flux low at short baseline}

{Fig. 1. -- {\em continued}}
\end{figure}
\clearpage
\begin{figure}
\spfig{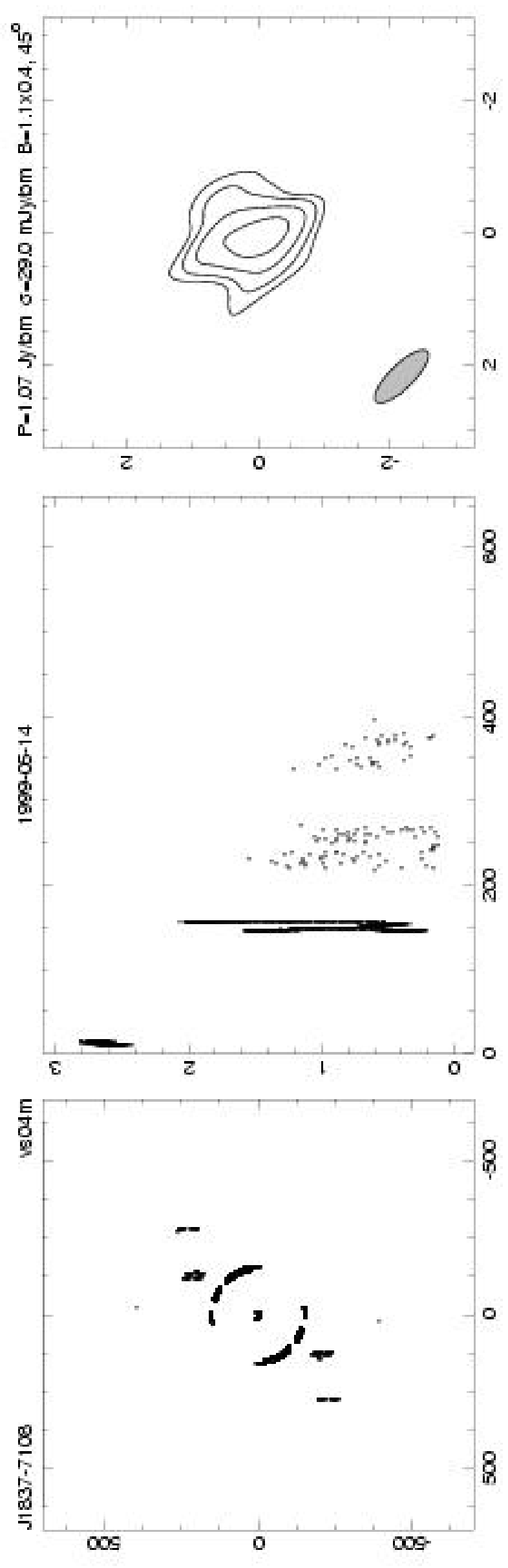} 
\spfig{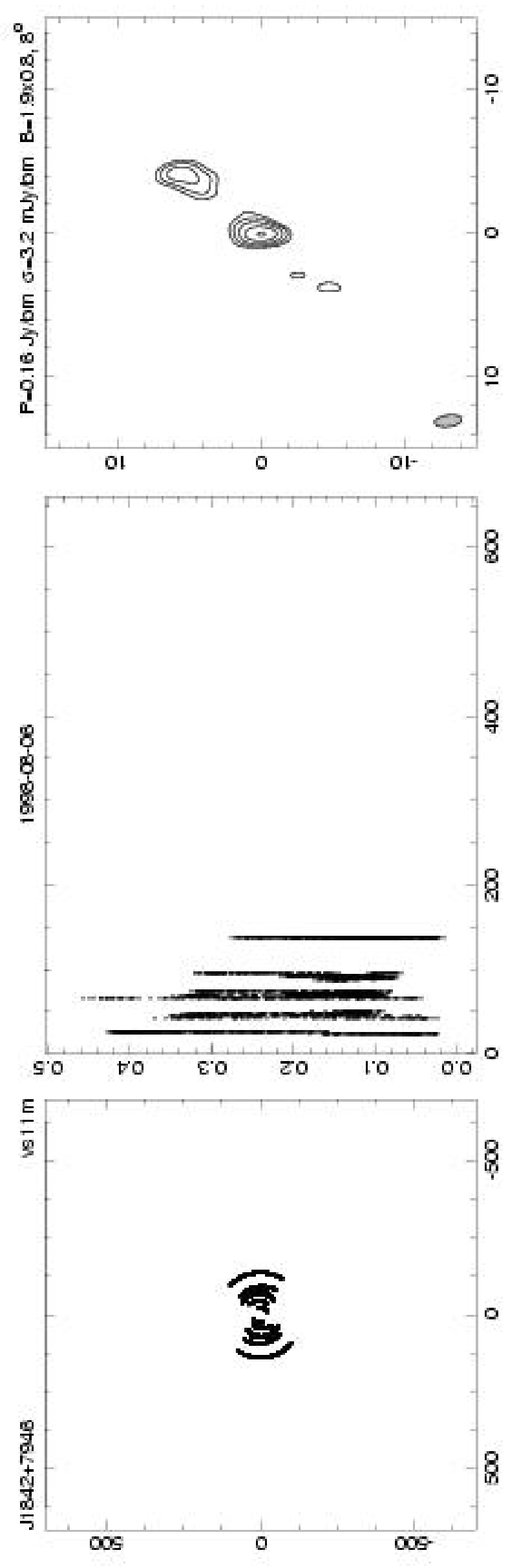}
\spfig{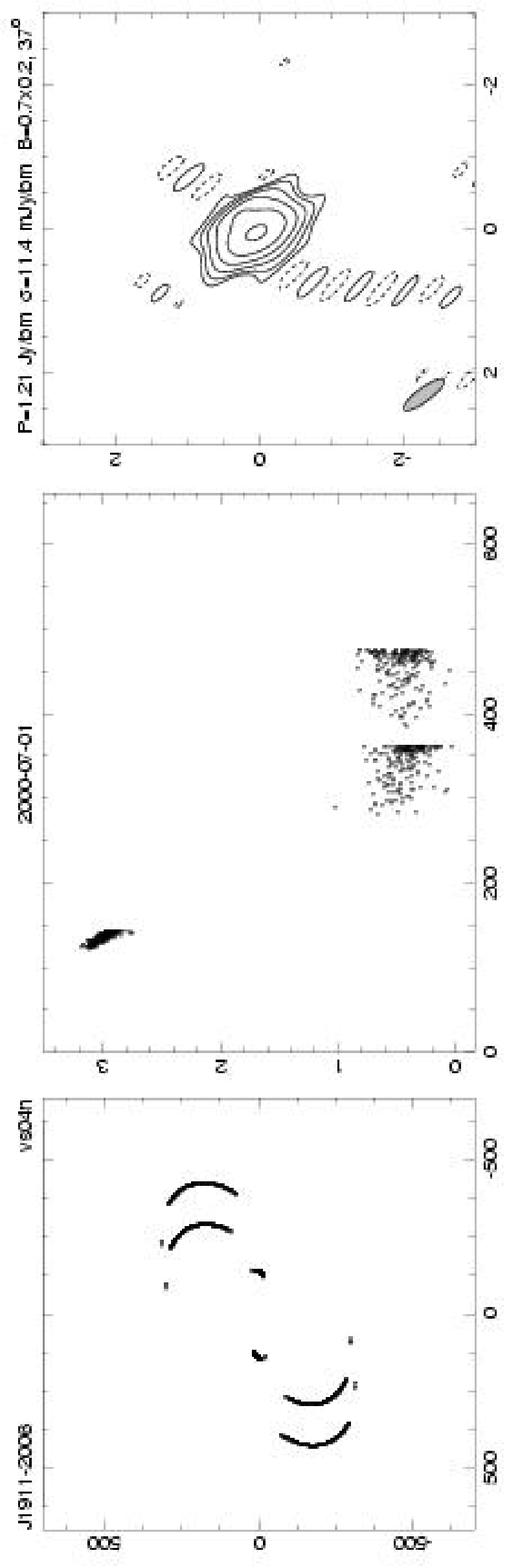}
\spfig{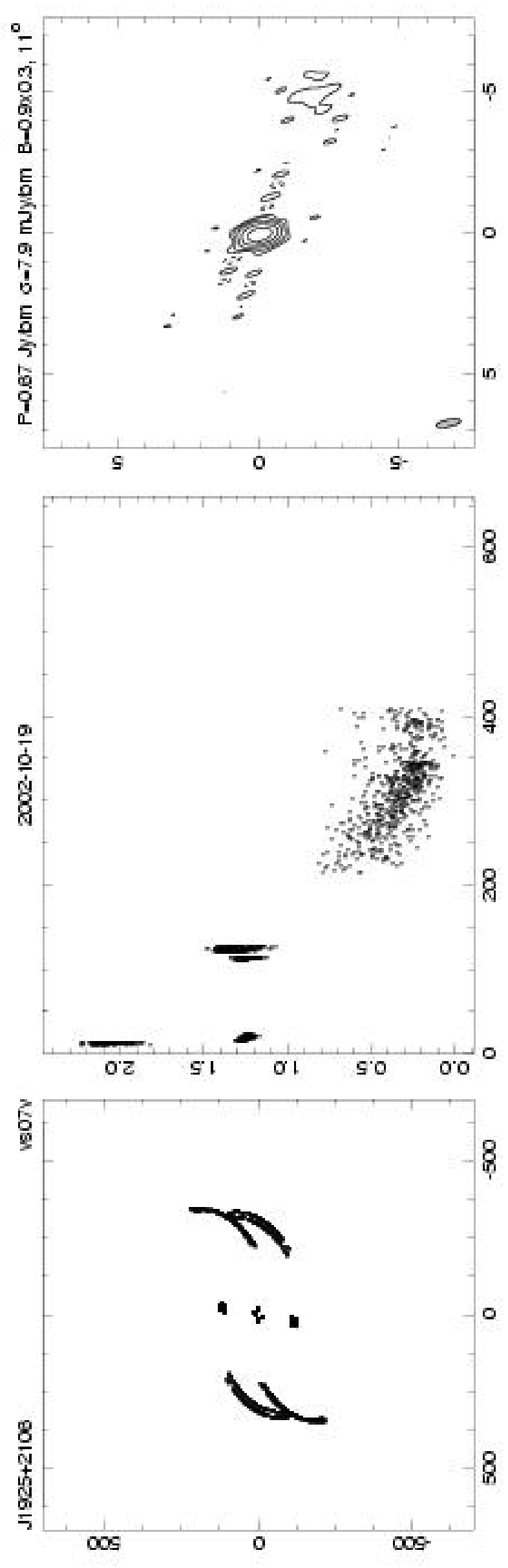}
\spfig{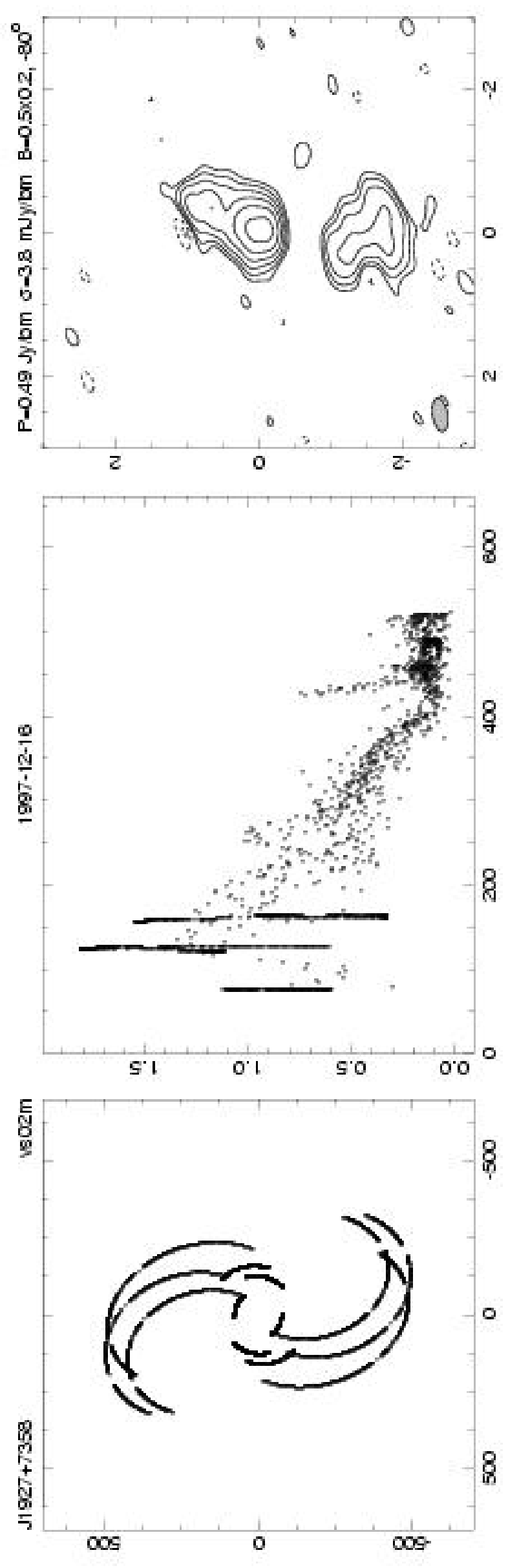}

{Fig. 1. -- {\em continued}}
\end{figure}
\clearpage
\begin{figure}
\spfig{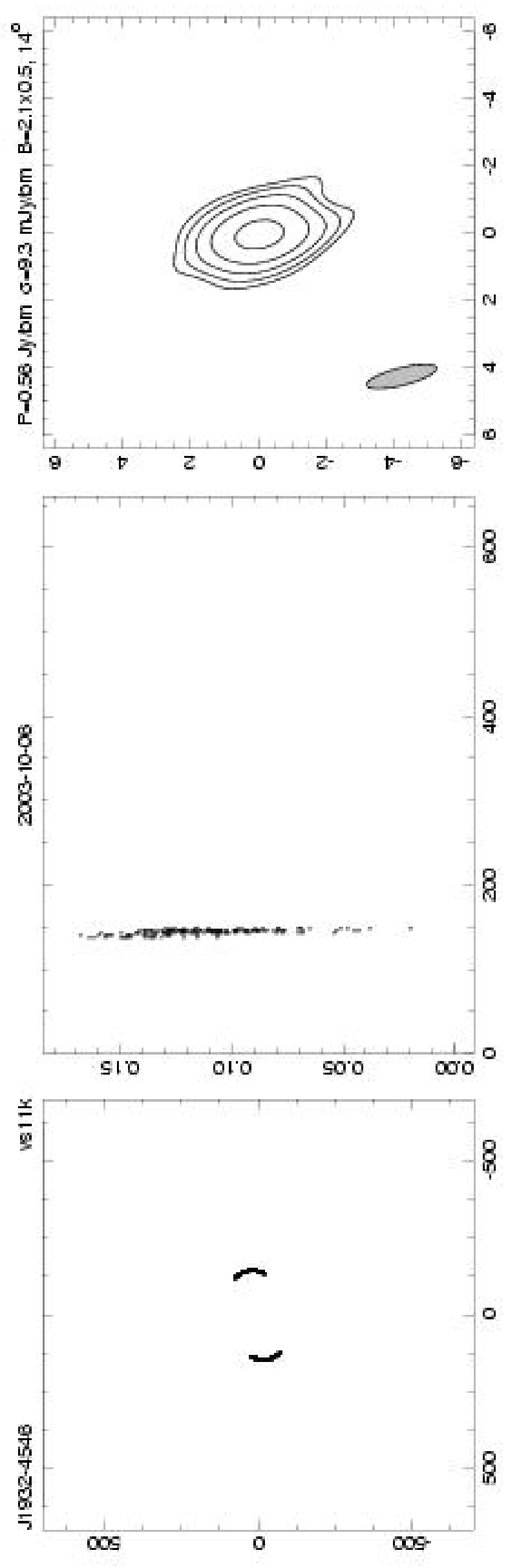}
\spfig{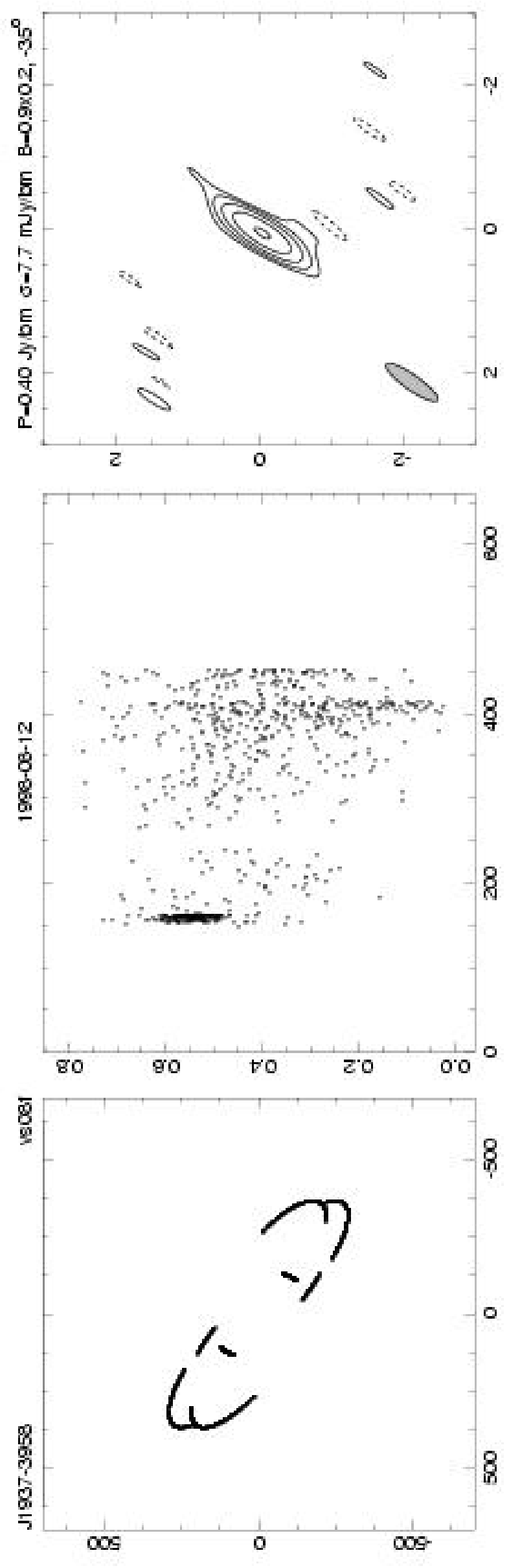}
\spfig{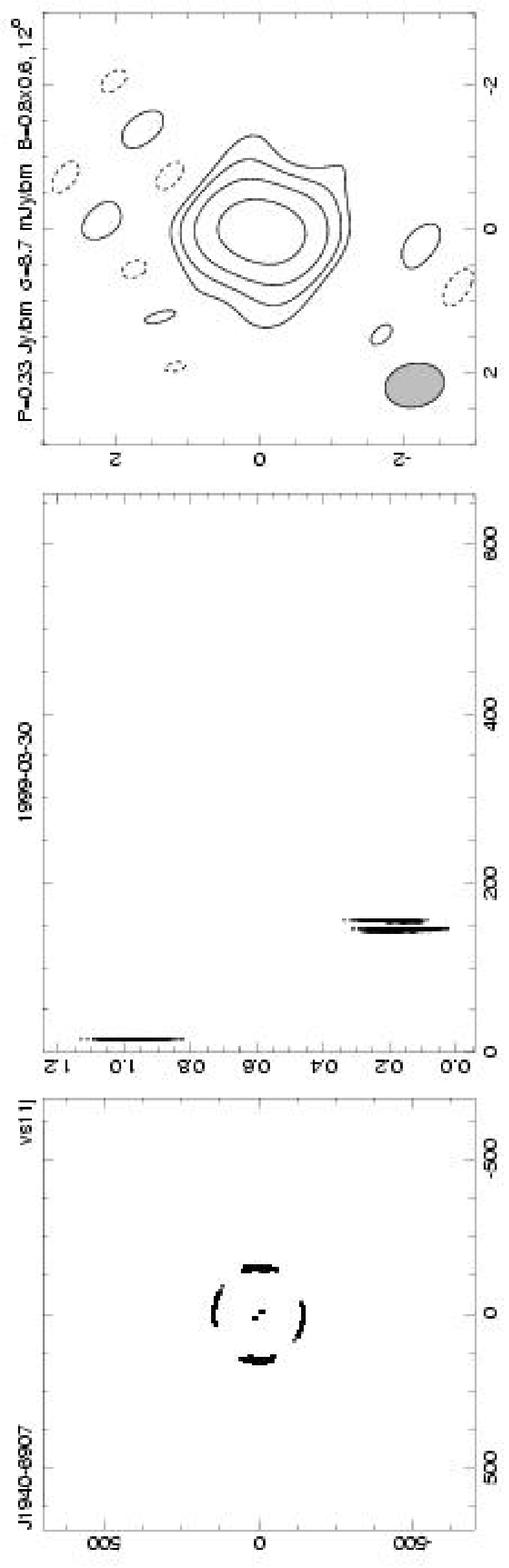}
\spfig{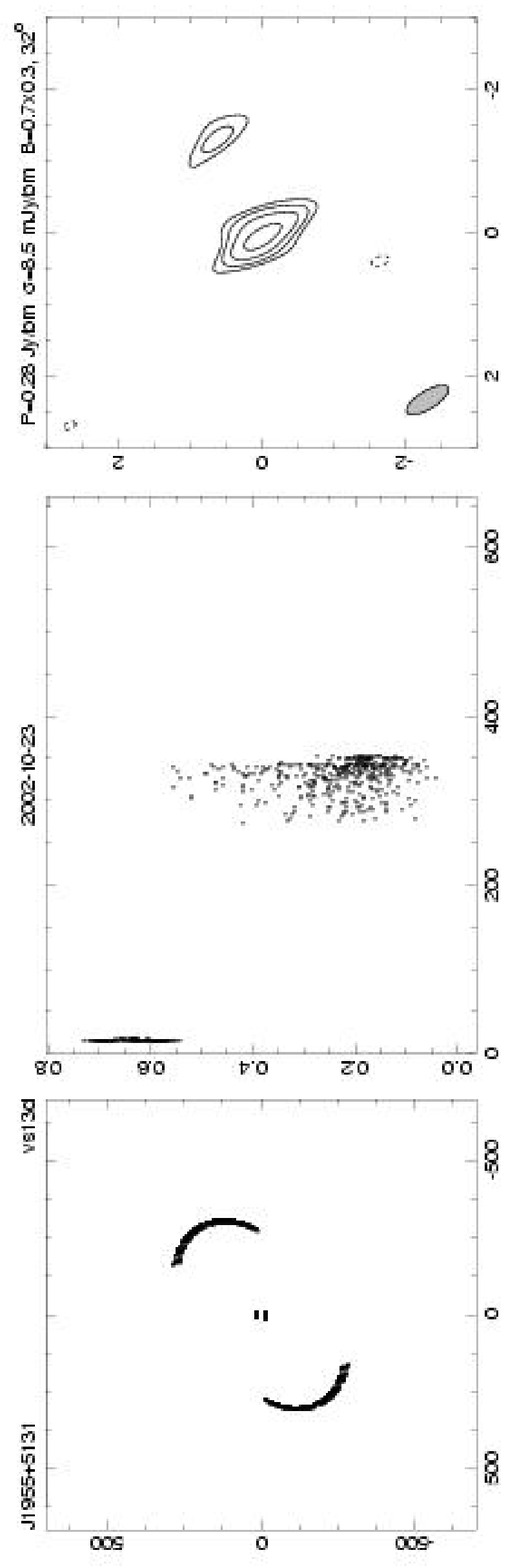}
\spfig{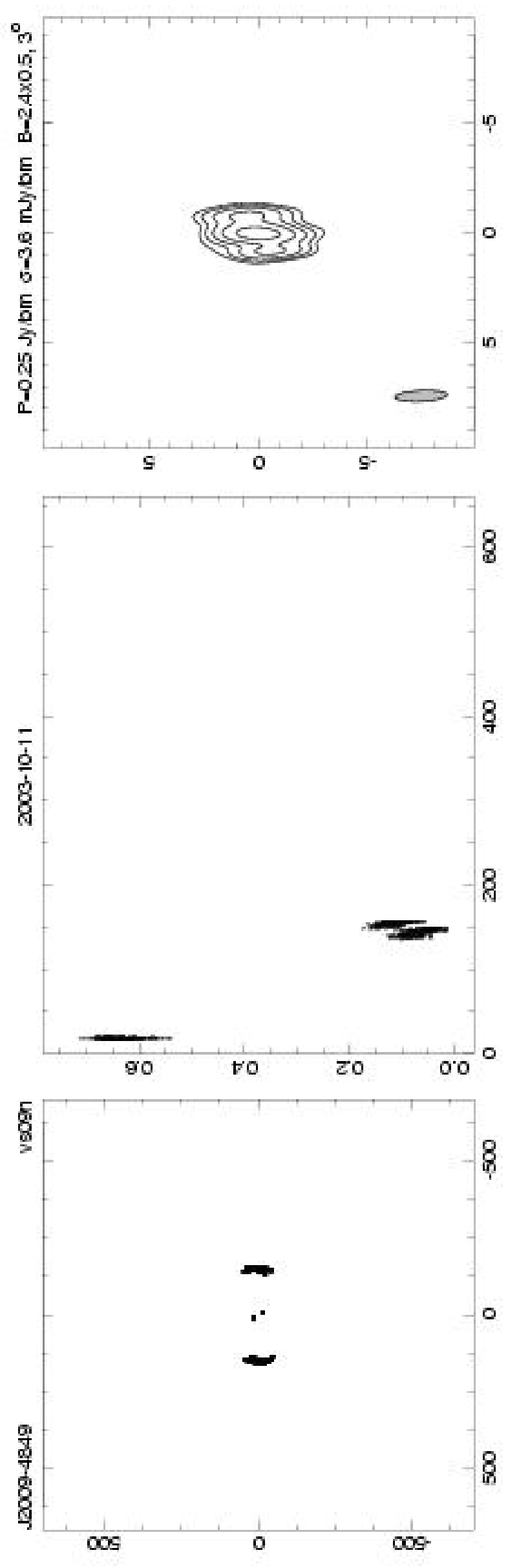}

{Fig. 1. -- {\em continued}}
\end{figure}
\clearpage
\begin{figure}
\spfig{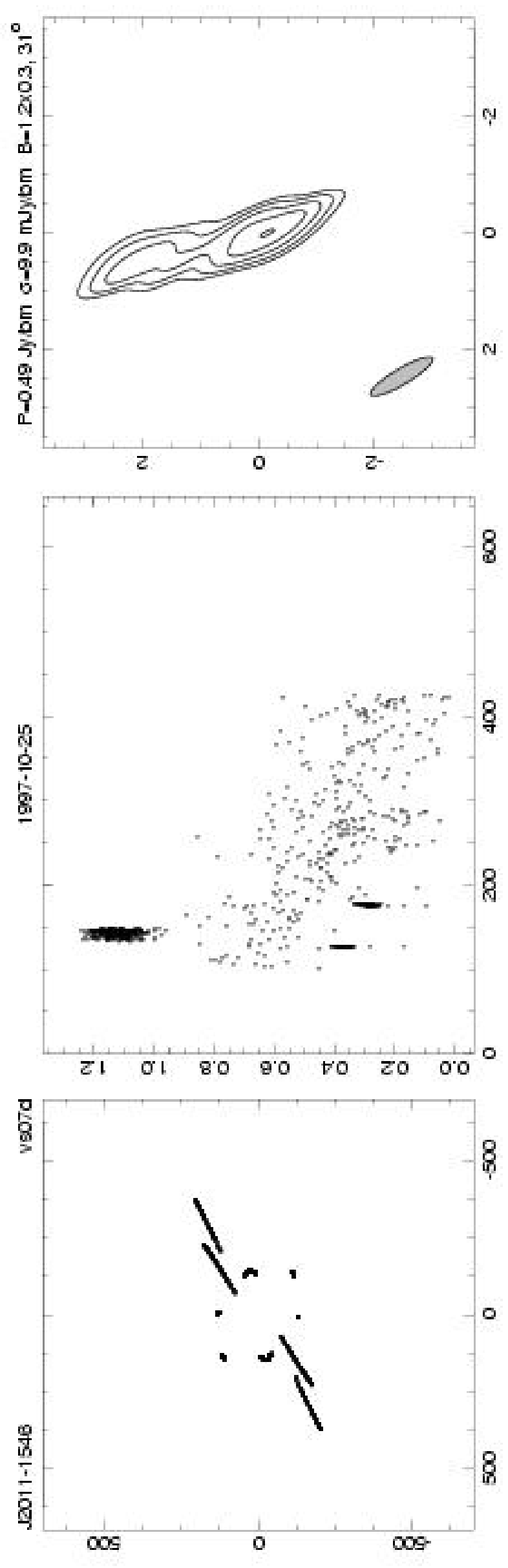}
\spfig{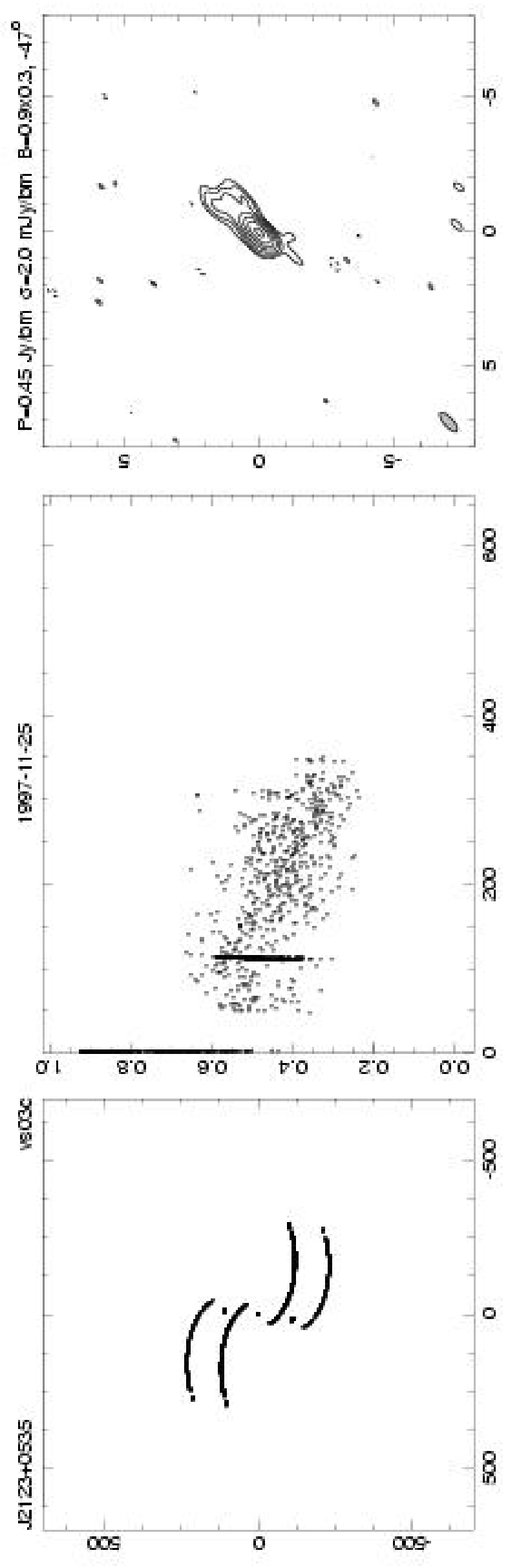}
\spfig{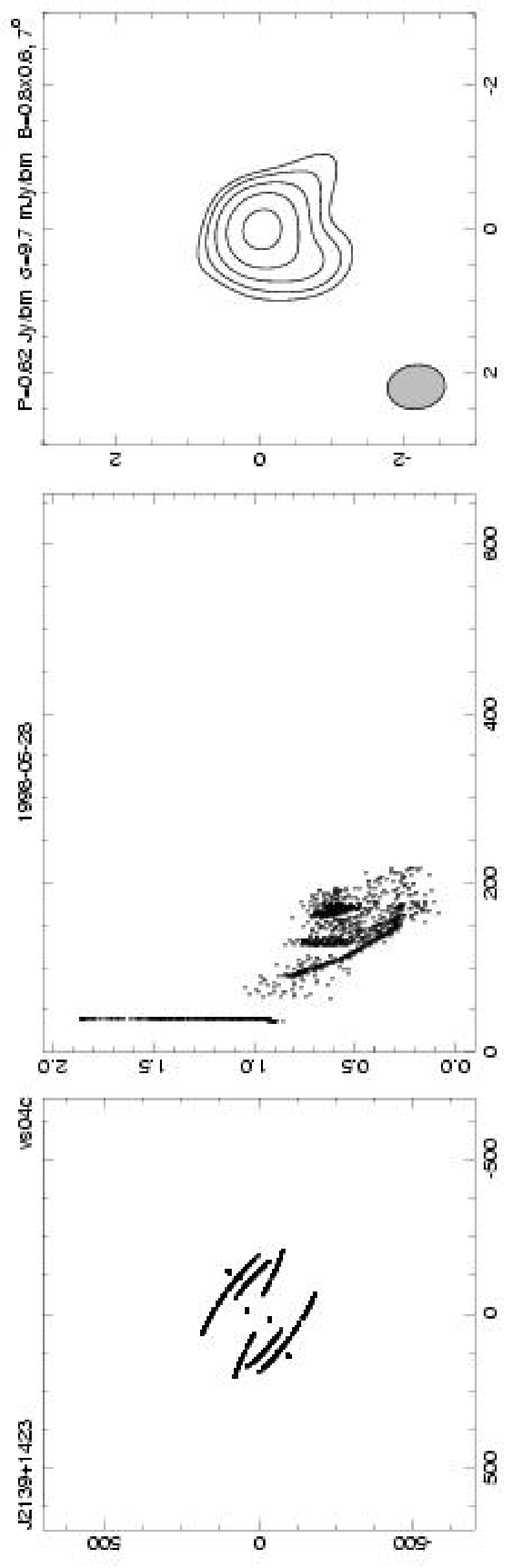}
\spfig{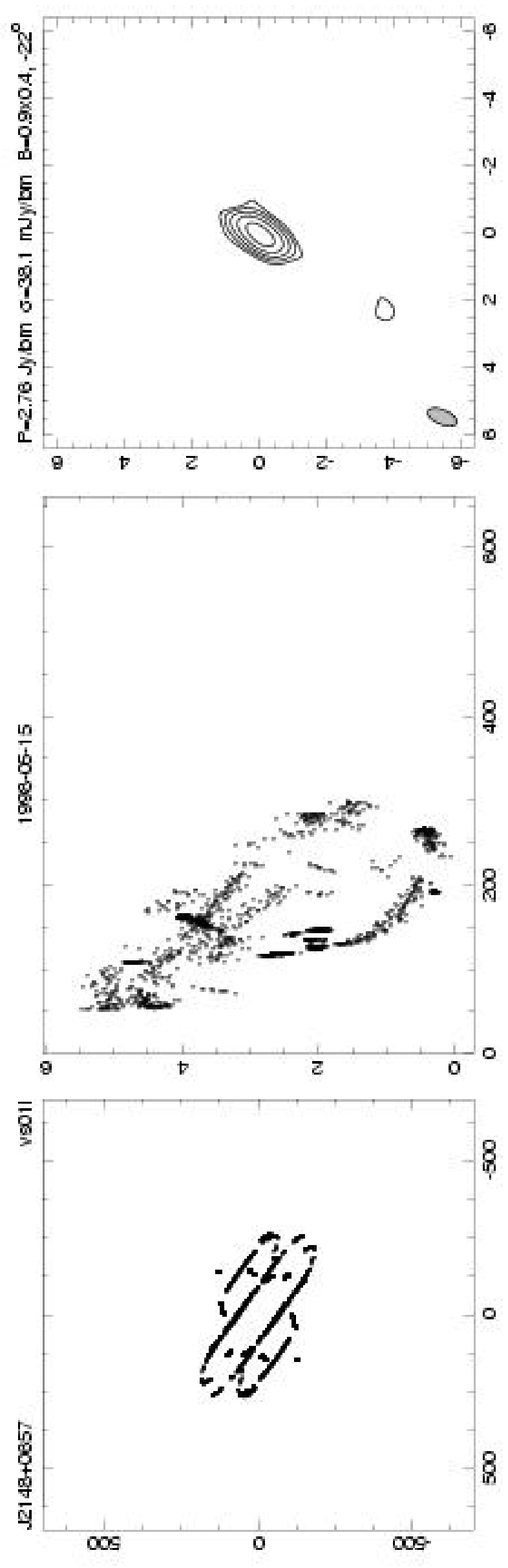}
\spfig{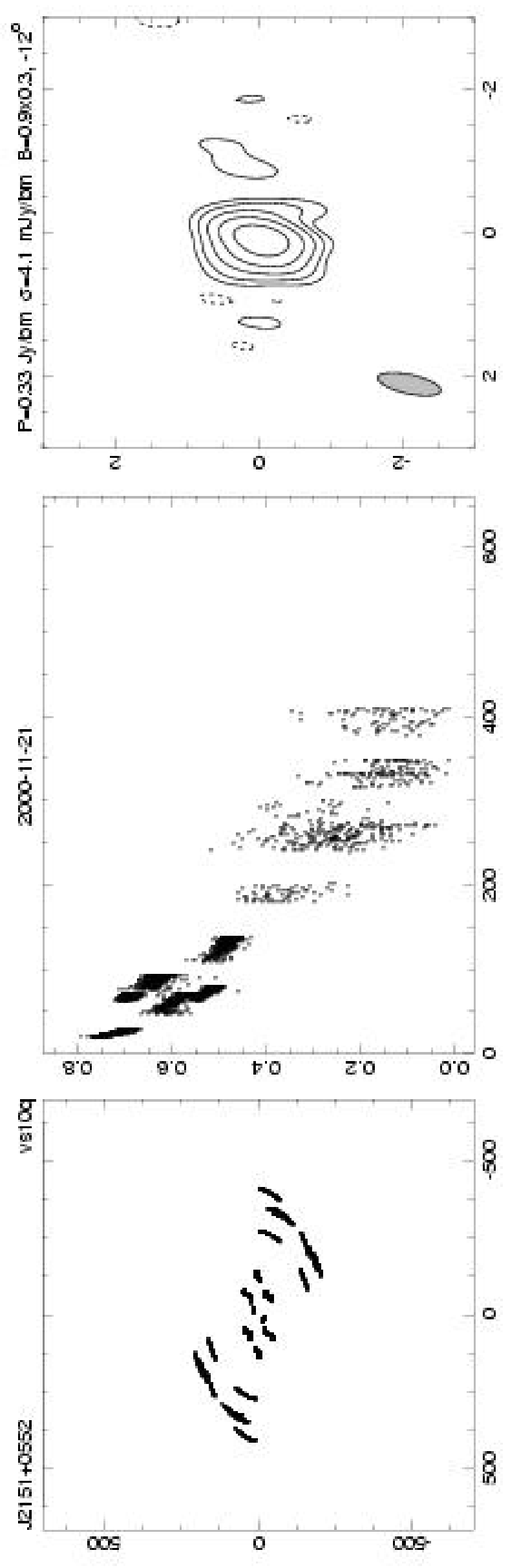}

{Fig. 1. -- {\em continued}}
\end{figure}
\clearpage
\begin{figure}
\spfig{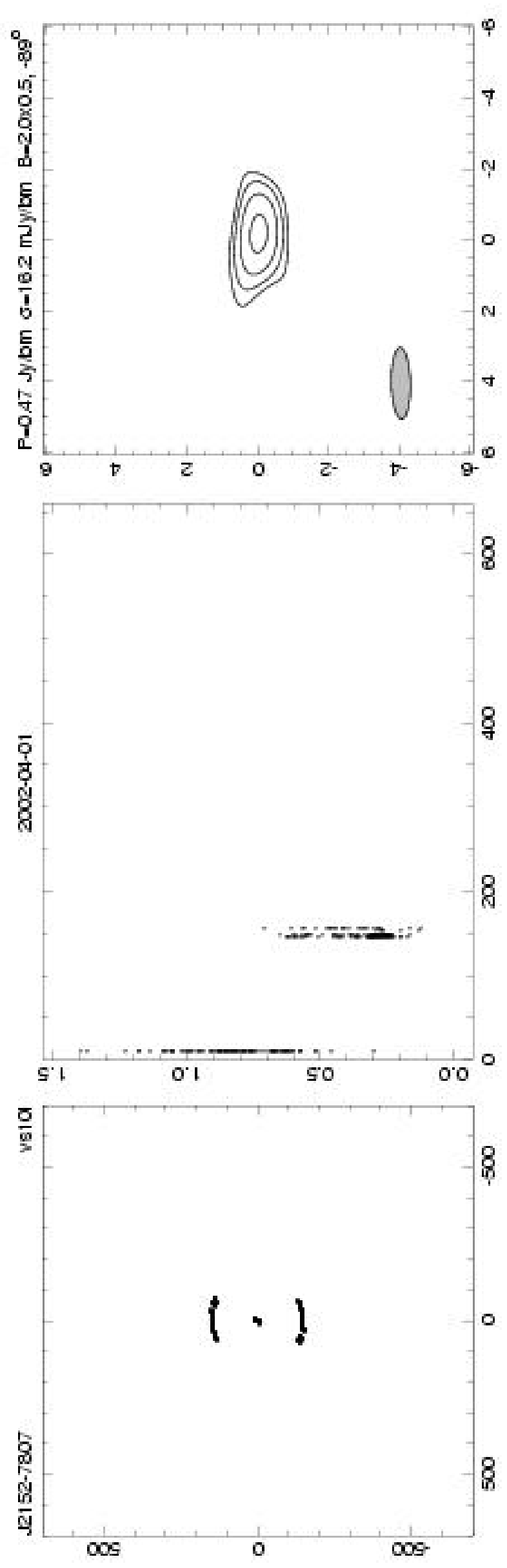}
\spfig{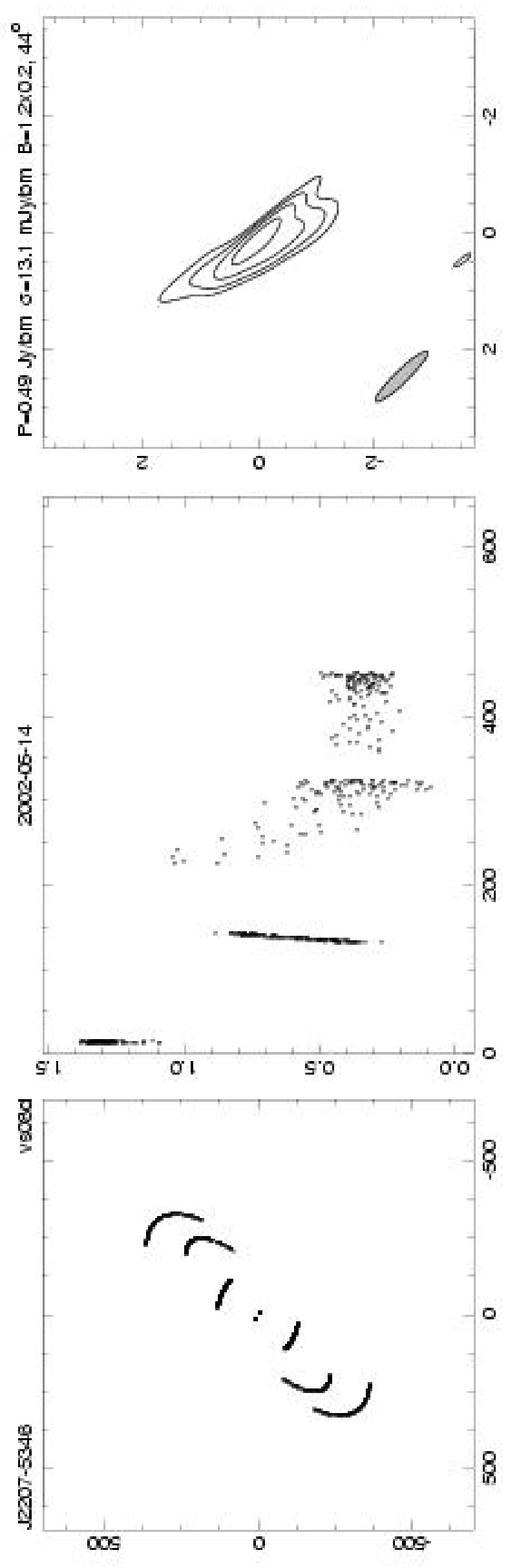}
\spfig{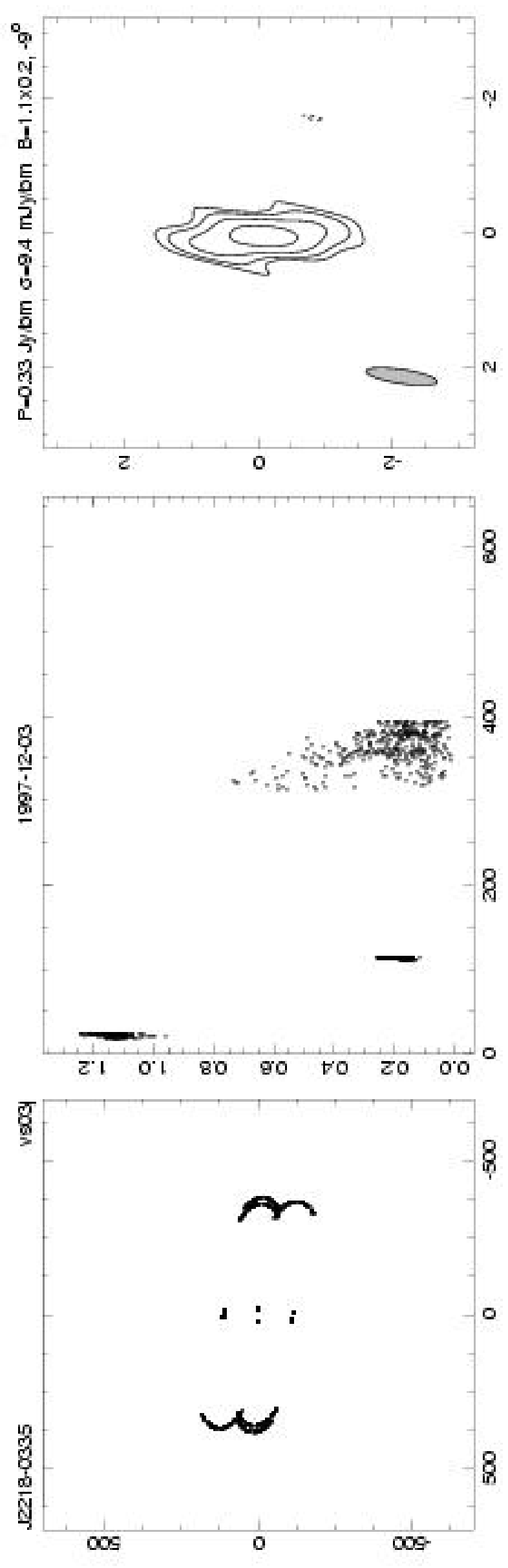}
\spfig{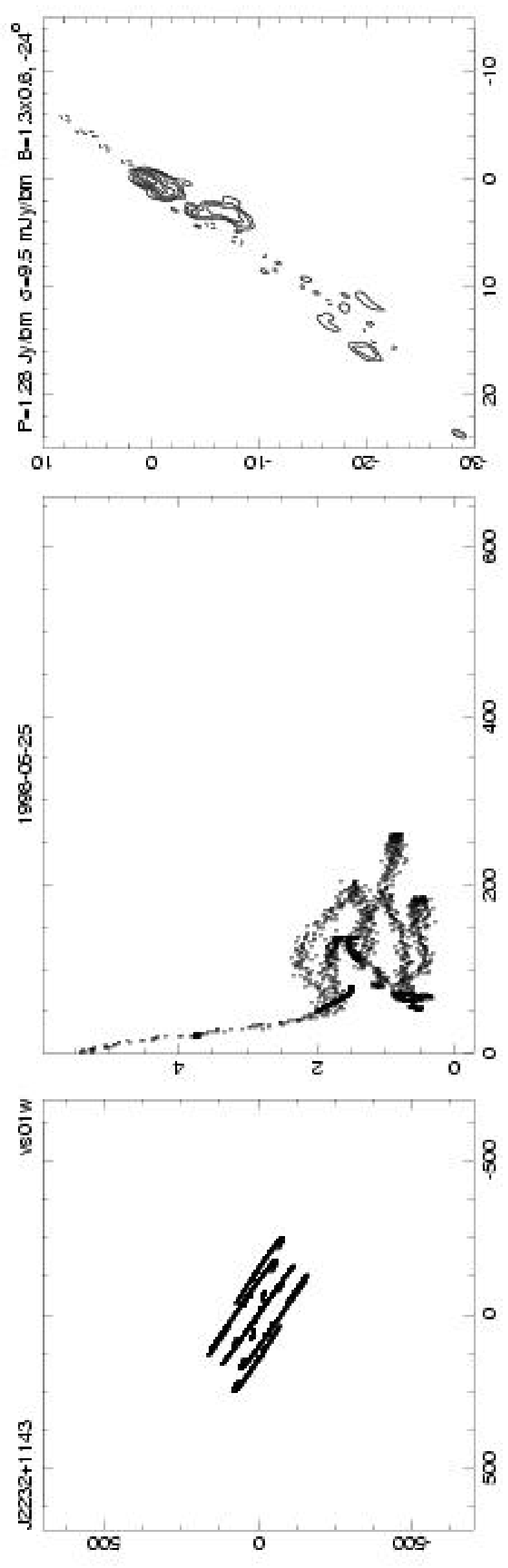} 
\spfig{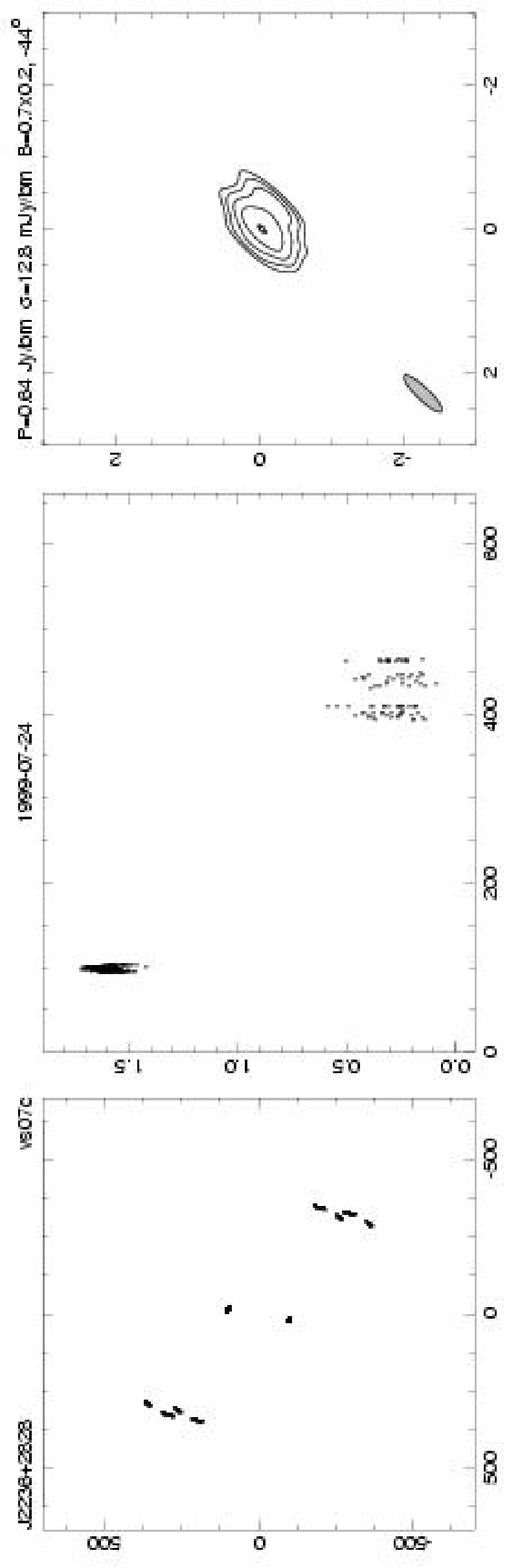}

{Fig. 1. -- {\em continued}}
\end{figure}
\clearpage
\begin{figure}
\spfig{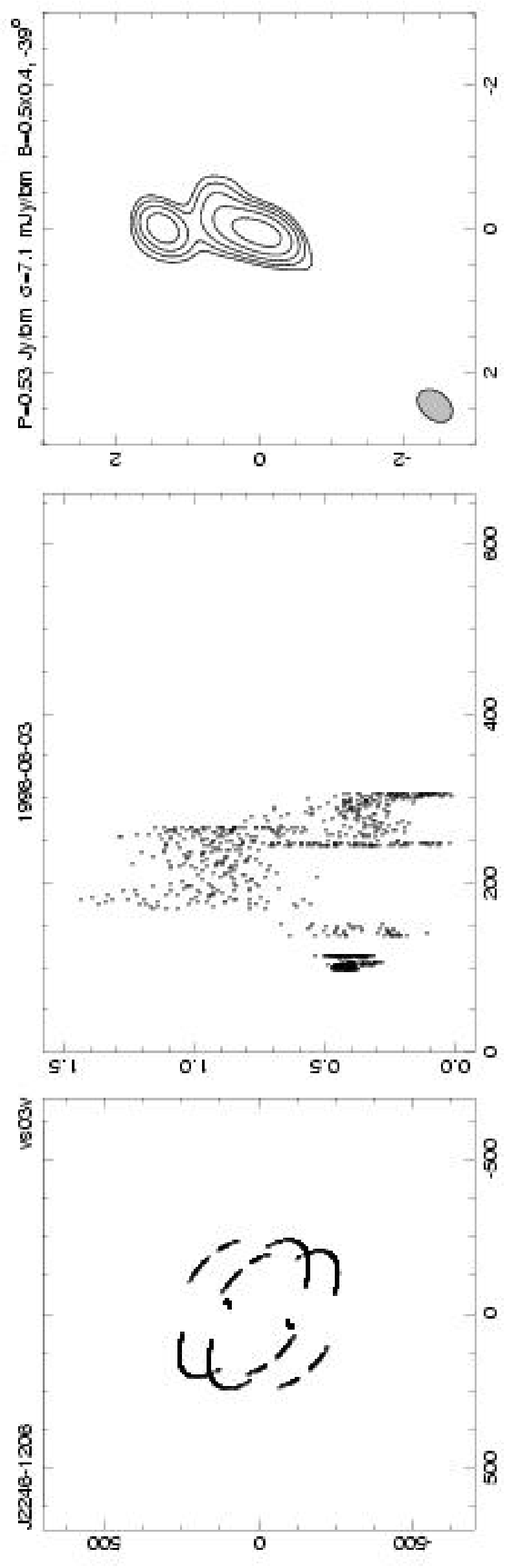}
\spfig{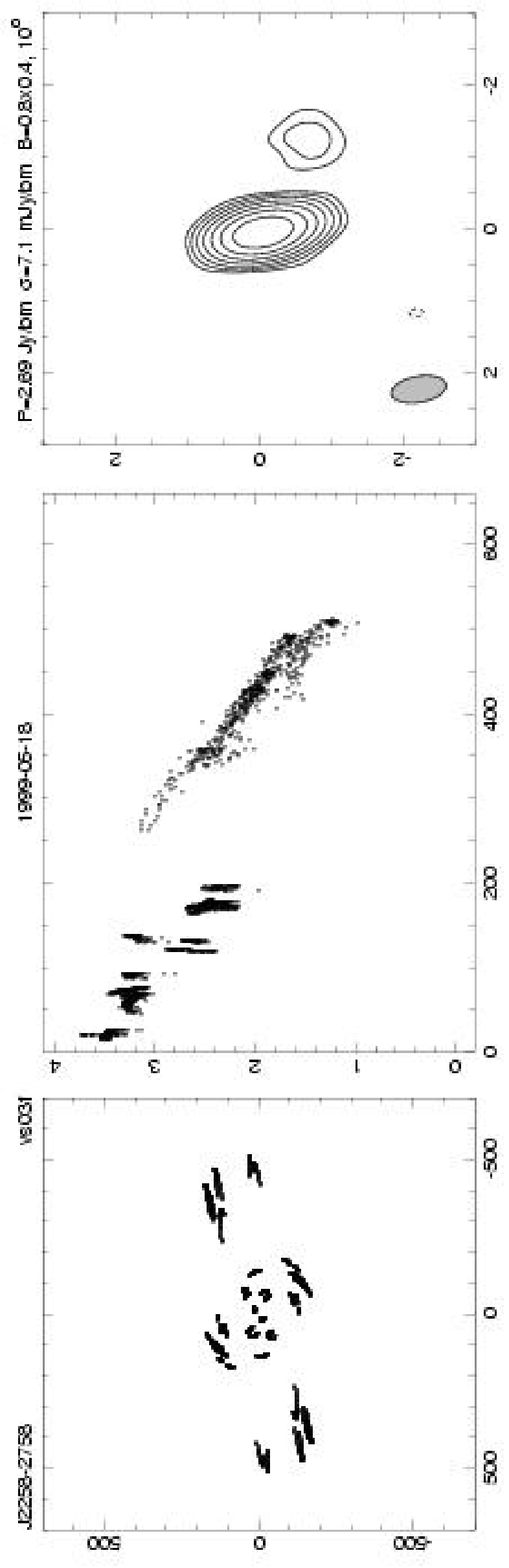}
\spfig{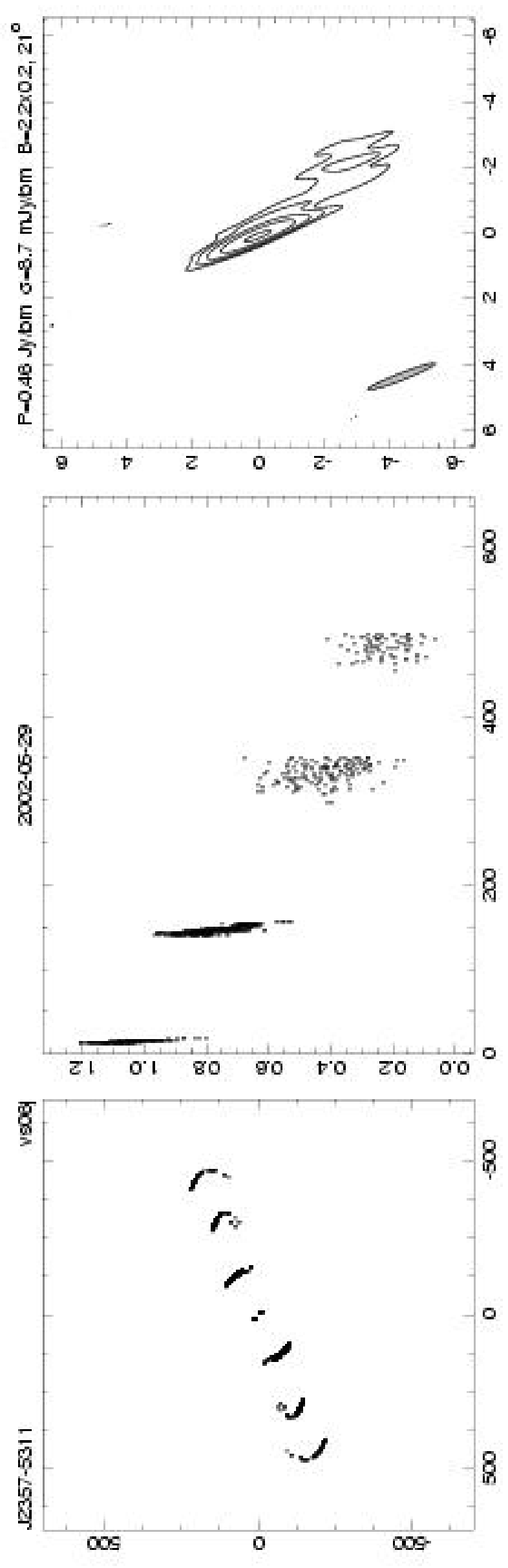}

{Fig. 1. -- {\em continued}}
\end{figure}
\clearpage

\clearpage

\begin{figure}
\plottwo{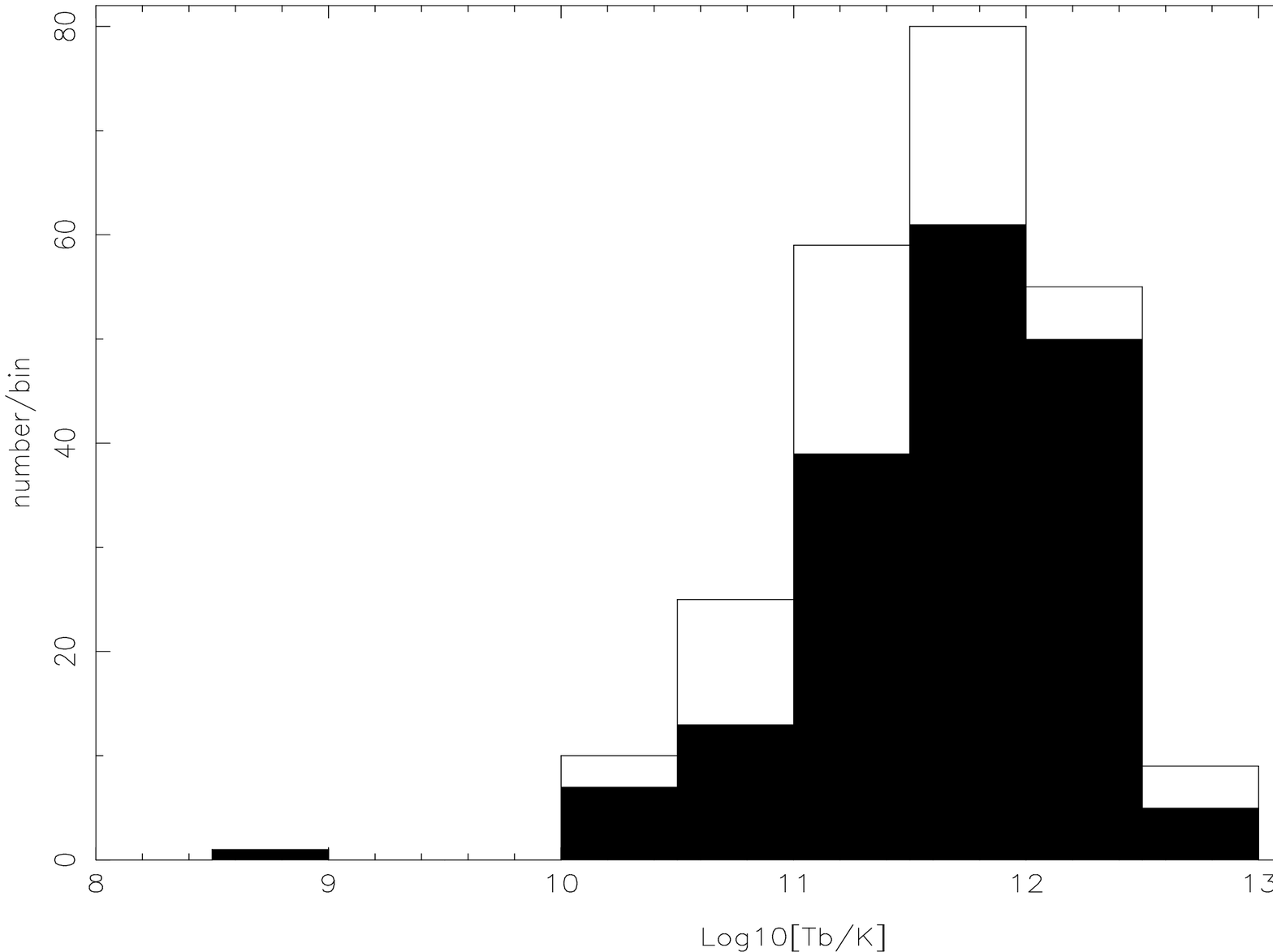}{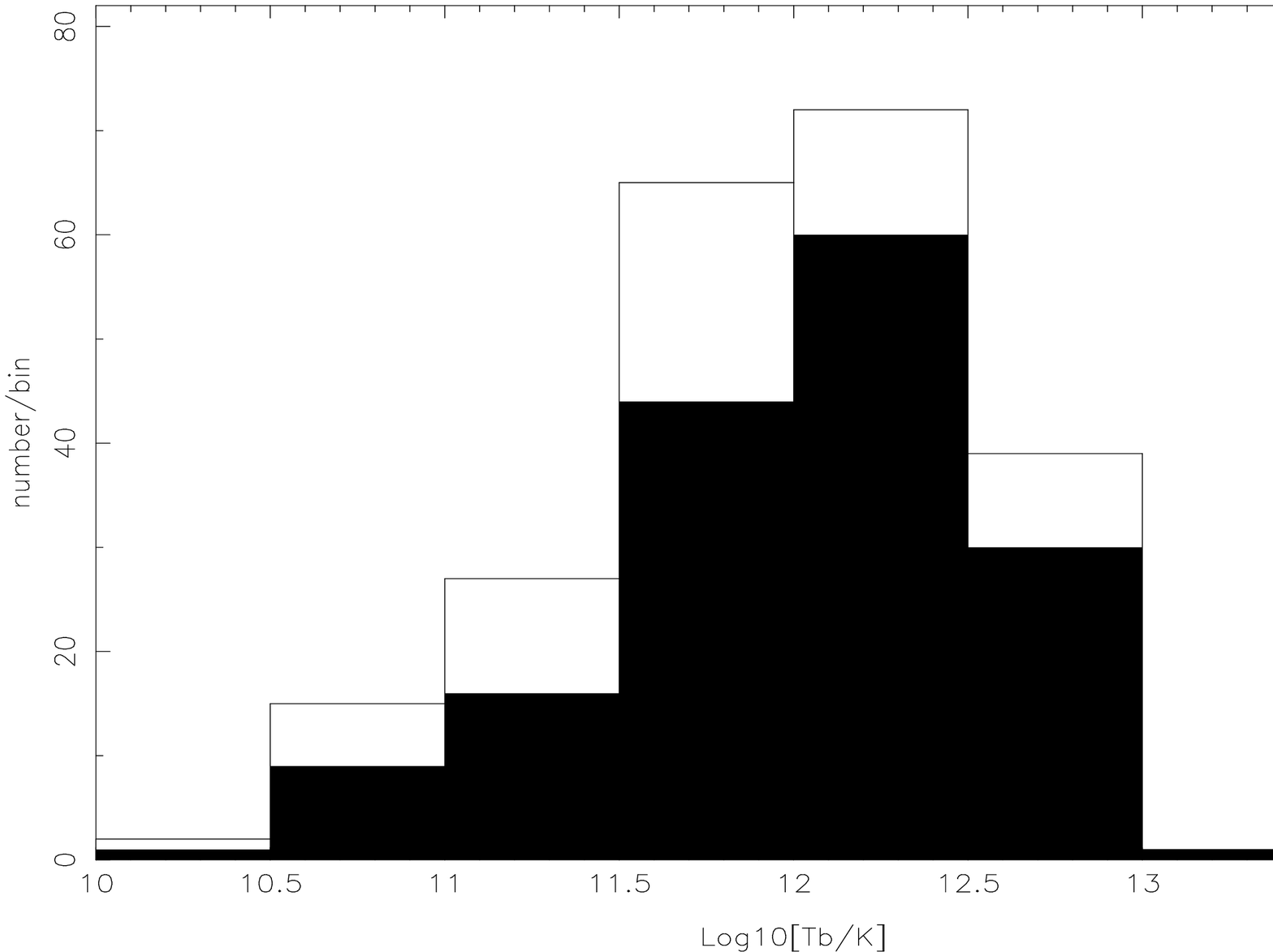}
\label{fig:hist}

{Fig. 2 -- Histogram of core brightness temperatures. The left
hand plot shows the core brightness temperature in the observer frame
for the 239 Survey sources with identified cores binned with
$\log_{10}(T_b)$ on the abscissa, and the number per bin as the
ordinate. We show the measured $T_b$ where the source
was resolvable with an filled bar, otherwise the lower limit to the
$T_b$ with open bar.
The right hand plot shows the $T_b$ in the source frame for
the subset of 222 sources which also have a measured redshift.}

\end{figure}


\end{document}